\expandafter\chardef\csname pre amssym.def at\endcsname=\the\catcode`\@
\catcode`\@=11

\def\undefine#1{\let#1\undefined}
\def\newsymbol#1#2#3#4#5{\let\next@\relax
 \ifnum#2=\@ne\let\next@\msafam@\else
 \ifnum#2=\tw@\let\next@\msbfam@\fi\fi
 \mathchardef#1="#3\next@#4#5}
\def\mathhexbox@#1#2#3{\relax
 \ifmmode\mathpalette{}{\m@th\mathchar"#1#2#3}%
 \else\leavevmode\hbox{$\m@th\mathchar"#1#2#3$}\fi}
\def\hexnumber@#1{\ifcase#1 0\or 1\or 2\or 3\or 4\or 5\or 6\or 7\or 8\or
 9\or A\or B\or C\or D\or E\or F\fi}

\font\tenmsa=msam10
\font\sevenmsa=msam7
\font\fivemsa=msam5
\newfam\msafam
\textfont\msafam=\tenmsa
\scriptfont\msafam=\sevenmsa
\scriptscriptfont\msafam=\fivemsa
\edef\msafam@{\hexnumber@\msafam}
\mathchardef\dabar@"0\msafam@39
\def\dashrightarrow{\mathrel{\dabar@\dabar@\mathchar"0\msafam@4B}}
\def\dashleftarrow{\mathrel{\mathchar"0\msafam@4C\dabar@\dabar@}}

\def\ulcorner{\delimiter"4\msafam@70\msafam@70 }
\def\urcorner{\delimiter"5\msafam@71\msafam@71 }
\def\llcorner{\delimiter"4\msafam@78\msafam@78 }
\def\lrcorner{\delimiter"5\msafam@79\msafam@79 }
\def\yen{{\mathhexbox@\msafam@55 }}
\def\checkmark{{\mathhexbox@\msafam@58 }}
\def\circledR{{\mathhexbox@\msafam@72 }}
\def\maltese{{\mathhexbox@\msafam@7A }}

\font\tenmsb=msbm10
\font\sevenmsb=msbm7
\font\fivemsb=msbm5
\newfam\msbfam
\textfont\msbfam=\tenmsb
\scriptfont\msbfam=\sevenmsb
\scriptscriptfont\msbfam=\fivemsb
\edef\msbfam@{\hexnumber@\msbfam}

\catcode`\@=\csname pre amssym.def at\endcsname

\expandafter\ifx\csname pre amssym.tex at\endcsname\relax \else \endinput\fi
\expandafter\chardef\csname pre amssym.tex at\endcsname=\the\catcode`\@
\catcode`\@=11
\newsymbol\boxdot 1200
\newsymbol\boxplus 1201
\newsymbol\boxtimes 1202
\newsymbol\square 1003
\newsymbol\blacksquare 1004
\newsymbol\centerdot 1205
\newsymbol\lozenge 1006
\newsymbol\blacklozenge 1007
\newsymbol\circlearrowright 1308
\newsymbol\circlearrowleft 1309
\undefine\rightleftharpoons
\newsymbol\rightleftharpoons 130A
\newsymbol\leftrightharpoons 130B
\newsymbol\boxminus 120C
\newsymbol\Vdash 130D
\newsymbol\Vvdash 130E
\newsymbol\vDash 130F
\newsymbol\twoheadrightarrow 1310
\newsymbol\twoheadleftarrow 1311
\newsymbol\leftleftarrows 1312
\newsymbol\rightrightarrows 1313
\newsymbol\upuparrows 1314
\newsymbol\downdownarrows 1315
\newsymbol\upharpoonright 1316
 
\newsymbol\downharpoonright 1317
\newsymbol\upharpoonleft 1318
\newsymbol\downharpoonleft 1319
\newsymbol\rightarrowtail 131A
\newsymbol\leftarrowtail 131B
\newsymbol\leftrightarrows 131C
\newsymbol\rightleftarrows 131D
\newsymbol\Lsh 131E
\newsymbol\Rsh 131F
\newsymbol\rightsquigarrow 1320
\newsymbol\leftrightsquigarrow 1321
\newsymbol\looparrowleft 1322
\newsymbol\looparrowright 1323
\newsymbol\circeq 1324
\newsymbol\succsim 1325
\newsymbol\gtrsim 1326
\newsymbol\gtrapprox 1327
\newsymbol\multimap 1328
\newsymbol\therefore 1329
\newsymbol\because 132A
\newsymbol\doteqdot 132B
 
\newsymbol\triangleq 132C
\newsymbol\precsim 132D
\newsymbol\lesssim 132E
\newsymbol\lessapprox 132F
\newsymbol\eqslantless 1330
\newsymbol\eqslantgtr 1331
\newsymbol\curlyeqprec 1332
\newsymbol\curlyeqsucc 1333
\newsymbol\preccurlyeq 1334
\newsymbol\leqq 1335
\newsymbol\leqslant 1336
\newsymbol\lessgtr 1337
\newsymbol\backprime 1038
\newsymbol\risingdotseq 133A
\newsymbol\fallingdotseq 133B
\newsymbol\succcurlyeq 133C
\newsymbol\geqq 133D
\newsymbol\geqslant 133E
\newsymbol\gtrless 133F
\newsymbol\sqsubset 1340
\newsymbol\sqsupset 1341
\newsymbol\vartriangleright 1342
\newsymbol\vartriangleleft 1343
\newsymbol\trianglerighteq 1344
\newsymbol\trianglelefteq 1345
\newsymbol\bigstar 1046
\newsymbol\between 1347
\newsymbol\blacktriangledown 1048
\newsymbol\blacktriangleright 1349
\newsymbol\blacktriangleleft 134A
\newsymbol\vartriangle 134D
\newsymbol\blacktriangle 104E
\newsymbol\triangledown 104F
\newsymbol\eqcirc 1350
\newsymbol\lesseqgtr 1351
\newsymbol\gtreqless 1352
\newsymbol\lesseqqgtr 1353
\newsymbol\gtreqqless 1354
\newsymbol\Rrightarrow 1356
\newsymbol\Lleftarrow 1357
\newsymbol\veebar 1259
\newsymbol\barwedge 125A
\newsymbol\doublebarwedge 125B
\undefine\angle
\newsymbol\angle 105C
\newsymbol\measuredangle 105D
\newsymbol\sphericalangle 105E
\newsymbol\varpropto 135F
\newsymbol\smallsmile 1360
\newsymbol\smallfrown 1361
\newsymbol\Subset 1362
\newsymbol\Supset 1363
\newsymbol\Cup 1264
 
\newsymbol\Cap 1265
 
\newsymbol\curlywedge 1266
\newsymbol\curlyvee 1267
\newsymbol\leftthreetimes 1268
\newsymbol\rightthreetimes 1269
\newsymbol\subseteqq 136A
\newsymbol\supseteqq 136B
\newsymbol\bumpeq 136C
\newsymbol\Bumpeq 136D
\newsymbol\lll 136E
 
\newsymbol\ggg 136F
 
\newsymbol\circledS 1073
\newsymbol\pitchfork 1374
\newsymbol\dotplus 1275
\newsymbol\backsim 1376
\newsymbol\backsimeq 1377
\newsymbol\complement 107B
\newsymbol\intercal 127C
\newsymbol\circledcirc 127D
\newsymbol\circledast 127E
\newsymbol\circleddash 127F
\newsymbol\lvertneqq 2300
\newsymbol\gvertneqq 2301
\newsymbol\nleq 2302
\newsymbol\ngeq 2303
\newsymbol\nless 2304
\newsymbol\ngtr 2305
\newsymbol\nprec 2306
\newsymbol\nsucc 2307
\newsymbol\lneqq 2308
\newsymbol\gneqq 2309
\newsymbol\nleqslant 230A
\newsymbol\ngeqslant 230B
\newsymbol\lneq 230C
\newsymbol\gneq 230D
\newsymbol\npreceq 230E
\newsymbol\nsucceq 230F
\newsymbol\precnsim 2310
\newsymbol\succnsim 2311
\newsymbol\lnsim 2312
\newsymbol\gnsim 2313
\newsymbol\nleqq 2314
\newsymbol\ngeqq 2315
\newsymbol\precneqq 2316
\newsymbol\succneqq 2317
\newsymbol\precnapprox 2318
\newsymbol\succnapprox 2319
\newsymbol\lnapprox 231A
\newsymbol\gnapprox 231B
\newsymbol\nsim 231C
\newsymbol\ncong 231D
\newsymbol\diagup 231E
\newsymbol\diagdown 231F
\newsymbol\varsubsetneq 2320
\newsymbol\varsupsetneq 2321
\newsymbol\nsubseteqq 2322
\newsymbol\nsupseteqq 2323
\newsymbol\subsetneqq 2324
\newsymbol\supsetneqq 2325
\newsymbol\varsubsetneqq 2326
\newsymbol\varsupsetneqq 2327
\newsymbol\subsetneq 2328
\newsymbol\supsetneq 2329
\newsymbol\nsubseteq 232A
\newsymbol\nsupseteq 232B
\newsymbol\nparallel 232C
\newsymbol\nmid 232D
\newsymbol\nshortmid 232E
\newsymbol\nshortparallel 232F
\newsymbol\nvdash 2330
\newsymbol\nVdash 2331
\newsymbol\nvDash 2332
\newsymbol\nVDash 2333
\newsymbol\ntrianglerighteq 2334
\newsymbol\ntrianglelefteq 2335
\newsymbol\ntriangleleft 2336
\newsymbol\ntriangleright 2337
\newsymbol\nleftarrow 2338
\newsymbol\nrightarrow 2339
\newsymbol\nLeftarrow 233A
\newsymbol\nRightarrow 233B
\newsymbol\nLeftrightarrow 233C
\newsymbol\nleftrightarrow 233D
\newsymbol\divideontimes 223E
\newsymbol\varnothing 203F
\newsymbol\nexists 2040
\newsymbol\Finv 2060
\newsymbol\Game 2061
\newsymbol\mho 2066
\newsymbol\eth 2067
\newsymbol\eqsim 2368
\newsymbol\beth 2069
\newsymbol\gimel 206A
\newsymbol\daleth 206B
\newsymbol\lessdot 236C
\newsymbol\gtrdot 236D
\newsymbol\ltimes 226E
\newsymbol\rtimes 226F
\newsymbol\shortmid 2370
\newsymbol\shortparallel 2371
\newsymbol\smallsetminus 2272
\newsymbol\thicksim 2373
\newsymbol\thickapprox 2374
\newsymbol\approxeq 2375
\newsymbol\succapprox 2376
\newsymbol\precapprox 2377
\newsymbol\curvearrowleft 2378
\newsymbol\curvearrowright 2379
\newsymbol\digamma 207A
\newsymbol\varkappa 207B
\newsymbol\Bbbk 207C
\newsymbol\hslash 207D
\undefine\hbar
\newsymbol\hbar 207E
\newsymbol\backepsilon 237F
\catcode`\@=\csname pre amssym.tex at\endcsname

\magnification=1200
\hsize=468truept
\vsize=646truept
\voffset=-10pt
\parskip=4pt
\baselineskip=14truept
\count0=1

\dimen100=\hsize

\def\leftill#1#2#3#4{
\medskip
\line{$
\vcenter{
\hsize = #1truept \hrule\hbox{\vrule\hbox to  \hsize{\hss \vbox{\vskip#2truept
\hbox{{\copy100 \the\count105}: #3}\vskip2truept}\hss }
\vrule}\hrule}
\dimen110=\dimen100
\advance\dimen110 by -36truept
\advance\dimen110 by -#1truept
\hss \vcenter{\hsize = \dimen110
\medskip
\noindent { #4\par\medskip}}$}
\advance\count105 by 1
}
\def\rightill#1#2#3#4{
\medskip
\line{
\dimen110=\dimen100
\advance\dimen110 by -36truept
\advance\dimen110 by -#1truept
$\vcenter{\hsize = \dimen110
\medskip
\noindent { #4\par\medskip}}
\hss \vcenter{
\hsize = #1truept \hrule\hbox{\vrule\hbox to  \hsize{\hss \vbox{\vskip#2truept
\hbox{{\copy100 \the\count105}: #3}\vskip2truept}\hss }
\vrule}\hrule}
$}
\advance\count105 by 1
}
\def\midill#1#2#3{\medskip
\line{$\hss
\vcenter{
\hsize = #1truept \hrule\hbox{\vrule\hbox to  \hsize{\hss \vbox{\vskip#2truept
\hbox{{\copy100 \the\count105}: #3}\vskip2truept}\hss }
\vrule}\hrule}
\dimen110=\dimen100
\advance\dimen110 by -36truept
\advance\dimen110 by -#1truept
\hss $}
\advance\count105 by 1
}
\def\insectnum{\copy110\the\count120
\advance\count120 by 1
}

\font\ninerm=cmr9

\font\tenrm=cmr10 at 10pt

\font\sc=cmcsc10

\def\msb{\fam\msbfam\tenmsb}

\def\bba{{\msb A}}
\def\bbb{{\msb B}}
\def\bbc{{\msb C}}
\def\bbd{{\msb D}}

\def\bbh{{\msb H}}
\def\bbi{{\msb I}}

\def\bbo{{\msb O}}
\def\bbp{{\msb P}}
\def\bbq{{\msb Q}}
\def\bbr{{\msb R}}
\def\bbs{{\msb S}}
\def\bbt{{\msb T}}

\def\bbw{{\msb W}}

\def\bbz{{\msb Z}}
\def\grD{\Delta}

\def\grG{\Gamma}

\def\grL{\Lambda}
\def\grO{\Omega}

\def\grS{\Sigma}

\def\grU{\Upsilon}

\def\gra{\alpha}
\def\grb{\beta}

\def\grd{\delta}
\def\gre{\epsilon}

\def\grg{\gamma}
\def\gri{\iota}
\def\grk{\kappa}
\def\grl{\lambda}

\def\gro{\omega}

\def\grr{\rho}
\def\grs{\sigma}
\def\grt{\tau}

\font\svtnrm=cmr17

\font\aa=eufm10

\def\gs{{\Got s}}
\def\gg{{\Got g}}
\def\gp{{\Got p}}
\def\go{{\Got o}}

\def\gm{{\Got m}}

\def\gi{{\Got i}}
\def\gc{{\Got c}}
\def\gt{{\Got t}}
\def\Got#1{\hbox{\aa#1}}

\def\gsp1{{\Got s}{\Got p}(1)}

\def\bfu{{\bf u}}

\def\bfp{{\bf p}}

\def\bfb{{\bf b}}
\def\mh-1{\hat{\mu}^{-1}(0)}
\def\n-1c{\nu^{-1}(c)}
\def\m-1{\mu^{-1}(0)}
\def\p-1{\pi^{-1}}
\def\p'-1{\prime{\pi}^{-1}}

\def\cala{{\cal A}}

\def\calo{{\cal O}}

\def\cale{{\cal E}}
\def\calf{{\cal F}}
\def\calg{{\cal G}}
\def\calh{{\cal H}}
\def\cali{{\cal I}}

\def\call{{\cal L}}
\def\calm{{\cal M}}

\def\calr{{\cal R}}
\def\cals{{\cal S}}

\def\calv{{\cal V}}
\def\calw{{\cal W}}

\def\calz{{\cal Z}}

\def\la#1{\hbox to #1pc{\leftarrowfill}}
\def\ra#1{\hbox to #1pc{\rightarrowfill}}

\def\fract#1#2{\raise4pt\hbox{$ #1 \atop #2 $}}
\def\decdnar#1{\phantom{\hbox{$\scriptstyle{#1}$}}
\left\downarrow\vbox{\vskip15pt\hbox{$\scriptstyle{#1}$}}\right.}

\def\bowtie{\hbox to 1pt{\hss}\raise.66pt\hbox{$\scriptstyle{>}$}
\kern-4.9pt\triangleleft}
\def\hsmash{\triangleright\kern-4.4pt\raise.66pt\hbox{$\scriptstyle{<}$}}
\def\boxit#1{\vbox{\hrule\hbox{\vrule\kern3pt
\vbox{\kern3pt#1\kern3pt}\kern3pt\vrule}\hrule}}

\def\za{\vrule height6pt width4pt depth1pt}

\font\aa=eufm10

\def\Got#1{\hbox{\aa#1}}

\def\bfa{{\bf a}}
\def\bfb{{\bf b}}

\def\bfp{{\bf p}}
\def\bfq{{\bf q}}

\def\bfu{{\bf u}}

\def\cala{{\cal A}}

\def\calo{{\cal O}}

\def\cale{{\cal E}}
\def\calf{{\cal F}}
\def\calg{{\cal G}}
\def\calh{{\cal H}}
\def\cali{{\cal I}}

\def\call{{\cal L}}
\def\calm{{\cal M}}

\def\calr{{\cal R}}
\def\cals{{\cal S}}

\def\calv{{\cal V}}
\def\calw{{\cal W}}

\def\calz{{\cal Z}}

\def\gc{{\Got c}}

\def\gg{{\Got g}}

\def\gi{{\Got i}}

\def\gm{{\Got m}}

\def\go{{\Got o}}
\def\gp{{\Got p}}

\def\gs{{\Got s}}
\def\gt{{\Got t}}

\catcode`!=11 
 
  

\def\PiC{P\kern-.12em\lower.5ex\hbox{I}\kern-.075emC}
\def\PiCTeX{\PiC\kern-.11em\TeX}

\def\!ifnextchar#1#2#3{%
  \let\!testchar=#1%
  \def\!first{#2}%
  \def\!second{#3}%
  \futurelet\!nextchar\!testnext}
\def\!testnext{%
  \ifx \!nextchar \!spacetoken 
    \let\!next=\!skipspacetestagain
  \else
    \ifx \!nextchar \!testchar
      \let\!next=\!first
    \else 
      \let\!next=\!second 
    \fi 
  \fi
  \!next}
\def\\{\!skipspacetestagain} 
  \expandafter\def\\ {\futurelet\!nextchar\!testnext} 
\def\\{\let\!spacetoken= } \\  

\def\!tfor#1:=#2\do#3{%
  \edef\!fortemp{#2}%
  \ifx\!fortemp\!empty 
    \else
    \!tforloop#2\!nil\!nil\!!#1{#3}%
  \fi}
\def\!tforloop#1#2\!!#3#4{%
  \def#3{#1}%
  \ifx #3\!nnil
    \let\!nextwhile=\!fornoop
  \else
    #4\relax
    \let\!nextwhile=\!tforloop
  \fi 
  \!nextwhile#2\!!#3{#4}}

\def\!etfor#1:=#2\do#3{%
  \def\!!tfor{\!tfor#1:=}%
  \edef\!!!tfor{#2}%
  \expandafter\!!tfor\!!!tfor\do{#3}}

\def\!cfor#1:=#2\do#3{%
  \edef\!fortemp{#2}%
  \ifx\!fortemp\!empty 
  \else
    \!cforloop#2,\!nil,\!nil\!!#1{#3}%
  \fi}
\def\!cforloop#1,#2\!!#3#4{%
  \def#3{#1}%
  \ifx #3\!nnil
    \let\!nextwhile=\!fornoop 
  \else
    #4\relax
    \let\!nextwhile=\!cforloop
  \fi
  \!nextwhile#2\!!#3{#4}}

\def\!ecfor#1:=#2\do#3{%
  \def\!!cfor{\!cfor#1:=}%
  \edef\!!!cfor{#2}%
  \expandafter\!!cfor\!!!cfor\do{#3}}

\def\!empty{}
\def\!nnil{\!nil}
\def\!fornoop#1\!!#2#3{}

\def\!ifempty#1#2#3{%
  \edef\!emptyarg{#1}%
  \ifx\!emptyarg\!empty
    #2%
  \else
    #3%
  \fi}
 
\def\!getnext#1\from#2{%
  \expandafter\!gnext#2\!#1#2}%
\def\!gnext\\#1#2\!#3#4{%
  \def#3{#1}%
  \def#4{#2\\{#1}}%
  \ignorespaces}

%
\def\!getnextvalueof#1\from#2{%
  \expandafter\!gnextv#2\!#1#2}%
\def\!gnextv\\#1#2\!#3#4{%
  #3=#1%
  \def#4{#2\\{#1}}%
  \ignorespaces}

\def\!copylist#1\to#2{%
  \expandafter\!!copylist#1\!#2}
\def\!!copylist#1\!#2{%
  \def#2{#1}\ignorespaces}

\def\!wlet#1=#2{%
  \let#1=#2 
  \wlog{\string#1=\string#2}}
 
\def\!listaddon#1#2{%
  \expandafter\!!listaddon#2\!{#1}#2}
\def\!!listaddon#1\!#2#3{%
  \def#3{#1\\#2}}
 

\def\!rightappend#1\withCS#2\to#3{\expandafter\!!rightappend#3\!#2{#1}#3}
\def\!!rightappend#1\!#2#3#4{\def#4{#1#2{#3}}}

\def\!leftappend#1\withCS#2\to#3{\expandafter\!!leftappend#3\!#2{#1}#3}
\def\!!leftappend#1\!#2#3#4{\def#4{#2{#3}#1}}

\def\!lop#1\to#2{\expandafter\!!lop#1\!#1#2}
\def\!!lop\\#1#2\!#3#4{\def#4{#1}\def#3{#2}}



\def\!loop#1\repeat{\def\!body{#1}\!iterate}
\def\!iterate{\!body\let\!next=\!iterate\else\let\!next=\relax\fi\!next}
 
\def\!!loop#1\repeat{\def\!!body{#1}\!!iterate}
\def\!!iterate{\!!body\let\!!next=\!!iterate\else\let\!!next=\relax\fi\!!next}
 
\def\!removept#1#2{\edef#2{\expandafter\!!removePT\the#1}}
{\catcode`p=12 \catcode`t=12 \gdef\!!removePT#1pt{#1}}

\def\placevalueinpts of <#1> in #2 {%
  \!removept{#1}{#2}}
 
\def\!mlap#1{\hbox to 0pt{\hss#1\hss}}
\def\!vmlap#1{\vbox to 0pt{\vss#1\vss}}
 
\def\!not#1{%
  #1\relax
    \!switchfalse
  \else
    \!switchtrue
  \fi
  \if!switch
  \ignorespaces}


 

\let\!!!wlog=\wlog              
\def\wlog#1{}    

\newdimen\headingtoplotskip     
\newdimen\linethickness         
\newdimen\longticklength        
\newdimen\plotsymbolspacing     
\newdimen\shortticklength       
\newdimen\stackleading          
\newdimen\tickstovaluesleading  
\newdimen\totalarclength        
\newdimen\valuestolabelleading  

\newbox\!boxA                   
\newbox\!boxB                   
\newbox\!picbox                 
\newbox\!plotsymbol             
\newbox\!putobject              
\newbox\!shadesymbol            

\newcount\!countA               
\newcount\!countB               
\newcount\!countC               
\newcount\!countD               
\newcount\!countE               
\newcount\!countF               
\newcount\!countG               
\newcount\!fiftypt              
\newcount\!intervalno           
\newcount\!npoints              
\newcount\!nsegments            
\newcount\!ntemp                
\newcount\!parity               
\newcount\!scalefactor          
\newcount\!tfs                  
\newcount\!tickcase             

\newdimen\!Xleft                
\newdimen\!Xright               
\newdimen\!Xsave                
\newdimen\!Ybot                 
\newdimen\!Ysave                
\newdimen\!Ytop                 
\newdimen\!angle                
\newdimen\!arclength            
\newdimen\!areabloc             
\newdimen\!arealloc             
\newdimen\!arearloc             
\newdimen\!areatloc             
\newdimen\!bshrinkage           
\newdimen\!checkbot             
\newdimen\!checkleft            
\newdimen\!checkright           
\newdimen\!checktop             
\newdimen\!dimenA               
\newdimen\!dimenB               
\newdimen\!dimenC               
\newdimen\!dimenD               
\newdimen\!dimenE               
\newdimen\!dimenF               
\newdimen\!dimenG               
\newdimen\!dimenH               
\newdimen\!dimenI               
\newdimen\!distacross           
\newdimen\!downlength           
\newdimen\!dp                   
\newdimen\!dshade               
\newdimen\!dxpos                
\newdimen\!dxprime              
\newdimen\!dypos                
\newdimen\!dyprime              
\newdimen\!ht                   
\newdimen\!leaderlength         
\newdimen\!lshrinkage           
\newdimen\!midarclength         
\newdimen\!offset               
\newdimen\!plotheadingoffset    
\newdimen\!plotsymbolxshift     
\newdimen\!plotsymbolyshift     
\newdimen\!plotxorigin          
\newdimen\!plotyorigin          
\newdimen\!rootten              
\newdimen\!rshrinkage           
\newdimen\!shadesymbolxshift    
\newdimen\!shadesymbolyshift    
\newdimen\!tenAa                
\newdimen\!tenAc                
\newdimen\!tenAe                
\newdimen\!tshrinkage           
\newdimen\!uplength             
\newdimen\!wd                   
\newdimen\!wmax                 
\newdimen\!wmin                 
\newdimen\!xB                   
\newdimen\!xC                   
\newdimen\!xE                   
\newdimen\!xM                   
\newdimen\!xS                   
\newdimen\!xaxislength          
\newdimen\!xdiff                
\newdimen\!xleft                
\newdimen\!xloc                 
\newdimen\!xorigin              
\newdimen\!xpivot               
\newdimen\!xpos                 
\newdimen\!xprime               
\newdimen\!xright               
\newdimen\!xshade               
\newdimen\!xshift               
\newdimen\!xtemp                
\newdimen\!xunit                
\newdimen\!xxE                  
\newdimen\!xxM                  
\newdimen\!xxS                  
\newdimen\!xxloc                
\newdimen\!yB                   
\newdimen\!yC                   
\newdimen\!yE                   
\newdimen\!yM                   
\newdimen\!yS                   
\newdimen\!yaxislength          
\newdimen\!ybot                 
\newdimen\!ydiff                
\newdimen\!yloc                 
\newdimen\!yorigin              
\newdimen\!ypivot               
\newdimen\!ypos                 
\newdimen\!yprime               
\newdimen\!yshade               
\newdimen\!yshift               
\newdimen\!ytemp                
\newdimen\!ytop                 
\newdimen\!yunit                
\newdimen\!yyE                  
\newdimen\!yyM                  
\newdimen\!yyS                  
\newdimen\!yyloc                
\newdimen\!zpt                  

\newif\if!axisvisible           
\newif\if!gridlinestoo          
\newif\if!keepPO                
\newif\if!placeaxislabel        
\newif\if!switch                
\newif\if!xswitch               

\newtoks\!axisLaBeL             
\newtoks\!keywordtoks           

\newwrite\!replotfile           

\newhelp\!keywordhelp{The keyword mentioned in the error message in unknown. 
Replace NEW KEYWORD in the indicated response by the keyword that 
should have been specified.}    

\!wlet\!!origin=\!xM                   
\!wlet\!!unit=\!uplength               
\!wlet\!Lresiduallength=\!dimenG       
\!wlet\!Rresiduallength=\!dimenF       
\!wlet\!axisLength=\!distacross        
\!wlet\!axisend=\!ydiff                
\!wlet\!axisstart=\!xdiff              
\!wlet\!axisxlevel=\!arclength         
\!wlet\!axisylevel=\!downlength        
\!wlet\!beta=\!dimenE                  
\!wlet\!gamma=\!dimenF                 
\!wlet\!shadexorigin=\!plotxorigin     
\!wlet\!shadeyorigin=\!plotyorigin     
\!wlet\!ticklength=\!xS                
\!wlet\!ticklocation=\!xE              
\!wlet\!ticklocationincr=\!yE          
\!wlet\!tickwidth=\!yS                 
\!wlet\!totalleaderlength=\!dimenE     
\!wlet\!xone=\!xprime                  
\!wlet\!xtwo=\!dxprime                 
\!wlet\!ySsave=\!yM                    
\!wlet\!ybB=\!yB                       
\!wlet\!ybC=\!yC                       
\!wlet\!ybE=\!yE                       
\!wlet\!ybM=\!yM                       
\!wlet\!ybS=\!yS                       
\!wlet\!ybpos=\!yyloc                  
\!wlet\!yone=\!yprime                  
\!wlet\!ytB=\!xB                       
\!wlet\!ytC=\!xC                       
\!wlet\!ytE=\!downlength               
\!wlet\!ytM=\!arclength                
\!wlet\!ytS=\!distacross               
\!wlet\!ytpos=\!xxloc                  
\!wlet\!ytwo=\!dyprime                 

\!zpt=0pt                              
\!xunit=1pt
\!yunit=1pt
\!arearloc=\!xunit
\!areatloc=\!yunit
\!dshade=5pt
\!leaderlength=24in
\!tfs=256                              
\!wmax=5.3pt                           
\!wmin=2.7pt                           
\!xaxislength=\!xunit
\!xpivot=\!zpt
\!yaxislength=\!yunit 
\!ypivot=\!zpt
\plotsymbolspacing=.4pt
  \!dimenA=50pt \!fiftypt=\!dimenA     

\!rootten=3.162278pt                   
\!tenAa=8.690286pt                     
\!tenAc=2.773839pt                     
\!tenAe=2.543275pt                     

\def\!cosrotationangle{1}      
\def\!sinrotationangle{0}      
\def\!xpivotcoord{0}           
\def\!xref{0}                  
\def\!xshadesave{0}            
\def\!ypivotcoord{0}           
\def\!yref{0}                  
\def\!yshadesave{0}            
\def\!zero{0}                  

\let\wlog=\!!!wlog
%
  
\def\normalgraphs{%
  \longticklength=.4\baselineskip
  \shortticklength=.25\baselineskip
  \tickstovaluesleading=.25\baselineskip
  \valuestolabelleading=.8\baselineskip
  \linethickness=.4pt
  \stackleading=.17\baselineskip
  \headingtoplotskip=1.5\baselineskip
  \visibleaxes
  \ticksout
  \nogridlines
  \unloggedticks}
%
\def\setplotarea x from #1 to #2, y from #3 to #4 {%
  \!arealloc=\!M{#1}\!xunit \advance \!arealloc -\!xorigin
  \!areabloc=\!M{#3}\!yunit \advance \!areabloc -\!yorigin
  \!arearloc=\!M{#2}\!xunit \advance \!arearloc -\!xorigin
  \!areatloc=\!M{#4}\!yunit \advance \!areatloc -\!yorigin
  \!initinboundscheck
  \!xaxislength=\!arearloc  \advance\!xaxislength -\!arealloc
  \!yaxislength=\!areatloc  \advance\!yaxislength -\!areabloc
  \!plotheadingoffset=\!zpt
  \!dimenput {{\setbox0=\hbox{}\wd0=\!xaxislength\ht0=\!yaxislength\box0}}
     [bl] (\!arealloc,\!areabloc)}
%
\def\visibleaxes{%
  \def\!axisvisibility{\!axisvisibletrue}}

%

\def\!fixkeyword#1{%
  \errhelp=\!keywordhelp
  \errmessage{Unrecognized keyword `#1': \the\!keywordtoks{NEW KEYWORD}'}}

\!keywordtoks={enter `i\fixkeyword}

\def\fixkeyword#1{%
  \!nextkeyword#1 }


\def\axis {%
  \def\!nextkeyword##1 {%
    \expandafter\ifx\csname !axis##1\endcsname \relax
      \def\!next{\!fixkeyword{##1}}%
    \else
      \def\!next{\csname !axis##1\endcsname}%
    \fi
    \!next}%
  \!offset=\!zpt
  \!axisvisibility
  \!placeaxislabelfalse
  \!nextkeyword}

\def\!axisbottom{%
  \!axisylevel=\!areabloc
  \def\!tickxsign{0}%
  \def\!tickysign{-}%
  \def\!axissetup{\!axisxsetup}%
  \def\!axislabeltbrl{t}%
  \!nextkeyword}

\def\!axistop{%
  \!axisylevel=\!areatloc
  \def\!tickxsign{0}%
  \def\!tickysign{+}%
  \def\!axissetup{\!axisxsetup}%
  \def\!axislabeltbrl{b}%
  \!nextkeyword}

\def\!axisleft{%
  \!axisxlevel=\!arealloc
  \def\!tickxsign{-}%
  \def\!tickysign{0}%
  \def\!axissetup{\!axisysetup}%
  \def\!axislabeltbrl{r}%
  \!nextkeyword}

\def\!axisright{%
  \!axisxlevel=\!arearloc
  \def\!tickxsign{+}%
  \def\!tickysign{0}%
  \def\!axissetup{\!axisysetup}%
  \def\!axislabeltbrl{l}%
  \!nextkeyword}

\def\!axisshiftedto#1=#2 {%
  \if 0\!tickxsign
    \!axisylevel=\!M{#2}\!yunit
    \advance\!axisylevel -\!yorigin
  \else
    \!axisxlevel=\!M{#2}\!xunit
    \advance\!axisxlevel -\!xorigin
  \fi
  \!nextkeyword}

\def\!axisvisible{%
  \!axisvisibletrue  
  \!nextkeyword}

\def\!axisinvisible{%
  \!axisvisiblefalse
  \!nextkeyword}

\def\!axislabel#1 {%
  \!axisLaBeL={#1}%
  \!placeaxislabeltrue
  \!nextkeyword}

\expandafter\def\csname !axis/\endcsname{%
  \!axissetup 
  \if!placeaxislabel
    \!placeaxislabel
  \fi
  \if +\!tickysign 
    \!dimenA=\!axisylevel
    \advance\!dimenA \!offset 
    \advance\!dimenA -\!areatloc 
    \ifdim \!dimenA>\!plotheadingoffset
      \!plotheadingoffset=\!dimenA 
    \fi
  \fi}

\def\grid #1 #2 {%
  \!countA=#1\advance\!countA 1
  \axis bottom invisible ticks length <\!zpt> andacross quantity {\!countA} /
  \!countA=#2\advance\!countA 1
  \axis left   invisible ticks length <\!zpt> andacross quantity {\!countA} / }

\def\plotheading#1 {%
  \advance\!plotheadingoffset \headingtoplotskip
  \!dimenput {#1} [B] <.5\!xaxislength,\!plotheadingoffset>
    (\!arealloc,\!areatloc)}

\def\!axisxsetup{%
  \!axisxlevel=\!arealloc
  \!axisstart=\!arealloc
  \!axisend=\!arearloc
  \!axisLength=\!xaxislength
  \!!origin=\!xorigin
  \!!unit=\!xunit
  \!xswitchtrue
  \if!axisvisible 
    \!makeaxis
  \fi}

\def\!axisysetup{%
  \!axisylevel=\!areabloc
  \!axisstart=\!areabloc
  \!axisend=\!areatloc
  \!axisLength=\!yaxislength
  \!!origin=\!yorigin
  \!!unit=\!yunit
  \!xswitchfalse
  \if!axisvisible
    \!makeaxis
  \fi}

\def\!makeaxis{%
  \setbox\!boxA=\hbox{
    \beginpicture
      \!setdimenmode
      \setcoordinatesystem point at {\!zpt} {\!zpt}   
      \putrule from {\!zpt} {\!zpt} to
        {\!tickysign\!tickysign\!axisLength} 
        {\!tickxsign\!tickxsign\!axisLength}
    \endpicturesave <\!Xsave,\!Ysave>}%
    \wd\!boxA=\!zpt
    \!placetick\!axisstart}

\def\!placeaxislabel{%
  \advance\!offset \valuestolabelleading
  \if!xswitch
    \!dimenput {\the\!axisLaBeL} [\!axislabeltbrl]
      <.5\!axisLength,\!tickysign\!offset> (\!axisxlevel,\!axisylevel)
    \advance\!offset \!dp  
    \advance\!offset \!ht  
  \else
    \!dimenput {\the\!axisLaBeL} [\!axislabeltbrl]
      <\!tickxsign\!offset,.5\!axisLength> (\!axisxlevel,\!axisylevel)
  \fi
  \!axisLaBeL={}}

%


\def\arrow <#1> [#2,#3]{%
  \!ifnextchar<{\!arrow{#1}{#2}{#3}}{\!arrow{#1}{#2}{#3}<\!zpt,\!zpt> }}

\def\!arrow#1#2#3<#4,#5> from #6 #7 to #8 #9 {%
%
  \!xloc=\!M{#8}\!xunit   
  \!yloc=\!M{#9}\!yunit
  \!dxpos=\!xloc  \!dimenA=\!M{#6}\!xunit  \advance \!dxpos -\!dimenA
  \!dypos=\!yloc  \!dimenA=\!M{#7}\!yunit  \advance \!dypos -\!dimenA
  \let\!MAH=\!M
  \!setdimenmode
  \!xshift=#4\relax  \!yshift=#5\relax
  \!reverserotateonly\!xshift\!yshift
  \advance\!xshift\!xloc  \advance\!yshift\!yloc
%
  \!xS=-\!dxpos  \advance\!xS\!xshift
  \!yS=-\!dypos  \advance\!yS\!yshift
  \!start (\!xS,\!yS)
  \!ljoin (\!xshift,\!yshift)
%
  \!Pythag\!dxpos\!dypos\!arclength
  \!divide\!dxpos\!arclength\!dxpos  
  \!dxpos=32\!dxpos  \!removept\!dxpos\!!cos
  \!divide\!dypos\!arclength\!dypos  
  \!dypos=32\!dypos  \!removept\!dypos\!!sin
%
  \!halfhead{#1}{#2}{#3}
  \!halfhead{#1}{-#2}{-#3}
  \let\!M=\!MAH
  \ignorespaces}
%
  \def\!halfhead#1#2#3{%
    \!dimenC=-#1%
    \divide \!dimenC 2 
    \!dimenD=#2\!dimenC
    \!rotate(\!dimenC,\!dimenD)by(\!!cos,\!!sin)to(\!xM,\!yM)
    \!dimenC=-#1
    \!dimenD=#3\!dimenC
    \!dimenD=.5\!dimenD
    \!rotate(\!dimenC,\!dimenD)by(\!!cos,\!!sin)to(\!xE,\!yE)
    \!start (\!xshift,\!yshift)
    \advance\!xM\!xshift  \advance\!yM\!yshift
    \advance\!xE\!xshift  \advance\!yE\!yshift
    \!qjoin (\!xM,\!yM) (\!xE,\!yE) 
    \ignorespaces}

\def\betweenarrows #1#2 from #3 #4 to #5 #6 {%
  \!xloc=\!M{#3}\!xunit  \!xxloc=\!M{#5}\!xunit%
  \!yloc=\!M{#4}\!yunit  \!yyloc=\!M{#6}\!yunit%
  \!dxpos=\!xxloc  \advance\!dxpos by -\!xloc
  \!dypos=\!yyloc  \advance\!dypos by -\!yloc
  \advance\!xloc .5\!dxpos
  \advance\!yloc .5\!dypos
  \let\!MBA=\!M
  \!setdimenmode
  \ifdim\!dypos=\!zpt
    \ifdim\!dxpos<\!zpt \!dxpos=-\!dxpos \fi
    \put {\!lrarrows{\!dxpos}{#1}}#2{} at {\!xloc} {\!yloc}
  \else
    \ifdim\!dxpos=\!zpt
      \ifdim\!dypos<\!zpt \!dypos=-\!zpt \fi
      \put {\!udarrows{\!dypos}{#1}}#2{} at {\!xloc} {\!yloc}
    \fi
  \fi
  \let\!M=\!MBA
  \ignorespaces}

\def\!lrarrows#1#2{
  {\setbox\!boxA=\hbox{$\mkern-2mu\mathord-\mkern-2mu$}%
   \setbox\!boxB=\hbox{$\leftarrow$}\!dimenE=\ht\!boxB
   \setbox\!boxB=\hbox{}\ht\!boxB=2\!dimenE
   \hbox to #1{$\mathord\leftarrow\mkern-6mu
     \cleaders\copy\!boxA\hfil
     \mkern-6mu\mathord-$%
     \kern.4em $\vcenter{\box\!boxB}$$\vcenter{\hbox{#2}}$\kern.4em
     $\mathord-\mkern-6mu
     \cleaders\copy\!boxA\hfil
     \mkern-6mu\mathord\rightarrow$}}}

\def\!udarrows#1#2{
  {\setbox\!boxB=\hbox{#2}%
   \setbox\!boxA=\hbox to \wd\!boxB{\hss$\vert$\hss}%
   \!dimenE=\ht\!boxA \advance\!dimenE \dp\!boxA \divide\!dimenE 2
   \vbox to #1{\offinterlineskip
      \vskip .05556\!dimenE
      \hbox to \wd\!boxB{\hss$\mkern.4mu\uparrow$\hss}\vskip-\!dimenE
      \cleaders\copy\!boxA\vfil
      \vskip-\!dimenE\copy\!boxA
      \vskip\!dimenE\copy\!boxB\vskip.4em
      \copy\!boxA\vskip-\!dimenE
      \cleaders\copy\!boxA\vfil
      \vskip-\!dimenE \hbox to \wd\!boxB{\hss$\mkern.4mu\downarrow$\hss}
      \vskip .05556\!dimenE}}}

%

\def\putbar#1breadth <#2> from #3 #4 to #5 #6 {%
  \!xloc=\!M{#3}\!xunit  \!xxloc=\!M{#5}\!xunit%
  \!yloc=\!M{#4}\!yunit  \!yyloc=\!M{#6}\!yunit%
  \!dypos=\!yyloc  \advance\!dypos by -\!yloc
  \!dimenI=#2  
  \ifdim \!dimenI=\!zpt 
    \putrule#1from {#3} {#4} to {#5} {#6} 
  \else 
    \let\!MBar=\!M
    \!setdimenmode 
    \divide\!dimenI 2
    \ifdim \!dypos=\!zpt             
      \advance \!yloc -\!dimenI 
      \advance \!yyloc \!dimenI
    \else
      \advance \!xloc -\!dimenI 
      \advance \!xxloc \!dimenI
    \fi
    \putrectangle#1corners at {\!xloc} {\!yloc} and {\!xxloc} {\!yyloc}
    \let\!M=\!MBar 
  \fi
  \ignorespaces}

\def\setbars#1breadth <#2> baseline at #3 = #4 {%
  \edef\!barshift{#1}%
  \edef\!barbreadth{#2}%
  \edef\!barorientation{#3}%
  \edef\!barbaseline{#4}%
  \def\!bardobaselabel{\!bardoendlabel}%
  \def\!bardoendlabel{\!barfinish}%
  \let\!drawcurve=\!barcurve
  \!setbars}
\def\!setbars{%
  \futurelet\!nextchar\!!setbars}
\def\!!setbars{%
  \if b\!nextchar
    \def\!!!setbars{\!setbarsbget}%
  \else 
    \if e\!nextchar
      \def\!!!setbars{\!setbarseget}%
    \else
      \def\!!!setbars{\relax}%
    \fi
  \fi
  \!!!setbars}
\def\!setbarsbget baselabels (#1) {%
  \def\!barbaselabelorientation{#1}%
  \def\!bardobaselabel{\!!bardobaselabel}%
  \!setbars}
\def\!setbarseget endlabels (#1) {%
  \edef\!barendlabelorientation{#1}%
  \def\!bardoendlabel{\!!bardoendlabel}%
  \!setbars}

\def\!barcurve #1 #2 {%
  \if y\!barorientation
    \def\!basexarg{#1}%
    \def\!baseyarg{\!barbaseline}%
  \else
    \def\!basexarg{\!barbaseline}%
    \def\!baseyarg{#2}%
  \fi
  \expandafter\putbar\!barshift breadth <\!barbreadth> from {\!basexarg}
    {\!baseyarg} to {#1} {#2}
  \def\!endxarg{#1}%
  \def\!endyarg{#2}%
  \!bardobaselabel}

\def\!!bardobaselabel "#1" {%
  \put {#1}\!barbaselabelorientation{} at {\!basexarg} {\!baseyarg}
  \!bardoendlabel}
 
\def\!!bardoendlabel "#1" {%
  \put {#1}\!barendlabelorientation{} at {\!endxarg} {\!endyarg}
  \!barfinish}

\def\!barfinish{%
  \!ifnextchar/{\!finish}{\!barcurve}}

%
%
%
\def\putrectangle{%
  \!ifnextchar<{\!putrectangle}{\!putrectangle<\!zpt,\!zpt> }}
\def\!putrectangle<#1,#2> corners at #3 #4 and #5 #6 {%
%
  \!xone=\!M{#3}\!xunit  \!xtwo=\!M{#5}\!xunit%
  \!yone=\!M{#4}\!yunit  \!ytwo=\!M{#6}\!yunit%
  \ifdim \!xtwo<\!xone
    \!dimenI=\!xone  \!xone=\!xtwo  \!xtwo=\!dimenI
  \fi
  \ifdim \!ytwo<\!yone
    \!dimenI=\!yone  \!yone=\!ytwo  \!ytwo=\!dimenI
  \fi
  \!dimenI=#1\relax  \advance\!xone\!dimenI  \advance\!xtwo\!dimenI
  \!dimenI=#2\relax  \advance\!yone\!dimenI  \advance\!ytwo\!dimenI
  \let\!MRect=\!M
  \!setdimenmode
%
  \!shaderectangle
%
  \!dimenI=.5\linethickness
  \advance \!xone  -\!dimenI
  \advance \!xtwo   \!dimenI
  \putrule from {\!xone} {\!yone} to {\!xtwo} {\!yone} 
  \putrule from {\!xone} {\!ytwo} to {\!xtwo} {\!ytwo} 
%
  \advance \!xone   \!dimenI
  \advance \!xtwo  -\!dimenI%
  \advance \!yone  -\!dimenI
  \advance \!ytwo   \!dimenI
  \putrule from {\!xone} {\!yone} to {\!xone} {\!ytwo} 
  \putrule from {\!xtwo} {\!yone} to {\!xtwo} {\!ytwo} 
  \let\!M=\!MRect
  \ignorespaces}
 

\def\shaderectanglesoff{%
  \def\!shaderectangle{}%
  \ignorespaces}

\shaderectanglesoff
 
\def\!!shaderectangle{%
  \!dimenA=\!xtwo  \advance \!dimenA -\!xone
  \!dimenB=\!ytwo  \advance \!dimenB -\!yone
  \ifdim \!dimenA<\!dimenB
    \!startvshade (\!xone,\!yone,\!ytwo)
    \!lshade      (\!xtwo,\!yone,\!ytwo)
  \else
    \!starthshade (\!yone,\!xone,\!xtwo)
    \!lshade      (\!ytwo,\!xone,\!xtwo)
  \fi
  \ignorespaces}
  
\def\frame{%
  \!ifnextchar<{\!frame}{\!frame<\!zpt> }}
\long\def\!frame<#1> #2{%
  \beginpicture
    \setcoordinatesystem units <1pt,1pt> point at 0 0 
    \put {#2} [Bl] at 0 0 
    \!dimenA=#1\relax
    \!dimenB=\!wd \advance \!dimenB \!dimenA
    \!dimenC=\!ht \advance \!dimenC \!dimenA
    \!dimenD=\!dp \advance \!dimenD \!dimenA
    \let\!MFr=\!M
    \!setdimenmode
    \putrectangle corners at {-\!dimenA} {-\!dimenD} and {\!dimenB} {\!dimenC}
    \!setcoordmode
    \let\!M=\!MFr
  \endpicture
  \ignorespaces}
 
\def\rectangle <#1> <#2> {%
  \setbox0=\hbox{}\wd0=#1\ht0=#2\frame {\box0}}

%

\def\plot{%
  \!ifnextchar"{\!plotfromfile}{\!drawcurve}}
\def\!plotfromfile"#1"{%
  \expandafter\!drawcurve \input #1 /}

\def\setquadratic{%
  \let\!drawcurve=\!qcurve
  \let\!!Shade=\!!qShade
  \let\!!!Shade=\!!!qShade}

\def\setlinear{%
  \let\!drawcurve=\!lcurve
  \let\!!Shade=\!!lShade
  \let\!!!Shade=\!!!lShade}

\def\sethistograms{%
  \let\!drawcurve=\!hcurve}

\def\!qcurve #1 #2 {%
  \!start (#1,#2)
  \!Qjoin}
\def\!Qjoin#1 #2 #3 #4 {%
  \!qjoin (#1,#2) (#3,#4)             
  \!ifnextchar/{\!finish}{\!Qjoin}}

\def\!lcurve #1 #2 {%
  \!start (#1,#2)
  \!Ljoin}
\def\!Ljoin#1 #2 {%
  \!ljoin (#1,#2)                    
  \!ifnextchar/{\!finish}{\!Ljoin}}

\def\!finish/{\ignorespaces}

\def\!hcurve #1 #2 {%
  \edef\!hxS{#1}%
  \edef\!hyS{#2}%
  \!hjoin}
\def\!hjoin#1 #2 {%
  \putrectangle corners at {\!hxS} {\!hyS} and {#1} {#2}
  \edef\!hxS{#1}%
  \!ifnextchar/{\!finish}{\!hjoin}}

\def\vshade #1 #2 #3 {%
  \!startvshade (#1,#2,#3)
  \!Shadewhat}

\def\hshade #1 #2 #3 {%
  \!starthshade (#1,#2,#3)
  \!Shadewhat}

\def\!Shadewhat{%
  \futurelet\!nextchar\!Shade}
\def\!Shade{%
  \if <\!nextchar
    \def\!nextShade{\!!Shade}%
  \else
    \if /\!nextchar
      \def\!nextShade{\!finish}%
    \else
      \def\!nextShade{\!!!Shade}%
    \fi
  \fi
  \!nextShade}
\def\!!lShade<#1> #2 #3 #4 {%
  \!lshade <#1> (#2,#3,#4)                 
  \!Shadewhat}
\def\!!!lShade#1 #2 #3 {%
  \!lshade (#1,#2,#3)
  \!Shadewhat} 
\def\!!qShade<#1> #2 #3 #4 #5 #6 #7 {%
  \!qshade <#1> (#2,#3,#4) (#5,#6,#7)      
  \!Shadewhat}
\def\!!!qShade#1 #2 #3 #4 #5 #6 {%
  \!qshade (#1,#2,#3) (#4,#5,#6)
  \!Shadewhat} 

\setlinear

\def\setdashpattern <#1>{%
  \def\!Flist{}\def\!Blist{}\def\!UDlist{}%
  \!countA=0
  \!ecfor\!item:=#1\do{%
    \!dimenA=\!item\relax
    \expandafter\!rightappend\the\!dimenA\withCS{\\}\to\!UDlist%
    \advance\!countA  1
    \ifodd\!countA
      \expandafter\!rightappend\the\!dimenA\withCS{\!Rule}\to\!Flist%
      \expandafter\!leftappend\the\!dimenA\withCS{\!Rule}\to\!Blist%
    \else 
      \expandafter\!rightappend\the\!dimenA\withCS{\!Skip}\to\!Flist%
      \expandafter\!leftappend\the\!dimenA\withCS{\!Skip}\to\!Blist%
    \fi}%
  \!leaderlength=\!zpt
  \def\!Rule##1{\advance\!leaderlength  ##1}%
  \def\!Skip##1{\advance\!leaderlength  ##1}%
  \!Flist%
  \ifdim\!leaderlength>\!zpt 
  \else
    \def\!Flist{\!Skip{24in}}\def\!Blist{\!Skip{24in}}\ignorespaces
    \def\!UDlist{\\{\!zpt}\\{24in}}\ignorespaces
    \!leaderlength=24in
  \fi
  \!dashingon}

\def\!dashingon{%
  \def\!advancedashing{\!!advancedashing}%
  \def\!drawlinearsegment{\!lineardashed}%
  \def\!puthline{\!putdashedhline}%
  \def\!putvline{\!putdashedvline}%
  \ignorespaces}%
\def\!dashingoff{%
  \def\!advancedashing{\relax}%
  \def\!drawlinearsegment{\!linearsolid}%
  \def\!puthline{\!putsolidhline}%
  \def\!putvline{\!putsolidvline}%
  \ignorespaces}

\def\setdots{%
  \!ifnextchar<{\!setdots}{\!setdots<5pt>}}
\def\!setdots<#1>{%
  \!dimenB=#1\advance\!dimenB -\plotsymbolspacing
  \ifdim\!dimenB<\!zpt
    \!dimenB=\!zpt
  \fi
\setdashpattern <\plotsymbolspacing,\!dimenB>}
 
\def\setdotsnear <#1> for <#2>{%
  \!dimenB=#2\relax  \advance\!dimenB -.05pt  
  \!dimenC=#1\relax  \!countA=\!dimenC 
  \!dimenD=\!dimenB  \advance\!dimenD .5\!dimenC  \!countB=\!dimenD
  \divide \!countB  \!countA
  \ifnum 1>\!countB 
    \!countB=1
  \fi
  \divide\!dimenB  \!countB
  \setdots <\!dimenB>}
 
\def\setdashes{%
  \!ifnextchar<{\!setdashes}{\!setdashes<5pt>}}
\def\!setdashes<#1>{\setdashpattern <#1,#1>}
 
\def\setdashesnear <#1> for <#2>{%
  \!dimenB=#2\relax  
  \!dimenC=#1\relax  \!countA=\!dimenC 
  \!dimenD=\!dimenB  \advance\!dimenD .5\!dimenC  \!countB=\!dimenD
  \divide \!countB  \!countA
  \ifodd \!countB 
  \else 
    \advance \!countB  1
  \fi
  \divide\!dimenB  \!countB
  \setdashes <\!dimenB>}
 
\def\setsolid{%
  \def\!Flist{\!Rule{24in}}\def\!Blist{\!Rule{24in}}%
  \def\!UDlist{\\{24in}\\{\!zpt}}%
  \!dashingoff}  
\setsolid


 
  
 
\def\!divide#1#2#3{%
  \!dimenB=#1
  \!dimenC=#2
  \!dimenD=\!dimenB
  \divide \!dimenD \!dimenC
  \!dimenA=\!dimenD
  \multiply\!dimenD \!dimenC
  \advance\!dimenB -\!dimenD
  \!dimenD=\!dimenC
    \ifdim\!dimenD<\!zpt \!dimenD=-\!dimenD 
  \fi
  \ifdim\!dimenD<64pt
    \!divstep[\!tfs]\!divstep[\!tfs]%
  \else 
    \!!divide
  \fi
  #3=\!dimenA\ignorespaces}

\def\!!divide{%
  \ifdim\!dimenD<256pt
    \!divstep[64]\!divstep[32]\!divstep[32]%
  \else 
    \!divstep[8]\!divstep[8]\!divstep[8]\!divstep[8]\!divstep[8]%
    \!dimenA=2\!dimenA
  \fi}

\def\!divstep[#1]{
  \!dimenB=#1\!dimenB
  \!dimenD=\!dimenB
    \divide \!dimenD by \!dimenC
  \!dimenA=#1\!dimenA
    \advance\!dimenA by \!dimenD%
  \multiply\!dimenD by \!dimenC
    \advance\!dimenB by -\!dimenD}
 
\def\Divide <#1> by <#2> forming <#3> {%
  \!divide{#1}{#2}{#3}}

 
 

 

\def\ellipticalarc axes ratio #1:#2 #3 degrees from #4 #5 center at #6 #7 {%
  \!angle=#3pt\relax
  \ifdim\!angle>\!zpt 
    \def\!sign{}
  \else 
    \def\!sign{-}\!angle=-\!angle
  \fi
  \!xxloc=\!M{#6}\!xunit
  \!yyloc=\!M{#7}\!yunit     
  \!xxS=\!M{#4}\!xunit
  \!yyS=\!M{#5}\!yunit
  \advance\!xxS -\!xxloc
  \advance\!yyS -\!yyloc
  \!divide\!xxS{#1pt}\!xxS 
  \!divide\!yyS{#2pt}\!yyS 
  \let\!MC=\!M
  \!setdimenmode
  \!xS=#1\!xxS  \advance\!xS\!xxloc
  \!yS=#2\!yyS  \advance\!yS\!yyloc
  \!start (\!xS,\!yS)%
  \!loop\ifdim\!angle>14.9999pt
    \!rotate(\!xxS,\!yyS)by(\!cos,\!sign\!sin)to(\!xxM,\!yyM) 
    \!rotate(\!xxM,\!yyM)by(\!cos,\!sign\!sin)to(\!xxE,\!yyE)
    \!xM=#1\!xxM  \advance\!xM\!xxloc  \!yM=#2\!yyM  \advance\!yM\!yyloc
    \!xE=#1\!xxE  \advance\!xE\!xxloc  \!yE=#2\!yyE  \advance\!yE\!yyloc
    \!qjoin (\!xM,\!yM) (\!xE,\!yE)
    \!xxS=\!xxE  \!yyS=\!yyE 
    \advance \!angle -15pt
  \repeat
  \ifdim\!angle>\!zpt
    \!angle=100.53096\!angle
    \divide \!angle 360 
    \!sinandcos\!angle\!!sin\!!cos
    \!rotate(\!xxS,\!yyS)by(\!!cos,\!sign\!!sin)to(\!xxM,\!yyM) 
    \!rotate(\!xxM,\!yyM)by(\!!cos,\!sign\!!sin)to(\!xxE,\!yyE)
    \!xM=#1\!xxM  \advance\!xM\!xxloc  \!yM=#2\!yyM  \advance\!yM\!yyloc
    \!xE=#1\!xxE  \advance\!xE\!xxloc  \!yE=#2\!yyE  \advance\!yE\!yyloc
    \!qjoin (\!xM,\!yM) (\!xE,\!yE)
  \fi
  \let\!M=\!MC
  \ignorespaces}

\def\!rotate(#1,#2)by(#3,#4)to(#5,#6){%
  \!dimenA=#3#1\advance \!dimenA -#4#2
  \!dimenB=#3#2\advance \!dimenB  #4#1
  \divide \!dimenA 32  \divide \!dimenB 32 
  #5=\!dimenA  #6=\!dimenB
  \ignorespaces}
\def\!sin{4.17684}
\def\!cos{31.72624}

\def\!sinandcos#1#2#3{%
 \!dimenD=#1
 \!dimenA=\!dimenD
 \!dimenB=32pt
 \!removept\!dimenD\!value
 \!dimenC=\!dimenD
 \!dimenC=\!value\!dimenC \divide\!dimenC by 64 
 \advance\!dimenB by -\!dimenC
 \!dimenC=\!value\!dimenC \divide\!dimenC by 96 
 \advance\!dimenA by -\!dimenC
 \!dimenC=\!value\!dimenC \divide\!dimenC by 128 
 \advance\!dimenB by \!dimenC%
 \!removept\!dimenA#2
 \!removept\!dimenB#3
 \ignorespaces}




\def\putrule#1from #2 #3 to #4 #5 {%
  \!xloc=\!M{#2}\!xunit  \!xxloc=\!M{#4}\!xunit%
  \!yloc=\!M{#3}\!yunit  \!yyloc=\!M{#5}\!yunit%
  \!dxpos=\!xxloc  \advance\!dxpos by -\!xloc
  \!dypos=\!yyloc  \advance\!dypos by -\!yloc
  \ifdim\!dypos=\!zpt
    \def\!!Line{\!puthline{#1}}\ignorespaces
  \else
    \ifdim\!dxpos=\!zpt
      \def\!!Line{\!putvline{#1}}\ignorespaces
    \else 
       \def\!!Line{}
    \fi
  \fi
  \let\!ML=\!M
  \!setdimenmode
  \!!Line%
  \let\!M=\!ML
  \ignorespaces}

\def\!putsolidhline#1{%
  \ifdim\!dxpos>\!zpt 
    \put{\!hline\!dxpos}#1[l] at {\!xloc} {\!yloc}
  \else 
    \put{\!hline{-\!dxpos}}#1[l] at {\!xxloc} {\!yyloc}
  \fi
  \ignorespaces}
 
\def\!putsolidvline#1{%
  \ifdim\!dypos>\!zpt 
    \put{\!vline\!dypos}#1[b] at {\!xloc} {\!yloc}
  \else 
    \put{\!vline{-\!dypos}}#1[b] at {\!xxloc} {\!yyloc}
  \fi
  \ignorespaces}
 
\def\!hline#1{\hbox to #1{\leaders \hrule height\linethickness\hfill}}
\def\!vline#1{\vbox to #1{\leaders \vrule width\linethickness\vfill}}

\def\!putdashedhline#1{%
  \ifdim\!dxpos>\!zpt 
    \!DLsetup\!Flist\!dxpos
    \put{\hbox to \!totalleaderlength{\!hleaders}\!hpartialpattern\!Rtrunc}
      #1[l] at {\!xloc} {\!yloc} 
  \else 
    \!DLsetup\!Blist{-\!dxpos}
    \put{\!hpartialpattern\!Ltrunc\hbox to \!totalleaderlength{\!hleaders}}
      #1[r] at {\!xloc} {\!yloc} 
  \fi
  \ignorespaces}
 
\def\!putdashedvline#1{%
  \!dypos=-\!dypos
  \ifdim\!dypos>\!zpt 
    \!DLsetup\!Flist\!dypos 
    \put{\vbox{\vbox to \!totalleaderlength{\!vleaders}
      \!vpartialpattern\!Rtrunc}}#1[t] at {\!xloc} {\!yloc} 
  \else 
    \!DLsetup\!Blist{-\!dypos}
    \put{\vbox{\!vpartialpattern\!Ltrunc
      \vbox to \!totalleaderlength{\!vleaders}}}#1[b] at {\!xloc} {\!yloc} 
  \fi
  \ignorespaces}

\def\!DLsetup#1#2{
  \let\!RSlist=#1
  \!countB=#2
  \!countA=\!leaderlength
  \divide\!countB by \!countA
  \!totalleaderlength=\!countB\!leaderlength
  \!Rresiduallength=#2%
  \advance \!Rresiduallength by -\!totalleaderlength
  \!Lresiduallength=\!leaderlength
  \advance \!Lresiduallength by -\!Rresiduallength
  \ignorespaces}
 
\def\!hleaders{%
  \def\!Rule##1{\vrule height\linethickness width##1}%
  \def\!Skip##1{\hskip##1}%
  \leaders\hbox{\!RSlist}\hfill}
 
\def\!hpartialpattern#1{%
  \!dimenA=\!zpt \!dimenB=\!zpt 
  \def\!Rule##1{#1{##1}\vrule height\linethickness width\!dimenD}%
  \def\!Skip##1{#1{##1}\hskip\!dimenD}%
  \!RSlist}
 
\def\!vleaders{%
  \def\!Rule##1{\hrule width\linethickness height##1}%
  \def\!Skip##1{\vskip##1}%
  \leaders\vbox{\!RSlist}\vfill}
 
\def\!vpartialpattern#1{%
  \!dimenA=\!zpt \!dimenB=\!zpt 
  \def\!Rule##1{#1{##1}\hrule width\linethickness height\!dimenD}%
  \def\!Skip##1{#1{##1}\vskip\!dimenD}%
  \!RSlist}
 
\def\!Rtrunc#1{\!trunc{#1}>\!Rresiduallength}
\def\!Ltrunc#1{\!trunc{#1}<\!Lresiduallength}
 
\def\!trunc#1#2#3{%
  \!dimenA=\!dimenB         
  \advance\!dimenB by #1%
  \!dimenD=\!dimenB  \ifdim\!dimenD#2#3\!dimenD=#3\fi
  \!dimenC=\!dimenA  \ifdim\!dimenC#2#3\!dimenC=#3\fi
  \advance \!dimenD by -\!dimenC}

\def\!start (#1,#2){%
  \!plotxorigin=\!xorigin  \advance \!plotxorigin by \!plotsymbolxshift
  \!plotyorigin=\!yorigin  \advance \!plotyorigin by \!plotsymbolyshift
  \!xS=\!M{#1}\!xunit \!yS=\!M{#2}\!yunit
  \!rotateaboutpivot\!xS\!yS
  \!copylist\!UDlist\to\!!UDlist
  \!getnextvalueof\!downlength\from\!!UDlist
  \!distacross=\!zpt
  \!intervalno=0 
  \global\totalarclength=\!zpt
  \ignorespaces}

\def\!ljoin (#1,#2){%
  \advance\!intervalno by 1
  \!xE=\!M{#1}\!xunit \!yE=\!M{#2}\!yunit
  \!rotateaboutpivot\!xE\!yE
  \!xdiff=\!xE \advance \!xdiff by -\!xS
  \!ydiff=\!yE \advance \!ydiff by -\!yS
  \!Pythag\!xdiff\!ydiff\!arclength
  \global\advance \totalarclength by \!arclength%
  \!drawlinearsegment
  \!xS=\!xE \!yS=\!yE
  \ignorespaces}

\def\!linearsolid{%
  \!npoints=\!arclength
  \!countA=\plotsymbolspacing
  \divide\!npoints by \!countA
  \ifnum \!npoints<1 
    \!npoints=1 
  \fi
  \divide\!xdiff by \!npoints
  \divide\!ydiff by \!npoints
  \!xpos=\!xS \!ypos=\!yS
  \loop\ifnum\!npoints>-1
    \!plotifinbounds
    \advance \!xpos by \!xdiff
    \advance \!ypos by \!ydiff
    \advance \!npoints by -1
  \repeat
  \ignorespaces}

\def\!lineardashed{%
  \ifdim\!distacross>\!arclength
    \advance \!distacross by -\!arclength  
  \else
    \loop\ifdim\!distacross<\!arclength
      \!divide\!distacross\!arclength\!dimenA
      \!removept\!dimenA\!t
      \!xpos=\!t\!xdiff \advance \!xpos by \!xS
      \!ypos=\!t\!ydiff \advance \!ypos by \!yS
      \!plotifinbounds
      \advance\!distacross by \plotsymbolspacing
      \!advancedashing
    \repeat  
    \advance \!distacross by -\!arclength
  \fi
  \ignorespaces}

\def\!!advancedashing{%
  \advance\!downlength by -\plotsymbolspacing
  \ifdim \!downlength>\!zpt
  \else
    \advance\!distacross by \!downlength
    \!getnextvalueof\!uplength\from\!!UDlist
    \advance\!distacross by \!uplength
    \!getnextvalueof\!downlength\from\!!UDlist
  \fi}

\def\inboundscheckoff{%
  \def\!plotifinbounds{\!plot(\!xpos,\!ypos)}%
  \def\!initinboundscheck{\relax}\ignorespaces}
 
\inboundscheckoff
 
\def\!!plotifinbounds{%
  \ifdim \!xpos<\!checkleft
  \else
    \ifdim \!xpos>\!checkright
    \else
      \ifdim \!ypos<\!checkbot
      \else
         \ifdim \!ypos>\!checktop
         \else
           \!plot(\!xpos,\!ypos)
         \fi 
      \fi
    \fi
  \fi}

\def\!!initinboundscheck{%
  \!checkleft=\!arealloc     \advance\!checkleft by \!xorigin
  \!checkright=\!arearloc    \advance\!checkright by \!xorigin
  \!checkbot=\!areabloc      \advance\!checkbot by \!yorigin
  \!checktop=\!areatloc      \advance\!checktop by \!yorigin}

%


\def\!logten#1#2{%
  \expandafter\!!logten#1\!nil
  \!removept\!dimenF#2%
  \ignorespaces}

\def\!!logten#1#2\!nil{%
  \if -#1%
    \!dimenF=\!zpt
    \def\!next{\ignorespaces}%
  \else
    \if +#1%
      \def\!next{\!!logten#2\!nil}%
    \else
      \if .#1%
        \def\!next{\!!logten0.#2\!nil}%
      \else
        \def\!next{\!!!logten#1#2..\!nil}%
      \fi
    \fi
  \fi
  \!next}

\def\!!!logten#1#2.#3.#4\!nil{%
  \!dimenF=1pt 
  \if 0#1%
    \!!logshift#3pt 
  \else 
    \!logshift#2/
    \!dimenE=#1.#2#3pt 
  \fi 
  \ifdim \!dimenE<\!rootten
    \multiply \!dimenE 10 
    \advance  \!dimenF -1pt
  \fi
  \!dimenG=\!dimenE
    \advance\!dimenG 10pt
  \advance\!dimenE -10pt 
  \multiply\!dimenE 10 
  \!divide\!dimenE\!dimenG\!dimenE
  \!removept\!dimenE\!t
  \!dimenG=\!t\!dimenE
  \!removept\!dimenG\!tt
  \!dimenH=\!tt\!tenAe
    \divide\!dimenH 100
  \advance\!dimenH \!tenAc
  \!dimenH=\!tt\!dimenH
    \divide\!dimenH 100   
  \advance\!dimenH \!tenAa
  \!dimenH=\!t\!dimenH
    \divide\!dimenH 100 
  \advance\!dimenF \!dimenH}

\def\!logshift#1{%
  \if #1/%
    \def\!next{\ignorespaces}%
  \else
    \advance\!dimenF 1pt 
    \def\!next{\!logshift}%
  \fi 
  \!next}
 
 \def\!!logshift#1{%
   \advance\!dimenF -1pt
   \if 0#1%
     \def\!next{\!!logshift}%
   \else
     \if p#1%
       \!dimenF=1pt
       \def\!next{\!dimenE=1p}%
     \else
       \def\!next{\!dimenE=#1.}%
     \fi
   \fi
   \!next}

\def\beginpicture{%
  \setbox\!picbox=\hbox\bgroup%
  \!xleft=\maxdimen  
  \!xright=-\maxdimen
  \!ybot=\maxdimen
  \!ytop=-\maxdimen}
 
\def\endpicture{%
  \ifdim\!xleft=\maxdimen
    \!xleft=\!zpt \!xright=\!zpt \!ybot=\!zpt \!ytop=\!zpt 
  \fi
  \global\!Xleft=\!xleft \global\!Xright=\!xright
  \global\!Ybot=\!ybot \global\!Ytop=\!ytop
  \egroup%
  \ht\!picbox=\!Ytop  \dp\!picbox=-\!Ybot
  \ifdim\!Ybot>\!zpt
  \else 
    \ifdim\!Ytop<\!zpt
      \!Ybot=\!Ytop
    \else
      \!Ybot=\!zpt
    \fi
  \fi
  \hbox{\kern-\!Xleft\lower\!Ybot\box\!picbox\kern\!Xright}}
 
\def\endpicturesave <#1,#2>{%
  \endpicture \global #1=\!Xleft \global #2=\!Ybot \ignorespaces}

\def\setcoordinatesystem{%
  \!ifnextchar{u}{\!getlengths }
    {\!getlengths units <\!xunit,\!yunit>}}
\def\!getlengths units <#1,#2>{%
  \!xunit=#1\relax
  \!yunit=#2\relax
  \!ifcoordmode 
    \let\!SCnext=\!SCccheckforRP
  \else
    \let\!SCnext=\!SCdcheckforRP
  \fi
  \!SCnext}
\def\!SCccheckforRP{%
  \!ifnextchar{p}{\!cgetreference }
    {\!cgetreference point at {\!xref} {\!yref} }}
\def\!cgetreference point at #1 #2 {%
  \edef\!xref{#1}\edef\!yref{#2}%
  \!xorigin=\!xref\!xunit  \!yorigin=\!yref\!yunit  
  \!initinboundscheck 
  \ignorespaces}
\def\!SCdcheckforRP{%
  \!ifnextchar{p}{\!dgetreference}%
    {\ignorespaces}}
\def\!dgetreference point at #1 #2 {%
  \!xorigin=#1\relax  \!yorigin=#2\relax
  \ignorespaces}

\long\def\put#1#2 at #3 #4 {%
  \!setputobject{#1}{#2}%
  \!xpos=\!M{#3}\!xunit  \!ypos=\!M{#4}\!yunit  
  \!rotateaboutpivot\!xpos\!ypos%
  \advance\!xpos -\!xorigin  \advance\!xpos -\!xshift
  \advance\!ypos -\!yorigin  \advance\!ypos -\!yshift
  \kern\!xpos\raise\!ypos\box\!putobject\kern-\!xpos%
  \!doaccounting\ignorespaces}
 
\long\def\multiput #1#2 at {%
  \!setputobject{#1}{#2}%
  \!ifnextchar"{\!putfromfile}{\!multiput}}
\def\!putfromfile"#1"{%
  \expandafter\!multiput \input #1 /}
\def\!multiput{%
  \futurelet\!nextchar\!!multiput}
\def\!!multiput{%
  \if *\!nextchar
    \def\!nextput{\!alsoby}%
  \else
    \if /\!nextchar
      \def\!nextput{\!finishmultiput}%
    \else
      \def\!nextput{\!alsoat}%
    \fi
  \fi
  \!nextput}
\def\!finishmultiput/{%
  \setbox\!putobject=\hbox{}%
  \ignorespaces}
 
\def\!alsoat#1 #2 {%
  \!xpos=\!M{#1}\!xunit  \!ypos=\!M{#2}\!yunit  
  \!rotateaboutpivot\!xpos\!ypos%
  \advance\!xpos -\!xorigin  \advance\!xpos -\!xshift
  \advance\!ypos -\!yorigin  \advance\!ypos -\!yshift
  \kern\!xpos\raise\!ypos\copy\!putobject\kern-\!xpos%
  \!doaccounting
  \!multiput}
 
\def\!alsoby*#1 #2 #3 {%
  \!dxpos=\!M{#2}\!xunit \!dypos=\!M{#3}\!yunit 
  \!rotateonly\!dxpos\!dypos
  \!ntemp=#1%
  \!!loop\ifnum\!ntemp>0
    \advance\!xpos by \!dxpos  \advance\!ypos by \!dypos
    \kern\!xpos\raise\!ypos\copy\!putobject\kern-\!xpos%
    \advance\!ntemp by -1
  \repeat
  \!doaccounting 
  \!multiput}
 
\def\accountingon{\def\!doaccounting{\!!doaccounting}\ignorespaces}

\accountingon
\def\!!doaccounting{%
  \!xtemp=\!xpos  
  \!ytemp=\!ypos
  \ifdim\!xtemp<\!xleft 
     \!xleft=\!xtemp 
  \fi
  \advance\!xtemp by  \!wd 
  \ifdim\!xright<\!xtemp 
    \!xright=\!xtemp
  \fi
  \advance\!ytemp by -\!dp
  \ifdim\!ytemp<\!ybot  
    \!ybot=\!ytemp
  \fi
  \advance\!ytemp by  \!dp
  \advance\!ytemp by  \!ht 
  \ifdim\!ytemp>\!ytop  
    \!ytop=\!ytemp  
  \fi}
 
\long\def\!setputobject#1#2{%
  \setbox\!putobject=\hbox{#1}%
  \!ht=\ht\!putobject  \!dp=\dp\!putobject  \!wd=\wd\!putobject
  \wd\!putobject=\!zpt
  \!xshift=.5\!wd   \!yshift=.5\!ht   \advance\!yshift by -.5\!dp
  \edef\!putorientation{#2}%
  \expandafter\!SPOreadA\!putorientation[]\!nil%
  \expandafter\!SPOreadB\!putorientation<\!zpt,\!zpt>\!nil\ignorespaces}
 
\def\!SPOreadA#1[#2]#3\!nil{\!etfor\!orientation:=#2\do\!SPOreviseshift}
 
\def\!SPOreadB#1<#2,#3>#4\!nil{\advance\!xshift by -#2\advance\!yshift by -#3}
 
\def\!SPOreviseshift{%
  \if l\!orientation 
    \!xshift=\!zpt
  \else 
    \if r\!orientation 
      \!xshift=\!wd
    \else 
      \if b\!orientation
        \!yshift=-\!dp
      \else 
        \if B\!orientation 
          \!yshift=\!zpt
        \else 
          \if t\!orientation 
            \!yshift=\!ht
          \fi 
        \fi
      \fi
    \fi
  \fi}

\long\def\!dimenput#1#2(#3,#4){%
  \!setputobject{#1}{#2}%
  \!xpos=#3\advance\!xpos by -\!xshift
  \!ypos=#4\advance\!ypos by -\!yshift
  \kern\!xpos\raise\!ypos\box\!putobject\kern-\!xpos%
  \!doaccounting\ignorespaces}

\def\!setdimenmode{%
  \let\!M=\!M!!\ignorespaces}
\def\!setcoordmode{%
  \let\!M=\!M!\ignorespaces}
\def\!ifcoordmode{%
  \ifx \!M \!M!}
\def\!ifdimenmode{%
  \ifx \!M \!M!!}
\def\!M!#1#2{#1#2} 
\def\!M!!#1#2{#1}
\!setcoordmode
\let\setdimensionmode=\!setdimenmode
\let\setcoordinatemode=\!setcoordmode




\def\!stack[#1]{%
  \let\!lglue=\hfill \let\!rglue=\hfill
  \expandafter\let\csname !#1glue\endcsname=\relax
  \!ifnextchar<{\!!stack}{\!!stack<\stackleading>}}
\def\!!stack<#1>#2{%
  \vbox{\def\!valueslist{}\!ecfor\!value:=#2\do{%
    \expandafter\!rightappend\!value\withCS{\\}\to\!valueslist}%
    \!lop\!valueslist\to\!value
    \let\\=\cr\lineskiplimit=\maxdimen\lineskip=#1%
    \baselineskip=-1000pt\halign{\!lglue##\!rglue\cr \!value\!valueslist\cr}}%
  \ignorespaces}


\def\!lines[#1]#2{%
  \let\!lglue=\hfill \let\!rglue=\hfill
  \expandafter\let\csname !#1glue\endcsname=\relax
  \vbox{\halign{\!lglue##\!rglue\cr #2\crcr}}%
  \ignorespaces}


\def\!Lines[#1]#2{%
  \let\!lglue=\hfill \let\!rglue=\hfill
  \expandafter\let\csname !#1glue\endcsname=\relax
  \vtop{\halign{\!lglue##\!rglue\cr #2\crcr}}%
  \ignorespaces}

 
 
 
\def\setplotsymbol(#1#2){%
  \!setputobject{#1}{#2}
  \setbox\!plotsymbol=\box\!putobject%
  \!plotsymbolxshift=\!xshift 
  \!plotsymbolyshift=\!yshift 
  \ignorespaces}
 
\setplotsymbol({\fiverm .})

 
\def\!!plot(#1,#2){%
  \!dimenA=-\!plotxorigin \advance \!dimenA by #1
  \!dimenB=-\!plotyorigin \advance \!dimenB by #2
  \kern\!dimenA\raise\!dimenB\copy\!plotsymbol\kern-\!dimenA%
  \ignorespaces}
 
\def\!!!plot(#1,#2){%
  \!dimenA=-\!plotxorigin \advance \!dimenA by #1
  \!dimenB=-\!plotyorigin \advance \!dimenB by #2
  \kern\!dimenA\raise\!dimenB\copy\!plotsymbol\kern-\!dimenA%
  \!countE=\!dimenA
  \!countF=\!dimenB
  \immediate\write\!replotfile{\the\!countE,\the\!countF.}%
  \ignorespaces}

\def\savelinesandcurves on "#1" {%
  \immediate\closeout\!replotfile
  \immediate\openout\!replotfile=#1%
  \let\!plot=\!!!plot}

\def\dontsavelinesandcurves {%
  \let\!plot=\!!plot}
\dontsavelinesandcurves

{\catcode`\%=11\xdef\!Commentsignal{
\def\writesavefile#1 {%
  \immediate\write\!replotfile{\!Commentsignal #1}%
  \ignorespaces}

\def\replot"#1" {%
  \expandafter\!replot\input #1 /}
\def\!replot#1,#2. {%
  \!dimenA=#1sp
  \kern\!dimenA\raise#2sp\copy\!plotsymbol\kern-\!dimenA
  \futurelet\!nextchar\!!replot}
\def\!!replot{%
  \if /\!nextchar 
    \def\!next{\!finish}%
  \else
    \def\!next{\!replot}%
  \fi
  \!next}


 
 
\def\!Pythag#1#2#3{%
  \!dimenE=#1\relax                                     
  \ifdim\!dimenE<\!zpt 
    \!dimenE=-\!dimenE 
  \fi
  \!dimenF=#2\relax
  \ifdim\!dimenF<\!zpt 
    \!dimenF=-\!dimenF 
  \fi
  \advance \!dimenF by \!dimenE
  \ifdim\!dimenF=\!zpt 
    \!dimenG=\!zpt
  \else 
    \!divide{8\!dimenE}\!dimenF\!dimenE
    \advance\!dimenE by -4pt
      \!dimenE=2\!dimenE
    \!removept\!dimenE\!!t
    \!dimenE=\!!t\!dimenE
    \advance\!dimenE by 64pt
    \divide \!dimenE by 2
    \!dimenH=7pt
    \!!Pythag\!!Pythag\!!Pythag
    \!removept\!dimenH\!!t
    \!dimenG=\!!t\!dimenF
    \divide\!dimenG by 8
  \fi
  #3=\!dimenG
  \ignorespaces}

\def\!!Pythag{
  \!divide\!dimenE\!dimenH\!dimenI
  \advance\!dimenH by \!dimenI
    \divide\!dimenH by 2}

\def\placehypotenuse for <#1> and <#2> in <#3> {%
  \!Pythag{#1}{#2}{#3}}

 
 
 
\def\!qjoin (#1,#2) (#3,#4){%
  \advance\!intervalno by 1
  \!ifcoordmode
    \edef\!xmidpt{#1}\edef\!ymidpt{#2}%
  \else
    \!dimenA=#1\relax \edef\!xmidpt{\the\!dimenA}%
    \!dimenA=#2\relax \edef\!ymidpt{\the\!dimenA}%
  \fi
  \!xM=\!M{#1}\!xunit  \!yM=\!M{#2}\!yunit   \!rotateaboutpivot\!xM\!yM
  \!xE=\!M{#3}\!xunit  \!yE=\!M{#4}\!yunit   \!rotateaboutpivot\!xE\!yE
%
  \!dimenA=\!xM  \advance \!dimenA by -\!xS
  \!dimenB=\!xE  \advance \!dimenB by -\!xM
  \!xB=3\!dimenA \advance \!xB by -\!dimenB
  \!xC=2\!dimenB \advance \!xC by -2\!dimenA
%
  \!dimenA=\!yM  \advance \!dimenA by -\!yS%
  \!dimenB=\!yE  \advance \!dimenB by -\!yM%
  \!yB=3\!dimenA \advance \!yB by -\!dimenB%
  \!yC=2\!dimenB \advance \!yC by -2\!dimenA%
%
  \!xprime=\!xB  \!yprime=\!yB
  \!dxprime=.5\!xC  \!dyprime=.5\!yC
  \!getf \!midarclength=\!dimenA
  \!getf \advance \!midarclength by 4\!dimenA
  \!getf \advance \!midarclength by \!dimenA
  \divide \!midarclength by 12
%
  \!arclength=\!dimenA
  \!getf \advance \!arclength by 4\!dimenA
  \!getf \advance \!arclength by \!dimenA
  \divide \!arclength by 12
  \advance \!arclength by \!midarclength
  \global\advance \totalarclength by \!arclength
%
%
  \ifdim\!distacross>\!arclength 
    \advance \!distacross by -\!arclength
  \else
    \!initinverseinterp
    \loop\ifdim\!distacross<\!arclength
      \!inverseinterp
      \!xpos=\!t\!xC \advance\!xpos by \!xB
        \!xpos=\!t\!xpos \advance \!xpos by \!xS
      \!ypos=\!t\!yC \advance\!ypos by \!yB
        \!ypos=\!t\!ypos \advance \!ypos by \!yS
      \!plotifinbounds
      \advance\!distacross \plotsymbolspacing
      \!advancedashing
    \repeat  
    \advance \!distacross by -\!arclength
  \fi
  \!xS=\!xE
  \!yS=\!yE
  \ignorespaces}

\def\!getf{\!Pythag\!xprime\!yprime\!dimenA%
  \advance\!xprime by \!dxprime
  \advance\!yprime by \!dyprime}

\def\!initinverseinterp{%
  \ifdim\!arclength>\!zpt
    \!divide{8\!midarclength}\!arclength\!dimenE
    \ifdim\!dimenE<\!wmin \!setinverselinear
    \else 
      \ifdim\!dimenE>\!wmax \!setinverselinear
      \else
        \def\!inverseinterp{\!inversequad}\ignorespaces
%
%
         \!removept\!dimenE\!Ew
         \!dimenF=-\!Ew\!dimenE
         \advance\!dimenF by 32pt
         \!dimenG=8pt 
         \advance\!dimenG by -\!dimenE
         \!dimenG=\!Ew\!dimenG
         \!divide\!dimenF\!dimenG\!beta
         \!gamma=1pt
         \advance \!gamma by -\!beta
      \fi
    \fi
  \fi
  \ignorespaces}

\def\!inversequad{%
  \!divide\!distacross\!arclength\!dimenG
  \!removept\!dimenG\!v
  \!dimenG=\!v\!gamma
  \advance\!dimenG by \!beta
  \!dimenG=\!v\!dimenG
  \!removept\!dimenG\!t}

\def\!setinverselinear{%
  \def\!inverseinterp{\!inverselinear}%
  \divide\!dimenE by 8 \!removept\!dimenE\!t
  \!countC=\!intervalno \multiply \!countC 2
  \!countB=\!countC     \advance \!countB -1
  \!countA=\!countB     \advance \!countA -1
  \wlog{\the\!countB th point (\!xmidpt,\!ymidpt) being plotted 
    doesn't lie in the}%
  \wlog{ middle third of the arc between the \the\!countA th 
    and \the\!countC th points:}%
  \wlog{ [arc length \the\!countA\space to \the\!countB]/[arc length 
    \the \!countA\space to \the\!countC]=\!t.}%
  \ignorespaces}
 
\def\!inverselinear{%
  \!divide\!distacross\!arclength\!dimenG
  \!removept\!dimenG\!t}

 

\def\startrotation{%
  \let\!rotateaboutpivot=\!!rotateaboutpivot
  \let\!rotateonly=\!!rotateonly
  \!ifnextchar{b}{\!getsincos }%
    {\!getsincos by {\!cosrotationangle} {\!sinrotationangle} }}
\def\!getsincos by #1 #2 {%
  \edef\!cosrotationangle{#1}%
  \edef\!sinrotationangle{#2}%
  \!ifcoordmode 
    \let\!ROnext=\!ccheckforpivot
  \else
    \let\!ROnext=\!dcheckforpivot
  \fi
  \!ROnext}
\def\!ccheckforpivot{%
  \!ifnextchar{a}{\!cgetpivot}%
    {\!cgetpivot about {\!xpivotcoord} {\!ypivotcoord} }}
\def\!cgetpivot about #1 #2 {%
  \edef\!xpivotcoord{#1}%
  \edef\!ypivotcoord{#2}%
  \!xpivot=#1\!xunit  \!ypivot=#2\!yunit
  \ignorespaces}
\def\!dcheckforpivot{%
  \!ifnextchar{a}{\!dgetpivot}{\ignorespaces}}
\def\!dgetpivot about #1 #2 {%
  \!xpivot=#1\relax  \!ypivot=#2\relax
  \ignorespaces}

\def\stoprotation{%
  \let\!rotateaboutpivot=\!!!rotateaboutpivot
  \let\!rotateonly=\!!!rotateonly
  \ignorespaces}
 
\def\!!rotateaboutpivot#1#2{%
  \!dimenA=#1\relax  \advance\!dimenA -\!xpivot
  \!dimenB=#2\relax  \advance\!dimenB -\!ypivot
  \!dimenC=\!cosrotationangle\!dimenA
    \advance \!dimenC -\!sinrotationangle\!dimenB
  \!dimenD=\!cosrotationangle\!dimenB
    \advance \!dimenD  \!sinrotationangle\!dimenA
  \advance\!dimenC \!xpivot  \advance\!dimenD \!ypivot
  #1=\!dimenC  #2=\!dimenD
  \ignorespaces}

\def\!!rotateonly#1#2{%
  \!dimenA=#1\relax  \!dimenB=#2\relax 
  \!dimenC=\!cosrotationangle\!dimenA
    \advance \!dimenC -\!rotsign\!sinrotationangle\!dimenB
  \!dimenD=\!cosrotationangle\!dimenB
    \advance \!dimenD  \!rotsign\!sinrotationangle\!dimenA
  #1=\!dimenC  #2=\!dimenD
  \ignorespaces}
\def\!rotsign{}
\def\!!!rotateaboutpivot#1#2{\relax}
\def\!!!rotateonly#1#2{\relax}
\stoprotation

\def\!reverserotateonly#1#2{%
  \def\!rotsign{-}%
  \!rotateonly{#1}{#2}%
  \def\!rotsign{}%
  \ignorespaces}

\def\!getspan span <#1>{%
  \!dshade=#1\relax
  \!ifcoordmode 
    \let\!GRnext=\!GRccheckforAP
  \else
    \let\!GRnext=\!GRdcheckforAP
  \fi
  \!GRnext}
\def\!GRccheckforAP{%
  \!ifnextchar{p}{\!cgetanchor }
    {\!cgetanchor point at {\!xshadesave} {\!yshadesave} }}
\def\!cgetanchor point at #1 #2 {%
  \edef\!xshadesave{#1}\edef\!yshadesave{#2}%
  \!xshade=\!xshadesave\!xunit  \!yshade=\!yshadesave\!yunit
  \ignorespaces}
\def\!GRdcheckforAP{%
  \!ifnextchar{p}{\!dgetanchor}%
    {\ignorespaces}}
\def\!dgetanchor point at #1 #2 {%
  \!xshade=#1\relax  \!yshade=#2\relax
  \ignorespaces}

\def\setshadesymbol{%
  \!ifnextchar<{\!setshadesymbol}{\!setshadesymbol<,,,> }}

\def\!setshadesymbol <#1,#2,#3,#4> (#5#6){%
  \!setputobject{#5}{#6}%
  \setbox\!shadesymbol=\box\!putobject%
  \!shadesymbolxshift=\!xshift \!shadesymbolyshift=\!yshift
%
  \!dimenA=\!xshift \advance\!dimenA \!smidge
  \!override\!dimenA{#1}\!lshrinkage%
  \!dimenA=\!wd \advance \!dimenA -\!xshift
    \advance\!dimenA \!smidge
    \!override\!dimenA{#2}\!rshrinkage
  \!dimenA=\!dp \advance \!dimenA \!yshift
    \advance\!dimenA \!smidge
    \!override\!dimenA{#3}\!bshrinkage
  \!dimenA=\!ht \advance \!dimenA -\!yshift
    \advance\!dimenA \!smidge
    \!override\!dimenA{#4}\!tshrinkage
  \ignorespaces}
\def\!smidge{-.2pt}%

\def\!override#1#2#3{%
  \edef\!!override{#2}%
  \ifx \!!override\empty
    #3=#1\relax
  \else
    \if z\!!override
      #3=\!zpt
    \else
      \ifx \!!override\!blankz
        #3=\!zpt
      \else
        #3=#2\relax
      \fi
    \fi
  \fi
  \ignorespaces}
\def\!blankz{ z}

\setshadesymbol ({\fiverm .})

\def\!startvshade#1(#2,#3,#4){%
  \let\!!xunit=\!xunit%
  \let\!!yunit=\!yunit%
  \let\!!xshade=\!xshade%
  \let\!!yshade=\!yshade%
  \def\!getshrinkages{\!vgetshrinkages}%
  \let\!setshadelocation=\!vsetshadelocation%
  \!xS=\!M{#2}\!!xunit
  \!ybS=\!M{#3}\!!yunit
  \!ytS=\!M{#4}\!!yunit
  \!shadexorigin=\!xorigin  \advance \!shadexorigin \!shadesymbolxshift
  \!shadeyorigin=\!yorigin  \advance \!shadeyorigin \!shadesymbolyshift
  \ignorespaces}
 
\def\!starthshade#1(#2,#3,#4){%
  \let\!!xunit=\!yunit%
  \let\!!yunit=\!xunit%
  \let\!!xshade=\!yshade%
  \let\!!yshade=\!xshade%
  \def\!getshrinkages{\!hgetshrinkages}%
  \let\!setshadelocation=\!hsetshadelocation%
  \!xS=\!M{#2}\!!xunit
  \!ybS=\!M{#3}\!!yunit
  \!ytS=\!M{#4}\!!yunit
  \!shadexorigin=\!xorigin  \advance \!shadexorigin \!shadesymbolxshift
  \!shadeyorigin=\!yorigin  \advance \!shadeyorigin \!shadesymbolyshift
  \ignorespaces}

\def\!lattice#1#2#3#4#5{%
  \!dimenA=#1
  \!dimenB=#2
  \!countB=\!dimenB
%
  \!dimenC=#3
  \advance\!dimenC -\!dimenA
  \!countA=\!dimenC
  \divide\!countA \!countB
  \ifdim\!dimenC>\!zpt
    \!dimenD=\!countA\!dimenB
    \ifdim\!dimenD<\!dimenC
      \advance\!countA 1 
    \fi
  \fi
  \!dimenC=\!countA\!dimenB
    \advance\!dimenC \!dimenA
  #4=\!countA
  #5=\!dimenC
  \ignorespaces}

\def\!qshade#1(#2,#3,#4)#5(#6,#7,#8){%
  \!xM=\!M{#2}\!!xunit
  \!ybM=\!M{#3}\!!yunit
  \!ytM=\!M{#4}\!!yunit
  \!xE=\!M{#6}\!!xunit
  \!ybE=\!M{#7}\!!yunit
  \!ytE=\!M{#8}\!!yunit
  \!getcoeffs\!xS\!ybS\!xM\!ybM\!xE\!ybE\!ybB\!ybC
  \!getcoeffs\!xS\!ytS\!xM\!ytM\!xE\!ytE\!ytB\!ytC
  \def\!getylimits{\!qgetylimits}%
  \!shade{#1}\ignorespaces}
 
\def\!lshade#1(#2,#3,#4){%
  \!xE=\!M{#2}\!!xunit
  \!ybE=\!M{#3}\!!yunit
  \!ytE=\!M{#4}\!!yunit
  \!dimenE=\!xE  \advance \!dimenE -\!xS
  \!dimenC=\!ytE \advance \!dimenC -\!ytS
  \!divide\!dimenC\!dimenE\!ytB
  \!dimenC=\!ybE \advance \!dimenC -\!ybS
  \!divide\!dimenC\!dimenE\!ybB
  \def\!getylimits{\!lgetylimits}%
  \!shade{#1}\ignorespaces}
 
\def\!getcoeffs#1#2#3#4#5#6#7#8{%
  \!dimenC=#4\advance \!dimenC -#2
  \!dimenE=#3\advance \!dimenE -#1
  \!divide\!dimenC\!dimenE\!dimenF
  \!dimenC=#6\advance \!dimenC -#4
  \!dimenH=#5\advance \!dimenH -#3
  \!divide\!dimenC\!dimenH\!dimenG
  \advance\!dimenG -\!dimenF
  \advance \!dimenH \!dimenE
  \!divide\!dimenG\!dimenH#8
  \!removept#8\!t
  #7=-\!t\!dimenE
  \advance #7\!dimenF
  \ignorespaces}

\def\!shade#1{%
  \!getshrinkages#1<,,,>\!nil
  \advance \!dimenE \!xS
  \!lattice\!!xshade\!dshade\!dimenE
    \!parity\!xpos
  \!dimenF=-\!dimenF
    \advance\!dimenF \!xE
  \!loop\!not{\ifdim\!xpos>\!dimenF}
    \!shadecolumn%
    \advance\!xpos \!dshade
    \advance\!parity 1
  \repeat
  \!xS=\!xE
  \!ybS=\!ybE
  \!ytS=\!ytE
  \ignorespaces}

\def\!vgetshrinkages#1<#2,#3,#4,#5>#6\!nil{%
  \!override\!lshrinkage{#2}\!dimenE
  \!override\!rshrinkage{#3}\!dimenF
  \!override\!bshrinkage{#4}\!dimenG
  \!override\!tshrinkage{#5}\!dimenH
  \ignorespaces}
\def\!hgetshrinkages#1<#2,#3,#4,#5>#6\!nil{%
  \!override\!lshrinkage{#2}\!dimenG
  \!override\!rshrinkage{#3}\!dimenH
  \!override\!bshrinkage{#4}\!dimenE
  \!override\!tshrinkage{#5}\!dimenF
  \ignorespaces}

\def\!shadecolumn{%
  \!dxpos=\!xpos
  \advance\!dxpos -\!xS
  \!removept\!dxpos\!dx
  \!getylimits
  \advance\!ytpos -\!dimenH
  \advance\!ybpos \!dimenG
  \!yloc=\!!yshade
  \ifodd\!parity 
     \advance\!yloc \!dshade
  \fi
  \!lattice\!yloc{2\!dshade}\!ybpos%
    \!countA\!ypos
  \!dimenA=-\!shadexorigin \advance \!dimenA \!xpos
  \loop\!not{\ifdim\!ypos>\!ytpos}
    \!setshadelocation
    \!rotateaboutpivot\!xloc\!yloc%
    \!dimenA=-\!shadexorigin \advance \!dimenA \!xloc
    \!dimenB=-\!shadeyorigin \advance \!dimenB \!yloc
    \kern\!dimenA \raise\!dimenB\copy\!shadesymbol \kern-\!dimenA
    \advance\!ypos 2\!dshade
  \repeat
  \ignorespaces}
 
\def\!qgetylimits{%
  \!dimenA=\!dx\!ytC              
  \advance\!dimenA \!ytB
  \!ytpos=\!dx\!dimenA
  \advance\!ytpos \!ytS
  \!dimenA=\!dx\!ybC              
  \advance\!dimenA \!ybB
  \!ybpos=\!dx\!dimenA
  \advance\!ybpos \!ybS}
 
\def\!lgetylimits{%
  \!ytpos=\!dx\!ytB
  \advance\!ytpos \!ytS
  \!ybpos=\!dx\!ybB
  \advance\!ybpos \!ybS}
 
\def\!vsetshadelocation{
  \!xloc=\!xpos
  \!yloc=\!ypos}
\def\!hsetshadelocation{
  \!xloc=\!ypos
  \!yloc=\!xpos}





\def\!axisticks {%
  \def\!nextkeyword##1 {%
    \expandafter\ifx\csname !ticks##1\endcsname \relax
      \def\!next{\!fixkeyword{##1}}%
    \else
      \def\!next{\csname !ticks##1\endcsname}%
    \fi
    \!next}%
  \!axissetup
    \def\!axissetup{\relax}%
  \edef\!ticksinoutsign{\!ticksinoutSign}%
  \!ticklength=\longticklength
  \!tickwidth=\linethickness
  \!gridlinestatus
  \!setticktransform
  \!maketick
  \!tickcase=0
  \def\!LTlist{}%
  \!nextkeyword}

\def\ticksout{%
  \def\!ticksinoutSign{+}}

\ticksout

\def\nogridlines{%
  \def\!gridlinestatus{\!gridlinestoofalse}}
\nogridlines

\def\loggedticks{%
  \def\!setticktransform{\let\!ticktransform=\!logten}}
\def\unloggedticks{%
  \def\!setticktransform{\let\!ticktransform=\!donothing}}
\def\!donothing#1#2{\def#2{#1}}
\unloggedticks

\expandafter\def\csname !ticks/\endcsname{%
  \!not {\ifx \!LTlist\empty}
    \!placetickvalues
  \fi
  \def\!tickvalueslist{}%
  \def\!LTlist{}%
  \expandafter\csname !axis/\endcsname}

\def\!maketick{%
  \setbox\!boxA=\hbox{%
    \beginpicture
      \!setdimenmode
      \setcoordinatesystem point at {\!zpt} {\!zpt}   
      \linethickness=\!tickwidth
      \ifdim\!ticklength>\!zpt
        \putrule from {\!zpt} {\!zpt} to
          {\!ticksinoutsign\!tickxsign\!ticklength}
          {\!ticksinoutsign\!tickysign\!ticklength}
      \fi
      \if!gridlinestoo
        \putrule from {\!zpt} {\!zpt} to
          {-\!tickxsign\!xaxislength} {-\!tickysign\!yaxislength}
      \fi
    \endpicturesave <\!Xsave,\!Ysave>}%
    \wd\!boxA=\!zpt}
  
\def\!ticksin{%
  \def\!ticksinoutsign{-}%
  \!maketick
  \!nextkeyword}

\def\!ticksout{%
  \def\!ticksinoutsign{+}%
  \!maketick
  \!nextkeyword}

\def\!tickslength<#1> {%
  \!ticklength=#1\relax
  \!maketick
  \!nextkeyword}

\def\!tickslong{%
  \!tickslength<\longticklength> }

\def\!ticksshort{%
  \!tickslength<\shortticklength> }

\def\!tickswidth<#1> {%
  \!tickwidth=#1\relax
  \!maketick
  \!nextkeyword}

\def\!ticksandacross{%
  \!gridlinestootrue
  \!maketick
  \!nextkeyword}

\def\!ticksbutnotacross{%
  \!gridlinestoofalse
  \!maketick
  \!nextkeyword}

\def\!tickslogged{%
  \let\!ticktransform=\!logten
  \!nextkeyword}

\def\!ticksunlogged{%
  \let\!ticktransform=\!donothing
  \!nextkeyword}

\def\!ticksunlabeled{%
  \!tickcase=0
  \!nextkeyword}

\def\!ticksnumbered{%
  \!tickcase=1
  \!nextkeyword}

\def\!tickswithvalues#1/ {%
  \edef\!tickvalueslist{#1! /}%
  \!tickcase=2
  \!nextkeyword}

\def\!ticksquantity#1 {%
  \ifnum #1>1
    \!updatetickoffset
    \!countA=#1\relax
    \advance \!countA -1
    \!ticklocationincr=\!axisLength
      \divide \!ticklocationincr \!countA
    \!ticklocation=\!axisstart
    \loop \!not{\ifdim \!ticklocation>\!axisend}
      \!placetick\!ticklocation
      \ifcase\!tickcase
          \relax 
        \or
          \relax 
        \or
          \expandafter\!gettickvaluefrom\!tickvalueslist
          \edef\!tickfield{{\the\!ticklocation}{\!value}}%
          \expandafter\!listaddon\expandafter{\!tickfield}\!LTlist%
      \fi
      \advance \!ticklocation \!ticklocationincr
    \repeat
  \fi
  \!nextkeyword}

\def\!ticksat#1 {%
  \!updatetickoffset
  \edef\!Loc{#1}%
  \if /\!Loc
    \def\next{\!nextkeyword}%
  \else
    \!ticksincommon
    \def\next{\!ticksat}%
  \fi
  \next}    
      
\def\!ticksfrom#1 to #2 by #3 {%
  \!updatetickoffset
  \edef\!arg{#3}%
  \expandafter\!separate\!arg\!nil
  \!scalefactor=1
  \expandafter\!countfigures\!arg/
  \edef\!arg{#1}%
  \!scaleup\!arg by\!scalefactor to\!countE
  \edef\!arg{#2}%
  \!scaleup\!arg by\!scalefactor to\!countF
  \edef\!arg{#3}%
  \!scaleup\!arg by\!scalefactor to\!countG
  \loop \!not{\ifnum\!countE>\!countF}
    \ifnum\!scalefactor=1
      \edef\!Loc{\the\!countE}%
    \else
      \!scaledown\!countE by\!scalefactor to\!Loc
    \fi
    \!ticksincommon
    \advance \!countE \!countG
  \repeat
  \!nextkeyword}

\def\!updatetickoffset{%
  \!dimenA=\!ticksinoutsign\!ticklength
  \ifdim \!dimenA>\!offset
    \!offset=\!dimenA
  \fi}

\def\!placetick#1{%
  \if!xswitch
    \!xpos=#1\relax
    \!ypos=\!axisylevel
  \else
    \!xpos=\!axisxlevel
    \!ypos=#1\relax
  \fi
  \advance\!xpos \!Xsave
  \advance\!ypos \!Ysave
  \kern\!xpos\raise\!ypos\copy\!boxA\kern-\!xpos
  \ignorespaces}

\def\!gettickvaluefrom#1 #2 /{%
  \edef\!value{#1}%
  \edef\!tickvalueslist{#2 /}%
  \ifx \!tickvalueslist\!endtickvaluelist
    \!tickcase=0
  \fi}
\def\!endtickvaluelist{! /}

\def\!ticksincommon{%
  \!ticktransform\!Loc\!t
  \!ticklocation=\!t\!!unit
  \advance\!ticklocation -\!!origin
  \!placetick\!ticklocation
  \ifcase\!tickcase
    \relax 
  \or 
    \ifdim\!ticklocation<-\!!origin
      \edef\!Loc{$\!Loc$}%
    \fi
    \edef\!tickfield{{\the\!ticklocation}{\!Loc}}%
    \expandafter\!listaddon\expandafter{\!tickfield}\!LTlist%
  \or 
    \expandafter\!gettickvaluefrom\!tickvalueslist
    \edef\!tickfield{{\the\!ticklocation}{\!value}}%
    \expandafter\!listaddon\expandafter{\!tickfield}\!LTlist%
  \fi}

\def\!separate#1\!nil{%
  \!ifnextchar{-}{\!!separate}{\!!!separate}#1\!nil}
\def\!!separate-#1\!nil{%
  \def\!sign{-}%
  \!!!!separate#1..\!nil}
\def\!!!separate#1\!nil{%
  \def\!sign{+}%
  \!!!!separate#1..\!nil}
\def\!!!!separate#1.#2.#3\!nil{%
  \def\!arg{#1}%
  \ifx\!arg\!empty
    \!countA=0
  \else
    \!countA=\!arg
  \fi
  \def\!arg{#2}%
  \ifx\!arg\!empty
    \!countB=0
  \else
    \!countB=\!arg
  \fi}
 
\def\!countfigures#1{%
  \if #1/%
    \def\!next{\ignorespaces}%
  \else
    \multiply\!scalefactor 10
    \def\!next{\!countfigures}%
  \fi
  \!next}

\def\!scaleup#1by#2to#3{%
  \expandafter\!separate#1\!nil
  \multiply\!countA #2\relax
  \advance\!countA \!countB
  \if -\!sign
    \!countA=-\!countA
  \fi
  #3=\!countA
  \ignorespaces}

\def\!scaledown#1by#2to#3{%
  \!countA=#1\relax
  \ifnum \!countA<0 
    \def\!sign{-}
    \!countA=-\!countA
  \else
    \def\!sign{}%
  \fi
  \!countB=\!countA
  \divide\!countB #2\relax
  \!countC=\!countB
    \multiply\!countC #2\relax
  \advance \!countA -\!countC
  \edef#3{\!sign\the\!countB.}
  \!countC=\!countA 
  \ifnum\!countC=0 
    \!countC=1
  \fi
  \multiply\!countC 10
  \!loop \ifnum #2>\!countC
    \edef#3{#3\!zero}%
    \multiply\!countC 10
  \repeat
  \edef#3{#3\the\!countA}
  \ignorespaces}

\def\!placetickvalues{%
  \advance\!offset \tickstovaluesleading
  \if!xswitch
    \setbox\!boxA=\hbox{%
      \def\\##1##2{%
        \!dimenput {##2} [B] (##1,\!axisylevel)}%
      \beginpicture 
        \!LTlist
      \endpicturesave <\!Xsave,\!Ysave>}%
    \!dimenA=\!axisylevel
      \advance\!dimenA -\!Ysave
      \advance\!dimenA \!tickysign\!offset
      \if -\!tickysign
        \advance\!dimenA -\ht\!boxA
      \else
        \advance\!dimenA  \dp\!boxA
      \fi
    \advance\!offset \ht\!boxA 
      \advance\!offset \dp\!boxA
    \!dimenput {\box\!boxA} [Bl] <\!Xsave,\!Ysave> (\!zpt,\!dimenA)
  \else
    \setbox\!boxA=\hbox{%
      \def\\##1##2{%
        \!dimenput {##2} [r] (\!axisxlevel,##1)}%
      \beginpicture 
        \!LTlist
      \endpicturesave <\!Xsave,\!Ysave>}%
    \!dimenA=\!axisxlevel
      \advance\!dimenA -\!Xsave
      \advance\!dimenA \!tickxsign\!offset
      \if -\!tickxsign
        \advance\!dimenA -\wd\!boxA
      \fi
    \advance\!offset \wd\!boxA
    \!dimenput {\box\!boxA} [Bl] <\!Xsave,\!Ysave> (\!dimenA,\!zpt)
  \fi}

\normalgraphs
\catcode`!=12 

\hfuzz=2pt
\font\svtnrm=cmr17

\def\tgrO{\tilde\grO}

\def\Got#1{\hbox{\aa#1}}
\def\bPhi{{\bf \Phi}}
\def\hi{{\bar{\Phi}}}
\def\Eta{{\Upsilon}}
\def\tU{\tilde{{U}}}
\def\tV{\tilde{{V}}}

\def\hX{\hat{X}}

\def\gsp1{{\Got s}{\Got p}(1)}

\def\tV{\tilde{V}}

\def\bdef{1}
\def\bfol{2}
\def\bhom{3}
\def\btop{4}
\def\bg2{5}
\def\bq{6}
\def\bex{7}
\def\bc{8}
\def\borb{A}
\centerline{\svtnrm 3-Sasakian Manifolds}
\bigskip
\centerline{\sc Charles P. Boyer and Krzysztof Galicki}
\footnote{}{\ninerm During the preparation of this work the authors
were supported by an NSF grant.}
\bigskip\bigskip
\baselineskip = 10 truept
\centerline{\bf An Introduction} 
\bigskip
We begin this review with a brief history of the subject for our
exposition shall have little to do with the chronology.
In 1960 Sasaki [Sas 1] introduced a geometric structure related to an
almost contact structure. This geometry became known 
as Sasakian geometry and has been studied 
extensively ever since. In 1970 Kuo [Kuo]
refined this notion and introduced manifolds with Sasakian $3$-structures
(see also [Kuo-Tach, Tach-Yu]). 
Independently, the same concept was invented by Udri\c ste [Ud].
Between 1970 and 1975 this new kind of 
geometry was investigated almost exclusively by
a group of Japanese geometers, including Ishihara, Kashiwada, Konishi, 
Kuo, Tachibana, Tanno, and Yu. Already in [Kuo] we learn that the
3-Sasakian geometry has some interesting topological implications. 
Using earlier results of Tachibana about the  harmonic forms
on compact Sasakian spaces [Tach], Kuo showed that odd Betti numbers up to the
middle dimension must be divisible by 4. In 1971 Kashiwada observed that
every 3-Sasakian manifold is Einstein with a positive Einstein
constant [Kas]. In the same year Tanno proved an interesting theorem about
the structure of the isometry group of every 3-Sasakian space [Tan 1].
In a related paper he studied a natural 3-dimensional foliation on such 
spaces showing that, if the foliation is regular, then the space of leaves
is an Einstein manifold of positive scalar curvature [Tan 2]. Tanno
clearly points to the importance of the 
analogy with the quaternionic Hopf fibration
$S^3\rightarrow S^7\rightarrow S^4,$ but does not go any further.  In fact,
Kashiwada's paper mentions a conjecture speculating that every 3-Sasakian
manifold is of constant curvature [Kas]. She attributed this conjecture
to Tanno and, at the time, these were the only known examples.

Very soon after, however, it became clear that such a conjecture could not
possibly be true. This is due to a couple of papers by Ishihara and Konishi
[I-Kon, Ish 1]. They made a fundamental observation that the space of leaves of
the natural 3-dimensional foliations mentioned above has a ``quaternionic
structure", part of which is the Einstein metric discovered by Tanno.  This
led Ishihara to an independent study of this ``sister geometry": quaternionic
K\"ahler manifolds [Ish 2].  His paper is very well-known and is almost always
cited as the source of the explicit coordinate description of quaternionic
K\"ahler geometry.  Among other results Ishihara showed that his definition
implies that the holonomy group of the metric is a subgroup of
$Sp(n)\!\cdot\!Sp(1),$ thus providing an important connection with the earlier
studies of such manifolds by Alekseevsky [Al 1], Bonan [Bon], Gray [Gra 1],
Kraines [Kra], and Wolf [Wol].  In 1975 Konishi [Kon] proved the existence of a
Sasakian $3$-structure on a natural principal $SO(3)$-bundle over any
quaternionic K\"ahler manifold of positive scalar curvature. This, with the
symmetric examples of Wolf, gives precisely all of the homogeneous 3-Sasakian
spaces.  Yet, at the time they did not appear explicitly and escaped any
systematic study until much later.

In fact, 1975 seems to be the year when 3-Sasakian manifolds are
relegated to an almost complete obscurity which lasted for about 15 years. 
From that point on the two
``sisters" fair very differently. The extent of this can be best
illustrated by the famous book on Einstein manifolds by Besse [Bes].
The book appeared in 1987 and provided the reader with
an excellent, up-to-date, and very complete account of
what was known about Einstein manifolds 10 years ago. But one is left
in the dark when trying to find 
references to any of the papers on 3-Sasakian manifolds we have cited;
3-Sasakian manifolds are never mentioned in Besse.  The other ``sister", on the
contrary, received a lot of space in a separate chapter.  Actually Einstein
metrics on Konishi's bundle do appear in Besse (see [Bes] 14.85, 14.86)
precisely in the context of the $SO(3)$-bundles over positive quaternionic
K\"ahler manifolds as a consequence of a theorem of B\'erard-Bergery ([Bes],
9.73). Obviously, the absence of 3-Sasakian spaces in Besse's book was the
result rather than the cause of this obscurity. One could even say it was
justified by the lack of any interesting examples. The authors have puzzled
over this phenomenon without any sound explanation. One can only speculate that
it is the holonomy reduction that made quaternionic K\"ahler manifolds so much
more attractive an object. Significantly, the holonomy group of a 3-Sasakian
manifold never reduces to a proper subgroup of the special orthogonal group.
And when in 1981 Salamon [Sal 1,2], independently with B\'erard-Bergery
[BeB\`er], generalized Penrose's twistor construction for self-dual 4-manifolds
introducing the twistor space over an arbitrary quaternionic K\"ahler manifold,
the research on quaternionic K\"ahler geometry flourished, fueled by powerful
tools from complex algebraic geometry. 

Finally, in the early nineties, 3-Sasakian manifolds start
a comeback. They begin to 
appear in two completely different contexts. First,
in the study of manifolds with real Killing spinors, Friedrich
and Kath notice that the existence of one such spinor
leads naturally  to a Sasakian-Einstein 
structure while three of them give the manifold
a 3-Sasakian structure [B-G-F-K, Fr-Kat 1]. 
Assuming regularity they are able to combine the result of Hitchin [Hit 1]
and Friedrich and Kurke [Fr-Kur] and obtain a classification
of all regular complete 7-manifolds with $3$-Sasakian structure [Fr-Kat 2].
This appears to be the first classification result about 3-Sasakian manifolds.
In 1993 the classification problem for manifolds admitting Killing spinors
found an elegant formulation in terms of holonomy groups [B\"ar].
B\"ar observes that if $(M,g)$ is a simply connected spin manifold with a 
non-trivial real
Killing spinor then the metric cone $(C(M),\bar{g})$ must admit a parallel 
spinor. In particular $(C(M),\bar{g})$ is Ricci-flat and
${\rm Hol}(\bar{g})$ is quite restricted so that only very few
groups can occur. One such possibility is 
${\rm Hol}(\bar{g})=Sp(m+1)$ which gives the cone a hyperk\"ahler structure. 
It easily follows that $M$ must be 3-Sasakian.

Independently, the hyperk\"ahler geometry of the cone $C(\cals)$
was the starting point of our 
research on 3-Sasakian manifold.
In 1991 the authors, together with Ben Mann, discovered that 3-Sasakian
manifolds appear naturally as levels sets of a certain moment map
on a hyperk\"ahler manifold with an isometric $SU(2)$-action rotating the
triple of complex structures [B-G-M 1]. In fact, if some obstructions for
the $SU(2)$-action vanish, then the hyperk\"ahler manifold is
precisely a cone on a 3-Sasakian space and, at the same time,
it is the Swann's bundle over the associated quaternionic K\"ahler orbifold
of positive scalar curvature [Sw]. We quickly realized that $\cals$ is
ultimately related to three other Einstein geometries: its hyperk\"ahler cone
$C(\cals)$, the associated twistor space $\calz$, and the associated
quaternionic K\"ahler orbifold $\calo$.  In this review we call the collection
of these four geometries together with all the relevant maps
$\diamondsuit(\cals)$. Thus, every $\cals$ comes together with a fundamental
diagram 
$$\matrix{&&C(\cals)&&\cr& \swarrow&&\searrow&\cr \calz &&\hskip
-15pt\la{4}\hskip -30pt\decdnar{}&& \cals.\cr & \searrow& &\swarrow&\cr
&&\calo&& \cr}$$ 
More importantly we also realized that, even when $\calo$ and
$\calz$ are compact Riemannian orbifolds, $\cals$ can be a smooth manifold.
This moment marks the beginning of our efforts to understand the geometry
and topology of 3-Sasakian manifolds. 

They have led us through the classification of all
3-Sasakian homogeneous spaces and a discovery of a new quotient construction
of infinitely many homotopy types of non-regular compact 3-Sasakian manifolds
[B-G-M-2].  In dimension 7 these examples turned out to be certain Eschenburg
bi-quotients of $U(3)$ by a 2-torus [Esch 1-2]. We gave a complete analysis of
the geometry and topology of such spaces [B-G-M 2]. The next important step was
the second author's work with Simon Salamon [G-Sal].  There we noticed that 
Kuo's theorem about odd Betti numbers of 3-Sasakian manifolds being divisible
by 4 missed a crucial point.  Because of the isometric $SU(2)$-action, all odd
Betti numbers up to the middle dimension must actually vanish. In the regular
case we were able to show that 3-Sasakian cohomology is just the primitive
cohomology of both $\calz$ and $\calo$.  These results were then extended to
the orbifold case in [B-G 1], where we also made a systematic study of the
orbifold twistor spaces $\calz$ and gave an orbifold extension of the LeBrun's
inversion theorem [Le 3].  Finally, the Vanishing Theorem for Betti numbers
provided us with the tools to study the geometry and topology of more
complicated examples.  This study [B-G-M-R 1,2] used a rational spectral
sequence and culminated in discovering that, in dimension 7, all rational
homology types not excluded by the Vanishing Theorem do occur and can be
constructed explicitly. These examples illustrate the richness of 3-Sasakian
geometry in dimension 7. For example, there is an infinite family of 3-Sasakian
7-manifolds that admit metrics of positive sectional curvature, while there is
another infinite family that can admit no metrics whose sectional curvature is
bounded below by an arbitrary fixed negative number! Later in [B-G-M 8] we
discovered how to handle the integral spectral sequence giving integral results
for our 7-dimensional examples up through the second homology group. We also
studied [B-G-M 7] the higher dimensional analogue showing that these meet with
an entirely different fate.

This review chapter is intended to give the reader a self-contained
account of everything we have learned about such spaces to date. 
We have tried to gather all the known results. In a chapter like this
it would be impossible to present every proof so we do quote some theorems
just referring to the literature. But we have tried to include
as many proofs as possible so that the review is not simply a long dry list of
theorems, propositions, and corollaries. When it comes to references we make
no claim of completeness, though we have tried to do our best. We apologize for
any omissions. At the end we hope to be able to convince our reader
that the 3-Sasakian geometry is at least as fascinating as any
other ``sister" geometry of the fundamental diagram $\diamondsuit(\cals)$.

Our review is organized as follows: We begin by setting up definitions,
notation, and describing elementary properties of Sasakian, Sasakian-Einstein,
and 3-Sasakian manifolds in Section \bdef. Next we discuss fundamentals
about the geometry of the associated foliations (arrows in the
diagram $\diamondsuit(\cals)$). We then give a classification of homogeneous
geometries in Section \bhom. Section \btop\  is all about Betti numbers
of Sasakian and 3-Sasakian manifolds while Section \bg2\ is a very brief 
look at the Killing spinors and $G_2$ structures. The following section
describes the geometry of the 3-Sasakian quotient construction. After this we
give a detailed study of ``toric" 3-Sasakian manifolds. We conclude with a
handful of open problems, questions, and some conjectures followed by an
appendix on fundamental properties of orbifolds.

\noindent{\sc Acknowledgments}: The authors would like thank Ben Mann
who is a friend and has been a collaborator on much of our work.
We also thank our other collaborators
Simon Salamon and Elmer Rees. We thank Roger Bielawski,
Alex Buium, Claude LeBrun, Liviu Ornea, and Uwe Semmelmann for
discussions and valuable comments. Last, but not least,
the second named author would like to than Max-Planck-Institute f\"ur
Mathematik in Bonn for support and hospitality. This review was written
during his stay in Bonn.
\vfil\eject
\centerline {\bf \bdef. Definitions and Basic Properties}
\medskip
In this section we introduce notation, definitions, and discuss some elementary
properties of Sasakian, Sasakian-Einstein, and 3-Sasakian manifolds.
Traditionally Sasakian structures were defined via contact structures by adding
a Riemannian metric with some additional conditions. We take a simpler and more
geometric approach that uses the holonomy reduction of the associated metric
cone.
\smallskip
\noindent
{\bf \bdef.1 Sasakian Manifolds} 
\smallskip
\noindent{\sc Definition \bdef.1.1}: \tensl
Let $(\cals,g)$ be a Riemannian manifold of real dimension $m$. 
We say that
$(\cals,g)$ is Sasakian if the holonomy group of the 
metric cone on $\cals$ \break $(C(\cals),{\bar g})=
(\bbr_+\times\cals, \ dr^2+r^2g)$ reduces to a subgroup of $U({m+1\over2})$.
In particular, $m=2n+1, n\geq1$ and $(C(\cals),\bar{g})$ is K\"ahler.
\tenrm

The following proposition provides three alternative characterizations of
the Sasakian property, the first one, perhaps, most in the spirit of the 
the original definition of Sasaki [Sas 1]:

\noindent{\sc Proposition \bdef.1.2}: \tensl
Let $(\cals,g)$ be a Riemannian manifold, $\nabla$ the Levi-Civita connection
of $g,$ and let $R(X,Y):\Gamma(T\cals)\rightarrow \Gamma(T\cals)$ denote
the Riemann curvature tensor of $\nabla.$ Then the following conditions
are equivalent:
\item{(i)}
There exists a Killing vector field $\xi$ of unit length on $\cals$
so that the tensor field $\Phi$ of type $(1,1)$, defined by
$\Phi(X) ~=~ \nabla_X \xi$,
satisfies the condition
$$(\nabla_X \Phi)(Y) ~=~ g(\xi,Y)X-g(X,Y)\xi\leqno{\bdef.1.3}$$
for any pair of vector fields $X$ and $Y$ on $\cals.$  
\item{(ii)} There exists a Killing vector field $\xi$ of unit length on $\cals$
so that the Riemann curvature satisfies the condition 
$$R(X,\xi)Y ~=~ g(\xi,Y)X-g(X,Y)\xi,\leqno{\bdef.1.4}$$
for any pair of vector fields $X$ and $Y$ on $\cals.$  
\item{(iii)} There exists a Killing vector field $\xi$ 
of unit length on $\cals$
so that the sectional curvature of every section containing $\xi$ equals one.
\item{(iv)} $(\cals, g)$ is Sasakian.
\tenrm

\noindent{\sc Proof}: We outline the proof of the equivalence of
(i) and (iv). The equivalence of (i) and (ii) is a simple calculation
relating $(\nabla_X \Phi)(Y)$ to $R(X,\xi)Y$ 
and is left to the reader (see [Y-K]). The equivalence of (ii) and (iii) is 
obvious.
  
We first show how (iv) implies (i).
Let $X,Y$ be any two vector fields on $\cals$ viewed as vector fields on
$C(\cals)$ and $\bar{\nabla}$ be the Levi-Civita connection of $\bar{g}$.
Then we have the following warped product formulas for the cone metric
connection [O'N, p. 206]:
$${\bar{\nabla}}_{\partial_r}\partial_r=0,\qquad
{\bar{\nabla}}_{\partial_r}X={\bar{\nabla}}_X\partial_r={1\over r}X,\qquad
{\bar{\nabla}}_XY=\nabla_XY-rg(X,Y)\partial_r.\leqno{\bdef.1.5}$$

Since the holonomy group of the cone $(C(\cals),\bar{g})$ reduces to
a subgroup of $U({m+1\over2})$ there is a parallel complex structure
$I$ on $C(\cals)$, i.e., $I$ commutes with
$\bar{\nabla}$. We can identify $\cals$ with $\cals\times\{1\}\subset C(\cals)$ 
and define
$$\xi=I(\partial_r),\qquad
\eta(Y)=g(\xi,Y), \qquad \Phi(Y)=\nabla_Y\xi\leqno{\bdef.1.6}$$
for any vector field $Y\in\Gamma(T\cals)$. It is then a simple calculation
to show that $\xi$ is actually a unit Killing vector field on $\cals$ and
it satisfies the curvature condition \bdef.1.3. Clearly, $\xi$ is unit
by definition and we have
$$g(\nabla_Y\xi,X)=\bar{g}({\bar{\nabla}}_Y\xi+g(\xi,Y)\partial_r,X)=
\bar{g}({\bar{\nabla}}_YI(\partial_r),X)=$$
$$=
\bar{g}(I({\bar{\nabla}}_Y\partial_r),X)=
\bar{g}(I(Y),X)$$ 
which is skew-symmetric in $X$ and $Y$.
The second condition follows from
$\bar{\nabla}I=0$, definition of $\Phi(Y)=\nabla_Y(I\partial_r)$, and
the formulas \bdef.1.5.
Conversely, we can 
construct a K\"ahler structure on
$C(\cals)$ as follows:  Let $\Psi = r\partial_ r$
denote the Euler field on $C(\cals)$ and define smooth section
of ${\rm End}~TC(\cals)$ by the formula
$$IY ~=~ \Phi (Y) - \eta(Y)\Psi,\qquad
I\Psi ~=~ \xi, \leqno{\bdef.1.7}$$
where $\eta(Y)=g(\xi,Y)$ is the dual 1-form of $\xi$.
It is easy to see that $I$ is an almost complex structure on $C(\cals)$ and the
metric $\bar{g}$ is Hermitian. To show that $C(\cals)$ is K\"ahler it is enough
to show that $\bar{\nabla}I=0.$ This is done by a direct calculation using
the definition of $I$ and equations \bdef.1.6.\hfill\za

The above discussion shows that there is a natural splitting of the tangent
bundle $TC(\cals)$ as
$TC(\cals) = \call_{\Psi}\oplus \call_\xi \oplus \calh$
where $\call_X$ denotes the trivial line bundle generated by the nowhere
vanishing vector field $X,$ and $\calh$ is a complement with respect to the
metric $\bar{g}.$ It follows immediately that the frame bundle of any
Sasakian manifold of dimension $2n+1$ reduces 
to the group $1\times U(n)$ [Sas 1]. It
follows that every Sasakian manifold has a canonical $\hbox{Spin}^c$
structure [Mor].

In view of the above proposition the triple $\{\xi,\eta,\Phi\}$ is called
a {\it Sasakian structure} on $(\cals, g),$ the Killing vector field $\xi$ and
the 1-form $\eta$ are called the {\it characteristic vector field} and the {\it
characteristic 1-form} of the Sasakian structure, respectively. We next give
some elementary properties of Sasakian structures.  All of them follow as an
immediate consequence of the definition and Proposition \bdef.1.2.

\noindent{\sc Proposition \bdef.1.8}: \tensl  Let $(\cals,g)$ be
a Sasakian manifold, $\{\xi,\eta,\Phi\}$ its
Sasakian structure, and $X$ and $Y$ any pair of vector fields on $\cals.$
Furthermore, let 
$N_{\Phi}(Y,X) ~=~ [\Phi Y,\Phi X]+\Phi^2[Y,X]-\Phi[Y,\Phi X]-\Phi[\Phi Y,X]$
be the Nijenhuis torsion tensor of $\Phi$.
Then
$$\Phi\!\circ\!\Phi(Y) ~=~ -Y+\eta(Y)\xi,\leqno{(i)}$$
$$\Phi\xi ~=~ 0,\qquad
\eta(\Phi Y) ~=~ 0,\leqno{(ii)}$$
$$g(X,\Phi Y) +g(\Phi X,Y)  ~=~ 0, \qquad
g(\Phi Y,\Phi X) ~=~ g(Y,X)-\eta(Y)\eta(X),\leqno{(iii)}$$
$$d\eta(Y,X) ~=~ 2g(\Phi Y,X),\qquad
N_{\Phi}(Y,X) ~=~ d\eta(Y,X)\otimes\xi.\leqno{(iv)}$$
\tenrm

A Sasakian manifold is not necessarily Einstein. 
As a simple consequence of the relation between Ricci curvature
of $\cals$ and its metric cone $C(\cals)$, the Einstein
condition can be expressed in terms of Ricci-flatness of the cone metric
$\bar{g}$ and we get

\noindent{\sc Proposition \bdef.1.9}: \tensl  Let $(\cals,g)$ be
be a Sasakian manifold of dimension $2n+1$. Then the metric $g$
is Einstein if and only if the cone metric $\bar{g}$ is Ricci-flat, i.e.,
$(C(\cals),\bar{g})$ 
is K\"ahler Ricci-flat (Calabi-Yau). In particular, it follows that the
restricted holonomy group ${\rm Hol}^0(\bar{g})\subset SU(n+1)$ and that the
Einstein constant of $g$ is positive and equals $2n$.\tenrm

An immediate consequence of the this proposition and Myers' Theorem is:

\noindent{\sc Corollary \bdef.1.10}: \tensl A complete Sasakian-Einstein
manifold is compact with diameter less than or equal to $\pi$ and with finite
fundamental group.  \tenrm

Now ${\rm Hol}^0(\bar{g})$ is the normal subgroup of the full holonomy group
${\rm Hol}(\bar{g})$ that is the component connected to the identity.  There is
a canonical epimorphism
$$\pi_1(\cals)=\pi_1(C(\cals))\ra{1.3} {\rm Hol}(\bar{g})/{\rm
Hol}^0(\bar{g}),$$ 
so if $\cals$ is simply-connected its structure group reduces to $1\times
SU(n)$ and it will admit a spin structure. We have

\noindent{\sc Corollary \bdef.1.11}: \tensl Let $\cals$ be a Sasakian-Einstein
manifold such that the full holonomy group of the cone metric ${\rm
Hol}(\bar{g})$ is contained in  $SU(m+1).$ Then $\cals$ admits a spin
structure.  In particular, every simply-connected Sasakian-Einstein manifold
admits a spin structure.  \tenrm

We give some examples that illustrate the complications in the presence of
fundamental group. The hypothesis of this corollary is not necessary as the
second example shows.

\noindent{\sc Examples \bdef.1.12}: The real projective space
$\cals=\bbr\bbp^{2n+1}$ with its canonical metric is Sasakian-Einstein, and the
cone $C(\cals)=(\bbc^{n+1}-\{0\})/\bbz_2$ with the usual antipodal
identification. We have ${\rm Hol}(\bar{g})\simeq \pi_1(\cals)\simeq \bbz_2.$
When $n$ is odd the antipodal map $\tau$ is in $SU(n+1),$ so
$\cals=\bbr\bbp^{2n+1}$ admits a spin structure. But when $n$ is even the
antipodal map $\tau$ does not lie in $SU(n+1),$ which obstructs a further
reduction of the structure group. In this case it is well-known that
$\cals=\bbr\bbp^{2n+1}$ does not admit a spin structure.  In fact the generator
of ${\rm Hol}(\bar{g})\simeq \bbz_2$ is the obstruction. There are many other
similar examples. An example that shows that the hypothesis in Corollary
\bdef.1.11 is not necessary is the following: Consider the lens space
$L(p;,q_1,\cdots,q_n)\simeq S^{2n+1}/\bbz_p$ where the $q_i$'s are relatively
prime to $p.$ The action on $\bbc^{n+1}-\{0\}$ is generated by
$(z_0,z_1,\cdots,z_n)\mapsto (\eta z_0,\eta^{q_1}z_1,\cdots,\eta^{q_n}z_n)$
where $\eta$ is a primitive $p$th root of unity. It is known [Fra] that if $p$
is odd, $L(p;q_1,\cdots,q_n)$ admits a spin structure. However, if $\sum_iq_i
+1$ is not divisible by $p,$ the holonomy group ${\rm Hol}(\bar{g})\simeq
\bbz_p$ does not lie in $SU(n+1).$

Let $\cals$ be a Sasakian manifold, suppose that the characteristic vector
field $\xi$ is complete. Since $\xi$ has unit norm, it defines a 1-dimensional
foliation $\calf$ on $\cals.$ We shall be interested in the case when all the
leaves of $\calf$ are compact. 

\noindent{\sc Definition \bdef.1.13}: \tensl Let $(\cals,g)$ be
be a compact Sasakian manifold and let $\calf$ be the 1-dimensional foliation
defined by $\xi$. We say that $\cals$ is {\it quasi-regular} if
the foliation $\calf$ is quasi-regular,
i.e., each point $p\in \cals$ has a cubical neighborhood $U$ such that any leaf
$\call$ of $\calf$ intersects a transversal through $p$ at most a finite number
of times $N(p).$ Furthermore, $\cals$ is called {\it regular} if  $N(p)=1$
for all $p\in \cals.$\tenrm

It is known that the quasi-regular property is equivalent to the condition that
all the leaves of the foliation are compact. In the regular case, the foliation
$\calf$ is simple, and defines a global submersion. In fact it defines a
principal $S^1$ bundle over its space of leaves. In the quasi-regular case it
is well-known [Tho, Mol] that $\xi$ generates a locally free circle action on
$\cals,$ and that the space of leaves is a compact orbifold (See the appendix
for a brief review of orbifolds and their relation to foliations, in particular
see \borb.3).  We shall denote the space of leaves of the
foliation $\calf$ on $\cals$ by $\calz.$ Then the natural projection
$\pi:\cals\ra{1.3} \calz$ is a Siefert fibration. It is an example of what we
call a principal V-bundle over $\calz.$ In Section 2 we shall study this
foliation in detail.
\smallskip
\noindent
{\bf \bdef.2 3-Sasakian Spaces}
\smallskip
Using all the definitions of the previous section we now 
describe a more specialized situation. Again, this can be done by an
additional holonomy reduction requirement.

\noindent{\sc Definition \bdef.2.1}: \tensl
Let $(\cals,g)$ be a Riemannian manifold of real dimension $m$.
We say that
$(\cals,g)$ is 3-Sasakian if the holonomy group of the
metric cone on $\cals$ \break $(C(\cals),{\bar g})=
(\bbr_+\times\cals, \ dr^2+r^2g)$ reduces to a subgroup of $Sp({m+1\over4})$.
In particular, $m=4n+3, n\geq1$ and $(C(\cals),\bar{g})$ is hyperk\"ahler.
\tenrm

Since $C(\cals)$ is hyperk\"ahler it has a hypercomplex structure $\{I^1, I^2, I^3\}$.
We can define $\xi^a=I^a(\partial_r)$ for each $a=1,2,3$. 
Then using the well-known properties of a hypercomplex structure together with
Proposition \bdef.1.2 gives:

\noindent{\sc Proposition \bdef.2.2}: \tensl
Let $(\cals,g)$ be a Riemannian manifold and let $\nabla$ denote the
Levi-Civita connection of $g$. Then $\cals$ is 3-Sasakian if and only if
it admits three characteristic vector fields $\{\xi^1,\xi^2,\xi^3\}$ (that is,
satisfying any of the corresponding conditions in Proposition \bdef.1.2) such
that $g(\xi^a,\xi^b)=\delta_{ab}$ and $[\xi^a,\xi^b]=2\epsilon_{abc}\xi^c$.
\tenrm

\noindent{\sc Remark \bdef.2.3}: By using Proposition \bdef.2.2 we can easily
generalize the definition of a 3-Sasakian structure to orbifolds. A Riemannian
orbifold $\cals$ is a {\it 3-Sasakian orbifold} if it admits three
characteristic vector fields satisfying the conditions of Proposition
\bdef.2.2, and if the action of the local uniformizing groups leaves the
characteristic vector fields $\{\xi^1,\xi^2,\xi^3\}$ invariant.

The triple $\{\xi^1,\xi^2,\xi^3\}$ defines
$\eta^a(Y)=g(\xi^a,Y)$ and $\Phi^a(Y)=\nabla_Y\xi^a$ for each $a=1,2,3$.
We call $\{\xi^a,\eta^a,\Phi^a\}_{a=1,2,3}$ the {\it 3-Sasakian structure}
on $(\cals,g)$. The hyperk\"ahler geometry of the cone $C(\cals)$ gives
$\cals$ a ``quaternionic structure" reflected by the composition laws
of the (1,1) tensors $\Phi^a$. The following proposition describes
additional properties of $\{\xi^a,\eta^a,\Phi^a\}$ not listed in Proposition
\bdef.1.8(i-iv).

\noindent{\sc Proposition \bdef.2.4}: \tensl
Let $(\cals,g)$ be a 3-Sasakian manifold and let
$\{\xi^a,\eta^a,\Phi^a\}_{a=1,2,3}$ be its 3-Sasakian structure.
Then
$$\eqalign{\eta^a(\xi^b) &~=~ \grd^{ab}, \cr
\Phi^a\xi^b &~=~ -\gre^{abc}\xi^c, \cr
\Phi^a\circ \Phi^b-\xi^a\otimes \eta^b &~=~  -\gre^{abc}\Phi^c
-\grd^{ab}\hbox{id}.\cr}$$
\tenrm

\noindent{\sc Remark \bdef.2.5}: For any $\tau=(\tau_1,\tau_2,\tau_3)\in
\bbr^3$ such that $\tau_1^2+\tau_3^2+\tau_3^2=1$ the vector field
$\xi(\tau)=\tau_1\xi^1+\tau_2\xi^2+\tau_3\xi^3$ has the Sasakian property.
Therefore a 3-Sasakian manifold has not just 3 but an $S^2$ worth of Sasakian
structures.  This is in complete analogy with the hyperk\"ahler case, and
perhaps the name {\it hypersasakian} would have been more consistent. However,
most of the existing literature uses the name {\it Sasakian 3-structure} or, as
we do, {\it 3-Sasakian structure}. Thus we have decided to stay with the
latter.

Since a hyperk\"ahler manifold is Ricci-flat, Proposition \bdef.1.9 and its
corollary immediately imply:

\noindent{\sc Corollary \bdef.2.6}: \tensl Every 3-Sasakian
manifold $(\cals, g)$ of dimension $4n+3$ is Einstein with
Einstein constant $\lambda=2(2n+1)$. Moreover, if $\cals$ is complete it is
compact with finite fundamental group. \tenrm

The important result that every 3-Sasakian manifold is Einstein was first
obtained by Kashiwada [Kas] using tensorial methods. One can also easily verify
the structure group of any 3-Sasakian manifold is reducible to $Sp(n)\times
\bbi_3$, where $\bbi_3$ denotes the three by three identity matrix [Kuo]. It
follows [B-G-M 2] that

\noindent{\sc Corollary \bdef.2.7}: \tensl Every 3-Sasakian
manifold $(\cals, g)$ is spin.\tenrm

If $(\cals, g)$ is compact the characteristic vector fields
$\{\xi^1,\xi^2,\xi^3\}$  are complete and define a 3-dimensional foliation
$\calf_3$ on $\cals.$ The leaves of this foliation are necessarily compact as
$\{\xi^1,\xi^2,\xi^3\}$ defines a locally free $Sp(1)$ action on $\cals$.
Hence, the foliation $\calf_3$ is automatically quasi-regular and the space of
leaves is a compact orbifold. We shall denote it by $\calo$.

\noindent{\sc Definition \bdef.2.8}: \tensl Let $(\cals,g)$ be
be a compact 3-Sasakian manifold of dimension 
$4n+3$, $n\geq1$, and let $\calf_3$ be the 3-dimensional foliation
defined by $\{\xi^1,\xi^2,\xi^3\}$. We say that $\cals$ is
{\it regular} if $\calf_3$ is regular. 
\tenrm

\noindent{\sc Remark \bdef.2.9}: When ${\rm dim}(\cals)=3$ the leaf
space of the foliation $\calf_3$ is a single point so it makes
no sense to talk about the regularity of $\calf_3$. In this case
we will say that $\cals$ is {\it regular} if the foliation $\calf_1$
defined by the characteristic vector field $\xi^1$ is regular. 

For any $\tau\in S^2$ we can consider again the characteristic
vector field $\xi(\tau)$ associated with the direction $\tau.$
This vector field defines a 1-dimensional foliation
$\calf_\tau\subset\calf_3\subset\cals$.  This foliation has compact leaves and
defines a locally free circle action $U(1)_\tau\subset Sp(1)$ on $\cals$. In
the next section we will describe the geometry of these foliations. Here, we
simply conclude by the following observation concerning regularity properties
of the foliations $\calf_\tau\subset\calf_3$ [Tan 2]:

\noindent{\sc Proposition \bdef.2.10}: \tensl 
Let $(\cals,g)$ 
be a compact 3-Sasakian manifold. If $\calf_3$ is regular then
$\calf_\tau$ is regular for all $\tau\in S^2$. Conversely, if
$\calf_\tau$ is regular for some $\tau=\tau_0\in S^2$ then it is regular
for all $\tau$ and, hence, $\calf_3$ is regular. Furthermore, if
$\calf_3$ is regular then either all the leaves are diffeomorphic to 
$SO(3)$ or all the leaves are diffeomorphic to $S^3.$  \tenrm 

Actually in the regular case it follows from a deeper result of Simon Salamon
[Sal 1] that all leaves are diffeomorphic to $S^3$ in precisely one case,
namely when $\cals = S^{4n+3}.$ (See the next section for further discussion.)

\noindent{\sc Remark \bdef.2.11}: Note that every Sasakian-Einstein
3-manifold must also have a 3-Sasakian structure. This is because in dimension
four Ricci-flat and K\"ahler is equivalent to hyperk\"ahler. Every
compact 3-Sasakian 3-manifold, by Proposition \bdef.1.2(iii), must be
a space of constant curvature 1. Hence, $\cals$ is covered by
a unit round 3-sphere and , in fact, it is always the homogeneous
spherical space form $S^3/\Gamma$, where $\Gamma$ is a discrete subgroup
of $Sp(1)$ [Sas 2]. The homogeneous spherical space forms in
dimension $3$ are well-known.  They are $Sp(1)/\grG$ where $\grG$ is one 
of the finite subgroups of $Sp(1)$, namely: 
$\grG= \bbz_m$ the cyclic group of order $m,$
$\grG= \bbd_m^{*}$ a binary dihedral group with $m$ is an integer
greater than 2,
$\grG= \bbt^{*}$ the binary tetrahedral group,
$\grG= \bbo^{*}$ the binary octahedral group,
$\grG= \bbi^{*}$ the binary icosahedral group. 
The only regular 3-Sasakian manifolds in dimension 3 are $S^3$ and
$SO(3)$. More generally, the diffeomorphism
classification of compact Sasakian 3-manifolds was recently completed
by Geiges [Gei]. In addition to $S^3/\Gamma$ one gets compact
quotients of the double cover of $PSL_2(\bbr)$ and the 3-dimensional
Heisenberg group.

\noindent{\sc Remark \bdef.2.12}: A Sasakian-Einstein
structure on a 3-Sasakian manifold does not have to be
a part of the 3-Sasakian structure. The simplest example when this
is the case is the lens
space $\bbz_k\backslash S^3$.
Consider the unit 3-sphere $S^3\simeq Sp(1)$ 
as the unit quaternion $\sigma\in\bbh$. 
Such a sphere has two 3-Sasakian structures
generated by the left and the right multiplication. 
Consider the homogeneous space
$\bbz_k\backslash S^3$, where the $\bbz_k$-action is given by
the multiplication from the left by $\rho\in Sp(1)$, $\rho^k=1$. 
The quotient still has the ``right" 3-Sasakian structure. But it also has
a  ``left" Sasakian structure (the centralizer of $\bbz_k$ in $Sp(1)$ is
an $S^1$ and it acts on the coset from the left).
This left Sasakian structure is actually regular while
none of the Sasakian structures of the right 3-Sasakian structure can be
regular unless $k=1,2$ [Tan 3].
\bigskip
\centerline{\bf \bfol. The Fundamental Foliations}
\medskip
In this section we discuss the foliations associated with Sasakian and
3-Sasakian manifolds and describe their consequences.
\smallskip
\noindent{\bf \bfol.1 The Sasakian Foliation} 
\smallskip
As mentioned in Section \bdef.1 a Sasakian manifold defines a Riemannian
foliation of dimension 1. Using the basic
properties described in Propositions \bdef.1.2 and \bdef.1.8. we have

\noindent{\sc Proposition \bfol.1.1}: \tensl Let $(\cals,g)$ be a Sasakian
manifold, and let $\calf$ denote the foliation defined by the characteristic
vector field $\xi.$ Then
\item{\rm(i)} The metric $g$ is bundle-like.
\item{\rm(ii)} The leaves of $\calf$ are totally geodesic. 
\item{\rm(iii)} The complementary vector bundle $\calh$ to the trivial line
subbundle of $T\cals$ generated by $\xi$ defines a strictly pseudoconvex CR
structure on $\cals$ with vanishing Webster torsion. \tenrm

In order to have a well behaved space of leaves we need a further assumption on
the foliation. We have a generalization of the well-known Boothby-Wang
fibration Theorem:

\noindent{\sc Theorem \bfol.1.2}: \tensl Let $\cals$ be a complete
quasi-regular Sasakian manifold. Then 
\item{\rm(i)} The leaves of $\calf$ are all diffeomorphic to circles with
cyclic leaf holonomy groups.
\item{\rm(ii)} The space of leaves $\calz =\cals/\calf$ has the
structure of a K\"ahler orbifold.

\noindent Suppose further that $(\cals,g)$ is Sasakian-Einstein. Then
\item{\rm(iii)} The leaf space $\calz$ is a simply-connected normal projective
algebraic variety with a K\"ahler-Einstein metric $h$ of positive scalar
curvature $4n(n+1)$ in such a way that $\pi:(\cals,g) \ra{1.3} (\calz,h)$ is
an orbifold Riemannian submersion.  
\item{(iv)} $\calz$ has the structure of a $\bbq$-factorial Fano variety.
Hence, it is uniruled with Kodaira dimension $\kappa(\calz)=-\infty.$ \tenrm

\noindent{\sc Proof}: Parts (i) and (ii) are straightforward generalizations of
the Boothby-Wang fibration in the Sasakian setting [Bl, K-Y] 
to the quasi-regular
case. The point is that the CR structure on $\cals$ pushes down to give a
complex structure on $\calz$ and the Sasakian nature of $\cals$ guarantees
that the complex structure will be K\"ahler. That $\calz$ is projective
algebraic is a consequence of Baily's version [Bai 2] of the Kodaira Embedding
Theorem.  Simple connectivity follows essentially from Kobayashi's argument in
the smooth case by using the singular version of the Riemann-Roch Theorem due
to Baum, Fulton, and Macpherson. The uniruledness is a result of Miyaoka and
Mori [Mi-Mo].  For details we refer the reader to [B-G 1, B-G 2].  \hfill\za

Let us recall that a complex variety $X$ is $\bbq$-factorial variety if for
every Weil divisor $D$ there exists a positive integer $m$ such that $mD$ is a
Cartier divisor. The smallest such integer $m(D)$ is called the order of $D.$
If $X$ is compact the least common multiple taken over all Weil divisors on
$X$ is the order of $X.$ Now on a compact complex orbifold Weil divisors
coincide with Baily divisors [B-G 1] and Baily divisors correspond to line
V-bundles. On $X$ we have the group $\hbox{Pic}^{orb}(X)$ of holomorphic line
V-bundles on $X$ and its subgroup $\hbox{Pic}(X)$ of holomorphic line
bundles or absolute line V-bundles in Baily's terminology [Bai 1, Bai 2]. It is
not difficult to prove [B-G 2]

\noindent{\sc Proposition \bfol.1.3}: \tensl Let $\cals$ be a complete
Sasakian-Einstein manifold, and let $\calz$ be the space of leaves of the
foliation $\calf$ on $\cals.$ Then $\hbox{Pic}(\calz)$ is free, and a subgroup
of $\hbox{Pic}^{orb}(\calz)$ which satisfies
\item{(i)} $\hbox{Pic}^{orb}(\calz)\otimes \bbq\simeq \hbox{Pic}(\calz)\otimes
\bbq.$
\item{(ii)} If $\pi_1^{orb}(\calz)\simeq 0,$ then
$\hbox{Pic}^{orb}(\calz)\simeq \hbox{Pic}(\calz).$
\tenrm

For an inversion theorem to Theorem \bfol.1.2 in the Sasakian-Einstein case and
the construction of many nontrivial examples the reader is referred to [B-G 2]
and [B-F-G-K] in the regular case. In particular, in dimension 5 we have

\noindent{\sc Theorem \bfol.1.4}[B-F-G-K]: \tensl Let $\cals$ be a
simply-connected regular Sasakian-Einstein manifold of dimension 5. Then
$\cals$ is one of the following: $S^5$, the Stiefel manifold $V_2(\bbr^4)$ of
2-frames in $\bbr^4,$ or the total space $S_k$ of the $S^1$ bundles
$S_k\rightarrow P_k$ for $3\leq k\leq 8$ where $P_k$ is a Del Pezzo surface
with a K\"ahler-Einstein metric [T-Y]. It is known that $S_k$ is diffeomorphic
to the $k$-fold connected sum $S^2\times S^3\# \cdots \# S^2\times S^3.$ \tenrm

\medskip
\noindent{\bf \bfol.2 The One Dimensional 3-Sasakian Foliation} 

Fixing a Sasakian structure, say $(\xi^1, \Phi^1,\eta^1)$ in the $3$-Sasakian
structure, we notice that subbundle $\calh = \hbox{ker}~\eta^1$ of $T\cals$
together with $I=-\Phi^1|\calh$ define the CR structure on $\cals.$ Actually a
$3$-Sasakian structure gives a special kind of CR structure, namely, a CR
structure with a compatible holomorphic contact structure. Notice that the
complex valued one form on $\cals$ defined by $\eta^+=\eta^2+i\eta^3$ is type
$(1,0)$ on $\cals.$ Moreover, one checks that $\eta^+$ is holomorphic with
respect to the CR structure $I.$  Although the 1-form $\eta^+$ is not invariant
under the circle action generated by $\xi^1,$ the trivial complex line bundle
$L^+$ generated by $\eta^+$ is invariant. Thus, the complex line bundle $L^+$
pushes down to a nontrivial complex V-line bundle $\call$ on $\calz.$ Let $V$
denote the one dimensional complex vector space generated by $L^+.$  Writing
the circle action as $\hbox{exp}~(i\phi\xi^1)$ shows that $V$ is the
representation with character $e^{-2i\phi},$ and since $\cals$ is a principal
$S^1$ V-bundle over $\calz,$ the twisted product $\call\simeq
\cals\times_{S^1}V$ is a complex line V-bundle on $\calz.$  Now we can define a
map of V-bundles $\theta :T^{(1,0)}\calz \ra{1.5} \call$ by
$$\theta(X)=\eta^+(\hX), \leqno{\bfol.2.1}$$ 
where $\hX$ is the horizontal lift of the vector field $X$ on $\calz.$ Notice
that $\theta(X)$ is not a function on $\calz$ but a section of $\call.$ Now a
straightforward computation shows that $\eta^+\wedge (d\eta^+)^n$ is a nowhere
vanishing section of $\grL^{(2n+1,0)}\calh$ on $\cals,$ and thus $\theta\wedge
(d\theta)^n$ is a nowhere vanishing section of $K\otimes \call^{n+1}$, where $K$
is the canonical V-line bundle (see the Appendix) on $\calz.$ Hence, in
$\hbox{Pic}^{orb}(\calz)$ we have the relation $\call^{n+1}\otimes K=1.$ So the
contact line V-bundle is $\call\simeq K^{-{1\over n+1}}$ in
$\hbox{Pic}^{orb}(\calz).$ Alternatively, the subbundle $\hbox{ker}~\theta$ is
a holomorphic subbundle of $T^{(1,0)}\calz$ which is maximally non-integrable.
This defines the complex contact structure on $\calz.$ Of course, this
construction depends on a choice of direction $\tau\in S^2$ in the 2-sphere of
complex structures.  However, the transitive action of $Sp(1)$ on $S^2$
guarantees that this structure is unique up to isomorphism as complex
contact manifolds.  We have [see B-G-M 1, B-G 1]:

\noindent{\sc Theorem \bfol.2.2} \tensl Let $\cals$ be a complete 3-Sasakian
manifold, choose a direction $\tau \in S^2,$ and let $\calz_\tau$ denote the
space of leaves of the corresponding foliation $\calf_\tau.$ Then $\calz_\tau$
is a compact $\bbq$-factorial contact Fano variety with a K\"ahler-Einstein
metric $h$ of scalar curvature \break 
$8(2n+1)(n+1)$ such that the natural projection
$\pi:\cals\ra{1.3} \calz_\tau$ is an orbifold Riemannian submersion with
respect to the Riemannian metrics $g$ on $\cals$ and $h$ on $\calz_\tau.$
\tenrm

We call the space $\calz_\tau,$ usually just written $\calz,$ the {\it twistor
space} associated to $\cals.$ Actually there is another object that could merit
the name the twistor space of $\cals,$ namely the trivial 2-sphere bundle
$S^2\times \cals$ with the structure induced from the twistor space $S^2\times
C(\cals)$ of the hyperk\"ahler cone. 

An important property of the twistor space in the case of quaternionic K\"ahler
manifolds is that it is ruled by rational curves. The same is true in our case
as long as one allows for singularities. We have

\noindent{\sc Proposition \bfol.2.3}: \tensl $\calz$ is ruled by a real family
of rational curves $C$ with possible singularities on the singular locus of
$\calz.$ All the curves $C$ are simply-connected, but $\pi_1^{orb}(C)$ 
can be a non-trivial cyclic group.\tenrm

For any line V-bundle $\call$ we let $\hat{\call}$ denote $\call$ minus its
zero section.

\noindent{\sc Proposition \bfol.2.4}: \tensl Let $\calz$ be the twistor space
of a $3$-Sasakian manifold $\cals$ of dimension $4n+3,$ and assume that
$\pi_1^{orb}(\calz)=0.$ If the contact line V-bundle $\call$ (or equivalently
its dual $\call^{-1}$) has a root in $\hbox{Pic}^{orb}(\calz),$ then it must be
a square root, namely $\call^{1\over 2}.$ Moreover, in this case if both
$\hat{\call}$ and $\hat{\call}^{1\over 2}$ are proper in the sense of
Kawasaki, then we must have $\calz =\bbp^{2n+1}.$ In particular, this holds if
the total space of $\hat{\call}$ is smooth. \tenrm

\noindent{\sc Proof}: By Proposition \bfol.1.3 $\hbox{Pic}^{orb}(\calz)$ is
torsion free. So the proof in [B-G 1] now goes through. By Proposition \bfol.2.3
$\calz$ is ruled by rational curves $C$ which on the singular locus take the
form $\Gamma\backslash \bbp^1.$ 
Now the restriction $\call^{-1}|C$ is $\calo(-2)$
which is a V-bundle if $C$ is singular. In either case it has only a square
root namely the tautological V-bundle $\calo(-1).$ Since these curves $C$ cover
$\calz$ this proves the first statement.  The second statement follows from a
modification of an argument due to Kobayashi and Ochiai [K-O] and used by
Salamon [Sal 1].  The main point is that since $\hat{\call},
\hat{\call}^{1\over 2}$ are proper and it follows that we can apply Kawasaki's
Riemann-Roch Theorem [Kaw 1] together with the Kodaira-Baily Vanishing Theorem
[Bai 2] to arbitrary powers of the line V-bundle $\call^{{1\over 2}}$ to give
$(n+1)(2n+3)$ infinitesimal automorphisms of the complex contact structure on
$\calz.$ Since $\pi_1^{orb}(\calz)=0,$ these integrate to global automorphisms
on $\calz$ and the result follows. See the Appendix and [B-G 1] for details.
\hfill\za

\noindent{\sc Remark} \bfol.2.5: There is an error in the statement of
Proposition 4.3 of [B-G 1]. The error is in leaving out the assumptions that
$\pi_1^{orb}(\calz)$ is trivial and that the contact line bundle is proper.
Example \bfol.2.6 below shows that the conclusion in Proposition \bfol.2.4 does
not necessarily hold if the hypothesis $\pi_1^{orb}(\calz)=0$ is omitted.
Likewise, Example \bfol.2.7 below gives a counterexample when the condition
that $\call$ be proper is omitted.

\noindent{\sc Example} \bfol.2.6: Consider the 3-Sasakian lens space $L(p;q)=
\bbz_p\backslash S^7$ constructed as follows: $S^7$ is the unit sphere in the
quaternionic vector space $\bbh^2$ with quaternionic coordinates $u_1, u_2.$
The action of $\bbz_p$ is the left action defined by 
$(u_1,u_2)\mapsto (\tau u_1,\tau^qu_2)$, where $\tau^p=1$ and
$p$ and $q$ are relatively
prime positive integers. If $p=2m$ for some integer $m$ then  $-\hbox{id}$ is
an element of $\bbz_{2m},$ so the 3-Sasakian manifolds $L(2m;q)$ and $L(m;q)$
both have the same twistor space, namely $\calz =\bbz_m\backslash \bbp^3,$ and
$\pi_1^{orb}(\calz)\simeq\bbz_m.$ There are clearly many similar examples in
all dimensions equal to $3$ mod $4.$

\noindent{\sc Example \bfol.2.7}: Consider the 3-Sasakian 7 manifolds
$\cals(p_1,p_2,p_3)$ described in Section \bex.4 below, where the $p_i$'s are
pairwise relatively prime, and precisely one of the $p_i$'s is even, say $p_1.$
$\cals(p_1,p_2,p_3)$ is simply-connected and its twistor space
$\calz(p_1,p_2,p_3)$ has $\pi_1^{orb}(\calz(p_1,p_2,p_3))=0.$ Now there is a
$\bbz_2$ acting on $\cals(p_1,p_2,p_3),$ but not freely, which acts
as the identity on $\calz(p_1,p_2,p_3).$ Thus, $\bbz_2\backslash
\cals(p_1,p_2,p_3)$ has $\calz(p_1,p_2,p_3)$ as its twistor space, and as a
V-bundle $\bbz_2\backslash \cals(p_1,p_2,p_3)\ra{1.3}
\calz(p_1,p_2,p_3)$ is not proper in the sense of Kawasaki [Kaw 2]. Thus, the
V-bundle $\hat{\call}$ is not proper, and Kawasaki's Riemann-Roch theorem
[Kaw 1] cannot be applied.

We now wish to formulate a converse to Theorem \bfol.2.2.

\noindent{\sc Definition \bfol.2.8}: \tensl A complete $\bbq$-factorial Fano
contact variety $\calz$ is said to be {\it good} if the total space of the
principal circle bundle $\cals$ associated with the contact V-line bundle
$\call$ is a smooth compact manifold.  \tenrm

Thus, for good $\bbq$-factorial Fano contact varieties, $\cals$ desingularizes
$\calz.$ As discussed in the Appendix
this happens precisely when all the leaf holonomy
groups inject into the group $S^1$ of the bundle.
Notice also that in this case
$\cals$ is necessarily compact.  We now are ready for:

\noindent{\sc Theorem \bfol.2.9}: \tensl A good $\bbq$-factorial Fano contact
variety $\calz$ is the twistor space associated to a compact $3$-Sasakian
manifold if and only if it admits a compatible K\"ahler-Einstein metric $h.$
\tenrm

\noindent{\sc Proof}: Let $\calz$ be a good $\bbq$-factorial Fano contact
variety with a compatible K\"ahler-Einstein metric $h.$ Choose the scale of
$h$ so that the scalar curvature is $8(2n+1)(n+1).$  Let $\pi:\cals\rightarrow
\calz$ denote the principal orbifold circle bundle associated to $\call.$ It is
a smooth compact submanifold embedded in the dual of the contact V-line bundle
$\call^{-1}.$ The K\"ahler-Einstein metric $h$ has Ricci form $\rho=4(n+1)\gro$,
where $\gro$ is the K\"ahler form on $\calz,$ and $\rho$ represents the first
Chern class of $K^{-1}.$  Let $\eta^1$ be the connection in
$\pi:\cals\rightarrow \calz$ with curvature form $2\pi^*\gro.$ Then the
Riemannian metric $g_{\cals}$ on $\cals$ can be defined by $g_{\cals}=\pi^*h +
(\eta^1)^2.$ It is standard (see the proof in Example 1 of Section 4.2 in 
[B-F-G-K])
that $g_{\cals}$ is Sasakian-Einstein. As in Proposition 2.2.4 of [Sw] the
V-bundle $\call\otimes \grL^{(1,0)}\calz$ has a section $\theta$ such that the
K\"ahler-Einstein metric $h$ decomposes as $h=|\theta|^2 +
h_D$, where $h_D$ is a metric in the V-bundle $D.$ Let us write $\pi^*\theta
=\eta^+.$ Since $\cals$ is a circle bundle in $\call^{-1},$ the contact bundle
$\call$ trivializes when pulled back to $\cals.$ This together with the
condition that $\theta\wedge (d\theta)^n$ is nowhere vanishing on $\calz$
implies that $\eta^+$ is a nowhere vanishing complex valued 1-form on $\cals.$
So the metric $g_{\cals}$ on $\cals$ can be written as
$$g_{\cals}=(\eta^1)^2 +|\eta^+|^2 +\pi^*h_D.$$ 
We claim that this metric is $3$-Sasakian.  To see this consider the total
space $M$ of the dual of the contact V-line bundle minus its $0$ section which
is $\cals\times \bbr^+.$ Put the cone metric $dr^2 + r^2g$ on $M.$ The natural
$\bbc^*$ action on $M$ induces homotheties of this metric. Now using a standard
Weitzenb\"ock argument, LeBrun [Le 3] shows that $M$ has a parallel holomorphic
symplectic structure and his argument works just as well in our case.  Let
$\vartheta$ denote the pullback of the contact form $\theta$ to $M$ which is a
holomorphic 1-form on $M$ that is homogeneous of degree 1 with respect to the
$\bbc^*$ action.  Thus $\grU = d\vartheta$ is a holomorphic symplectic form on
$M$ which is parallel with respect to the Levi-Civita connection of the cone
metric.  Hence, $(M,dr^2 + r^2g)$ is hyperk\"ahler.  Furthermore, if
$\{I^a\}_{a=1}^3$ denote hyperk\"ahler endomorphisms on $M,$
$\vartheta^2,\vartheta^3$ are the real and imaginary parts of $\vartheta,$ and
$\vartheta^1$ is the pullback of $\eta^1$ to $M,$ then LeBrun shows that
$$\vartheta^1I^1 =\vartheta^2I^2 = \vartheta^3I^3.$$ It then follows from our
previous work [B-G-M 1] that $g$ is $3$-Sasakian.  
But by construction $\calz$ is
the space of leaves of the foliation generated by $\xi^1,$ so $\calz$ must be
the twistor space of the compact $3$-Sasakian manifold $\cals.$ \hfill\za
\smallskip
\noindent{\bf \bfol.3 The Three Dimensional 3-Sasakian Foliation} 
\smallskip
Next we consider the three dimensional foliation $\calf_3$ discussed in Section
\bdef.2.

\noindent{\sc Proposition \bfol.3.1}: \tensl Let $(\cals,g)$ be a
3-Sasakian manifold such that the characteristic vector fields $\xi^a$ are
complete. Let $\calf_3$ denote the the canonical three dimensional foliation on
$\cals.$ Then
\item{\rm(i)} The metric $g$ is bundle-like.  
\item{\rm(ii)} The leaves of $\calf_3$ are totally geodesic spherical space
forms $\grG\backslash S^3$ of constant curvature one, 
where $\grG\subset Sp(1)=SU(2)$
is a finite subgroup. 
\item{\rm(iii)} The 3-Sasakian structure on $\cals$ restricts to a 3-Sasakian
structure on each leaf.
\item{\rm(iv)} The generic leaves are either $SU(2)$ or $SO(3).$ \tenrm

\noindent{\sc Proof}: The proof of (i), (ii), and (iii) follow from the basic
relations for 3-Sasakian manifolds as in Proposition \bfol.1.1. To prove (iv)
we notice that the foliation $\calf_3$ is regular restricted to the generic
stratum $\cals_0.$  By (ii) and regularity there is a finite subgroup
$\grG\subset SU(2)$ such that the leaves of this restricted foliation are all
diffeomorphic to $\grG\backslash S^3,$ which is 3-Sasakian 
by (iii). Now the regularity
of $\calf_3$ on $\cals_0$ implies that its leaves must all be regular with
respect to the foliation generated by $\xi^1.$  But a result of Tanno [Tan 1]
says that the only regular 3-Sasakian 3-manifolds have $\grG = \hbox{id}$ or
$\bbz_2,$ in which case (iv) follows.  \hfill\za

\noindent{\sc Example \bfol.3.2}: Consider the 3-Sasakian lens space $L(p;q)=
\bbz_p\backslash S^7$ of Example \bfol.2.6.
If $p$ is odd then $-\hbox{id}$ is not an element of
$\bbz_p$ so the generic leaf of the foliation $\calf_3$ is $S^3.$ The singular
stratum consists of two leaves both of the form $\bbz_p\backslash S^3$ with leaf
holonomy group $\bbz_p.$ These two leaves are described by $u_2=0$ and $u_1=0,$
respectively. If $p$ is even then $-\hbox{id}$ is an element of $\bbz_p,$ so
the generic leaf is $SU(2)/\bbz_2 = SO(3),$ and the leaf holonomy of the two
singular leaves is $\bbz_{p\over 2}.$

The next theorem was first proved by Ishihara [Ish 2] in the regular case using
slightly different methods. First we need to describe our structures in the
orbifold category. Recall that a quaternionic K\"ahler structure on a
Riemannian manifold $M$ is defined by 

\noindent{\sc Definition \bfol.3.3}: \tensl A Riemannian orbifold $\calo$ is
called a {\it quaternionic K\"ahler} orbifold if there is a rank 3 subbundle
$\calg$ of the endomorphism bundle $\hbox{End}~TM$ of $TM$ which is preserved
by the Levi-Civita connection and is locally generated by almost complex
structures $I,J,K$ that satisfy the algebra of the quaternions, and the action
of the local uniformizing groups preserves the bundle $\calg.$ An alternative
definition which works only in dimension greater than 4 is that $\calo$ is a
Riemannian orbifold whose holonomy group is a subgroup of $Sp(n)\!\cdot\!Sp(1).$
\tenrm

It is well-known that the strata of a quaternionic K\"ahler orbifold are not
necessarily quaternionic K\"ahler [G-L]. The strata will be quaternionic
K\"ahler if the local uniformizing groups act trivially on the fibres of
$\calg$ [D-Sw]. The group of the bundle $\calg$ is $SO(3)$ with the adjoint
representation. Thus, for each local uniformizing system on $\calo$ there is a
group homomorphism $\psi_i:\grG_i\ra{1.3} SO(3).$ 

\noindent{\sc Theorem \bfol.3.4}: \tensl Let $(\cals, g)$ be a 3-Sasakian
manifold of dimension $4n+3$ such that the characteristic vector fields $\xi^a$
are complete.  Then the space of leaves $\cals/\calf_3$ has the structure of a
quaternionic K\"ahler orbifold $(\calo,g_\calo)$ of dimension $4n$ such that
the natural projection $\pi:\cals\ra{1.3} \calo$ is a principal V-bundle with
group $SU(2)$ or $SO(3)$ and a Riemannian orbifold submersion such that the
scalar curvature of $g_\calo$ is $16n(n+2).$ \tenrm

\noindent{\sc Proof}: We can split $T\cals=\calv_3\oplus\calh$, where
$\calv_3$ is the subbundle spanned by the characteristic vector fields
$\{\xi^1,\xi^2,\xi^3\}$ and the ``horizontal" bundle is the orthogonal
complement $\calh=\calv_3^\perp$. Let $h\Phi^a=\Phi^a\mid_\calh$ be the 
restriction of characteristic endomorphisms. One can easily see that
$$h\Phi^a\circ h\Phi^b=-\delta^{ab}{\bf 1}+\sum_c\epsilon^{abc}h\Phi^c.$$
It follows that $\calh$ is pointwise a quaternionic vector space and
$\calo$ is a compact quaternionic orbifold. We must show that the metric
$g_\calo$ obtained from $g$ by the orbifold Riemannian submersion $\pi:\cals
\rightarrow\calo$ has its holonomy group reduced to a subgroup of 
$Sp(n)\!\cdot\! Sp(1)$. This can be done by constructing a parallel
4-form on $\calo$. Consider $\phi^a=d\eta^a$ and 
$$\hi^a=\phi^a+  
\sum_{b,c}\epsilon^{abc}\eta^b\wedge\eta^c.$$
It is easy to see that the 4-form
$\Omega=\sum_a\hi^a\wedge\hi^a$
is horizontal and
$Sp(1)$-invariant. It follows that there is a unique 4-form
$\hat\Omega$ on the
orbifold $\calo$ invariant under the action of the local uniformizing groups
such that $\pi^*\hat\Omega=\Omega$.  One can show that $\hat\Omega$ is parallel
on $\calo$ using standard tensor computation with O'Neill formulas [see B-G-M
1, G-L for details].  In the case $n=1$ the parallelism of the 4-form does not
further restrict Riemannian geometry of $\calo$. However, one can show that
$(\calo,g_\calo)$ is a compact self-dual Einstein orbifold.  Self-duality
follows easily from the fact that $\calo$ is quaternionic.  The fact that the
metric is Einstein is a simple computation and in the regular case can be found
in [Tan 2].  \hfill\za

There is an important inversion theorem of Theorem \bfol.3.4
originally in the regular case due to Konishi [Kon]. By now there are
several proofs of this, all of them related. Given a quaternionic K\"ahler
orbifold $\calo$ one can construct the Salamon twistor space $\calz$ and then
get $\cals$ from the inversion theorem of [B-G 1]. Another approach would be to
construct the orbifold version of Swann's bundle [Sw] on $\calo$ and then use
the results of [B-G-M 1] to obtain $\cals.$ Here our proof is essentially that
of Konishi's, only slightly modified to handle the orbifold situation. 

\noindent{\sc Theorem \bfol.3.5}: \tensl Let $(\calo,g_\calo)$ be a
quaternionic K\"ahler orbifold of dimension $4n$ with positive scalar curvature
$16n(n+2).$ Then there is a principal $SO(3)$ V-bundle over $\calo$ whose
total space $\cals$ admits a 3-Sasakian structure with scalar curvature
$2(2n+1)(4n+3).$ \tenrm

\noindent{\sc Proof}: Let $\calg$ denote the V-subbundle of $\hbox{End}~T\calo$
describing the quaternionic structure. Let $\{\tU_i\}$ be local uniformizing
neighborhoods that cover $\calo$ and $\cali^a_i$ a local framing of $\calg$ on
$\tU_i$ that satisfies
$$\cali^a_i\circ \cali^b_i= -\grd^{ab}\hbox{id} +\gre^{abc}\cali^c_i.$$
Since $\calo$ is quaternionic K\"ahler there are 1-forms $\tau^a_i$ on
each $\tU_i$ such that
$$\nabla \cali^a_i =\gre^{abc} \tau^b_i\otimes \cali^c_i.$$
Now the structure group of the V-bundle $\hbox{End}~T\calo$ is 
$Sp(n)\!\cdot\! Sp(1),$
and that of the V-subbundle $\calg$ is $SO(3).$  Let $\pi:\cals\ra{1.3}
\calo$ denote the principal $SO(3)$ V-bundle associated to $\calg.$ The local
1-forms $\tau^a_i$ are the components of an $\gs\go(3)$ connection
$\tau_i =\sum_{a=1}^3 \tau^a_ie_a,$ where $\{e_a\}$ denotes the standard basis
of $\gs\go(3)$ which satisfies the Lie bracket relations
$[e_a,e_b]=2\gre^{abc}e_c.$ The local connection forms satisfy the well-known
relations
$$\tau_i = \hbox{ad}_{g_{ij}}\tau_j + g^{-1}_{ij}dg_{ij}$$
in $\tU_i\cap \tU_j$ for some smooth map $g_{ij}: \tU_i\cap \tU_j\ra{1.3}
SO(3).$ Furthermore, from [Ish 2] one checks that the curvature forms
$$\gro^a_i = d\tau^a_i +\gre^{abc}\tau^b_i\wedge \tau^c_i \leqno{\bfol.3.6}$$
satisfy the relation
$2g_\calo(\cali_i^aX,Y)=\gro^a_i(X,Y).$
Now on each $\tU_i$ there exists a smooth local section $\grs_i:\tU_i\ra{1.3}
\cals$ and on $\cals$
there is a global 1-form $\eta^a$ such that $\tau^a_i =\grs^*_i\eta^a.$
On $\cals$ we define a Riemannian metric by
$$g= \pi^*g_\calo + \sum_{a=1}^3\eta^a\otimes \eta^a.$$
By construction the vector fields $\xi^a$ generating the $SO(3)$ action on
$\cals$ are dual to the forms $\eta^a$ with respect to this metric, viz
$g_\cals(X,\xi^a)=\eta^a(X)$ for any vector field $X$ on $\cals.$ Also by
construction the vector fields $\xi^a$ are Killing fields with respect to
the metric $g.$ Now define the $(1,1)$ tensor field $\Phi^a=\nabla
\xi^a.$ Since the $\xi^a$ are mutually orthogonal vector fields of unit length
on $\cals,$ one easily checks that $\Phi^a\xi^b =-\gre^{abc}\xi^c.$ Thus
$\Phi^a$ splits as
$$\Phi^a = h\Phi^a +\gre^{abc}\xi^b\otimes \eta^c.$$
One then checks using \bfol.3.6 that on each open set $\pi^{-1}(U_i),$
$h\Phi^a$ equals the horizontal component of $(\grs_i)_*\cali^a_i.$ From this
one then shows that 
$$\Phi^a\circ \Phi^b-\xi^a\otimes \eta^b ~=~  -\gre^{abc}\Phi^c
-\grd^{ab}\hbox{id}.$$
The result then follows by Propositions \bdef.2.2 and 1.2.3. \hfill\za

We mention that from the discussion in the Appendix it follows that if the
homomorphisms $\psi_i:\grG_i\ra{1.3} SO(3)$ are injective the total space
$\cals$ will be a smooth 3-Sasakian manifold.

Konishi's construction gives an $SO(3)$ bundle over $\calo.$ In the case that
$\calo$ is a smooth manifold there is a well-known obstruction [Sal 1] to
lifting this bundle to an $Sp(1)$ bundle, the Marchiafava-Romani class
$\varepsilon.$ Actually $\varepsilon$ is the obstruction to lifting the
principal $Sp(n)\!\cdot\!Sp(1)$ frame bundle to an $Sp(n)\!\times\!Sp(1)$ bundle
[Ma-Ro, Sal 1]. This obstruction also occurs when $\calo$ is an orbifold as
long as one uses Haefliger's orbifold cohomology (see Appendix). The class
$\varepsilon$ is the image of the connecting homomorphism 
$$\grd:H^1_{orb}(\calo,\calg)\ra{1.5}
H^2_{orb}(\calo,\bbz_2),\leqno{\bfol.3.7}$$
where $\calg$ is the sheaf of germs of smooth
orbifold maps from open sets of $\calo$ to the group $Sp(n)\!\cdot\!Sp(1).$ If
following Salamon [Sal 1] we write $T\calo\otimes \bbc\simeq {\bf E}\otimes {\bf
H},$ then $\varepsilon$ is the second Stiefel-Whitney class $w_2$ of the bundle
$S^2({\bf H})$ over $\calo.$ We have

\noindent{\sc Proposition \bfol.3.8}: \tensl The principal $SO(3)$ V-bundle
constructed in Theorem \bfol.3.5 lifts to a principal $Sp(1)$ V-bundle if and
only if $\varepsilon \in H^2_{orb}(\calo,\bbz_2)$ vanishes. Moreover, when
$\varepsilon =0$ the 3-Sasakian structure on the total space $\cals$ of the
$SO(3)$ V-bundle lifts to the total space $\cals'$ of the $Sp(1)$ V-bundle.
\tenrm

Thus, in the case that $\varepsilon =0$  there are precisely two 3-Sasakian
orbifolds $\cals,\cals'$ corresponding to the quaternionic K\"ahler orbifold
$\calo.$ Let $\calz$ denote the twistor space of the orbifold $\calo.$ Then
likewise, since $S^2\simeq SO(3)/S^1\simeq Sp(1)/S^1$ the two 3-Sasakian
orbifolds $\cals$ and $\cals'$ have the same twistor space $\calz.$ When
$\calo$ is a smooth manifold a result of Salamon [Sal 1] says that $\varepsilon
=0$ if and only if the quaternionic K\"ahler manifold $\calo$ is quaternionic
projective space. If we impose the condition that the orbifolds $\cals$ and
$\cals'$ are smooth manifolds, there is a similar result.

\noindent{\sc Theorem \bfol.3.9}: \tensl If two 3-Sasakian manifolds $\cals$
and $\cals'$ are associated to the same quaternionic K\"ahler orbifold $\calo$
or equivalently the same twistor space $\calz,$ then both $\cals$ and $\cals'$
have the same universal covering space $\tilde{\cals}$ and $\tilde{\cals}$ is a
standard 3-Sasakian sphere. \tenrm

\noindent{\sc Proof}: We work with the twistor space $\calz.$ Now $\cals$ and
$\cals'$ are unit circle bundles in the line V-bundles $\call^{-1}$ and
$\call'^{-1}$ respectively. Moreover, since $\cals'$ is a double cover of
$\cals,$ it follows that $\call =\call'^2.$ Consider the universal orbifold
cover $\tilde{\calz}$ of $\calz$  with $\pi_1^{orb}(\tilde{\calz})=0.$ Pull back
the V-bundles $\call$ and $\call'$ to V-bundles $\tilde{\call}$ and
$\tilde{\call'}$ on $\tilde{\calz}$ respectively. These bundles satisfy
$\tilde{\call'}=\tilde{\call^{1\over 2}}.$ By construction and naturality of
the covering maps $\tilde{\call}$ is the contact line bundle on
$\tilde{\calz}.$ Moreover, $\tilde{\cals}$ and $\tilde{\cals'}$ which are the
total spaces of the pullbacks of $\cals$ and $\cals'$ to $\tilde{\calz}$ are
both smooth manifolds since they cover smooth manifolds. Thus, by Proposition
\bfol.2.4 $\tilde{\calz}\simeq \bbp^{2n+1}.$ It follows that $\tilde{\cals'}
\simeq S^{4n+3}.$ \hfill\za

\noindent{\sc Remark \bfol.3.10}:  Konishi also considers the case when the
quaternionic K\"ahler manifold has negative scalar curvature.  This gives a
Sasakian $3$-structure on $\cals$ with indefinite signature $(3,4n)$.  

Finally, we give some general results concerning the curvature of any
$3$-Sasakian manifold.  Since the curvature of any Riemannian manifold is
completely determined by its sectional curvature and the sectional curvature
of any Sasakian manifold [Bl, Y-K] is completely determined by the
$\Phi$-sectional curvature, we essentially give the latter.  This shows that
the local geometry of any $3$-Sasakian manifold determines and is determined by
that of its associated quaternionic K\"ahler orbifold. 

\noindent{\sc Proposition} \bfol.3.11: \tensl  Let $(\cals,g,\xi^a)$ be a 
$3$-Sasakian manifold and let $K$ and $\check{K}$ denote the sectional 
curvatures of $g$ and its transverse component $\check{g}$, respectively.  
Then if $X$ is any horizontal vector field of unit length on $\cals$, we have
$K(X,\Phi^aX)=\check{K}(X,\Phi^aX)-3.$
\tenrm

\noindent{\sc Proof}: 
We first notice that [Bes: 9.29c] gives
$K(X,\Phi^aX)=\check{K}(X,\Phi^aX)-3|A_X\Phi^aX|^2$ where $A$ is O'Neill's
tensor [Bes] which is essentially the curvature of the $\gs\gp(1)$ valued
orbifold connection [B-G-M 1]. One then shows that for any horizontal vector
fields $X,Y$ on $\cals$ we have
$$A_X Y ~=~ \sum_{a=1}^3g(\Phi^a X,Y)\xi^a,\leqno{\bfol.3.12}$$
and the identity follows. \hfill\za
\smallskip
\noindent{\bf \bfol.4 The Second Einstein Metric}
\smallskip
Of course, by an Einstein metric we actually mean a homothety class of Einstein
metrics. In this section we shall show by using a theorem of Berard-Bergery
[Bes] that every 3-Sasakian manifold has at least two distinct homothety
classes of Einstein metrics. The method involves the canonical variation [Bes]
associated with Riemannian submersions. Due to the local nature of the
calculations involved this construction holds equally well for orbifold
Riemannian submersions. The canonical variation is constructed as follows
[Bes]:

\noindent Let $\pi:M\ra{1.3} B$ be an orbifold Riemannian submersion with $g$
the Riemannian metric on $M.$ Let $\calv$ and $\calh$ denote the vertical and
horizontal subbundles of the tangent bundle $TM.$ For each real number $t>0$ we
construct a one parameter family $g_t$ of Riemannian metrics on $M$ by defining
$$g_t|\calv =tg|\calv, \qquad g_t|\calh =g|\calh, \qquad g_t(\calv,\calh)=0.
\leqno{\bfol.4.1}$$
So for each $t>0$ we have an orbifold Riemannian submersion with the same base
space. Furthermore, if the fibers of $g$ are totally geodesic, so are the
fibers of $g_t.$ We apply the canonical variation to the orbifold Riemannian
submersion $\pi:\cals\ra{1.3} \calo.$ The metric as well as other objects on
$\calo$ will be denoted with a check such as $\check{g}.$

\noindent{\sc Theorem \bfol.4.2}: \tensl Every 3-Sasakian manifold admits a
second Einstein metric of positive scalar curvature. \tenrm

\noindent{\sc Proof}: We apply the canonical variation to the orbifold
Riemannian submersion $\pi:\cals\ra{1.3} \calo.$ According to the
B\'erard-Bergery Theorem [Bes: 9.73] there are several conditions to check.
First, the connection $\calh$ must be a Yang-Mills connection. The condition
for this is [Bes]:
$$\sum_ig((\nabla_{X_i}A)_{X_i}X,\xi^a)=0$$
for each $a=1,2,3$ and where $X_i$ is a local orthonormal frame of $\calh,$ $X$
is any horizontal vector field, and $A$ is O'Neill's tensor. Actually we can
use standard computations together with \bdef.1.3 and \bfol.3.11 to prove the
stronger condition $(\nabla_{X_i}A)_{X_i}X=0.$ Second $|A|^2$ must be constant.
To compute this notice that using [Bes] and \bfol.3.11 we find
$$g(A_{X_i},A_{X_j})= 3\grd_{ij}, \qquad g(A\xi^a,A\xi^b)=4n\grd^{ab}.
\leqno{\bfol.4.3}$$
This gives $|A|^2=12n.$ The final condition to be satisfied is 
$(\check{\grl})^2 -\hat{\grl}(12n+18)>0,$
where $\check{\grl}$ and $\hat{\grl}$ are the Einstein constants for $\calo$
and the fibers, respectively, and we have made use of \bfol.4.3. Since in our
case $\check{\grl}=4(n+2)$ and $\hat{\grl}=2,$ we see that the inequality is
satisfied. \hfill\za

The scalar curvature of any metric $g_t$ in the canonical variation of the
metric $g$ is given by the formula 
$s_t =16n(n+2) +{6/t} -12nt$ [Bes].
Moreover, the value of $t$ that gives the second Einstein metric is
$t_0={\hat{\grl}\over \check{\grl}-\hat{\grl}}={1\over 2n+3}.$
The ratio of the two metrics depends only on the homothety class and is given
by
$${s_{1\over 2n+3}\over s_1}=1+{6(n+1)\over (2n+3)(2n+1)}. \leqno{\bfol.4.4}$$

In the special case $n=1$ that is $\dim~\cals =7,$ both the 3-Sasakian metric
and the second Einstein metric have weak $G_2$ holonomy [F-K-M-S, G-Sal]. See
Theorem \bg2.2.9 below.
\smallskip
\noindent{\bf \bfol.5 Invariants and the Classification of 3-Sasakian
Structures} 
\smallskip
We consider the question of equivalence of 3-Sasakian manifolds. A 3-Sasakian
structure $\{\xi^a,\eta^a,\Phi^a\}_{a=1}^3$ on a manifold $(\cals,g)$ is
determined completely by the metric $g$ and the characteristic vector fields
$\xi^a.$ 

\noindent{\sc Definition \bfol.5.1}: \tensl Two 3-Sasakian manifolds
$(\cals,g)$ and $(\cals',g')$ are said to be {\it
isomorphic} if there exist a diffeomorphism $F:\cals\ra{1.3} \cals'$ and an
$\phi\in Sp(1)$ such that $F^*g'=g$ and $\acute{\xi}^a =(\hbox{Ad}_\phi)_*
F_*\xi^a,$ where $\hbox{Ad}$ denotes the adjoint action of $Sp(1)$ on its Lie
algebra $\gs\gp(1).$ \tenrm

In practice we shall always choose a basis $\acute{\xi}^a$ of the 3-Sasakian
structure on $\cals'$ so that $F_*\xi^a=\acute{\xi}^a.$ Now given such a
diffeomorphism $F:\cals\ra{1.3} \cals'$ it is clear that the corresponding
foliations are $F$-related, that is that $F_*\calf_1=\calf'_1$ and $F_*\calf_3
=\calf'_3.$ This induces a commutative diagram of orbifold diffeomorphisms
$$\matrix{{\cal S}&\fract{F}{\ra{1.5}}&\cals'\cr &&\cr
\decdnar{}&&\decdnar{}\cr &&\cr 
{\cal Z}&\fract{F_1}{\ra{1.5}}&{\cal Z}'\cr &&\cr
\decdnar{}&&\decdnar{}\cr &&\cr 
{\cal O}&\fract{F_3}{\ra{1.5}}& {\cal O}'.\cr}\leqno{\bfol.5.2}$$
This implies that if $L_x$ is the leaf of $\calf_3$ at $x\in \calo,$ then
$F(L_x)$ is the leaf at $F_3(x)\in \calo',$ that is, $L'_{F_3(x)}=F(L_x).$ Let
$G(L)$ denote the leaf holonomy group of the leaf $L.$ Then we have
$G(L'_{F_3(x)})\approx G(L_x).$ More generally let $\calg(\cals)$ denote the
holonomy groupoid [Moo-Sch] 
of the foliation $\calf_3,$ that is the set of triples
$(x,y,[\gra])$ where $x,y\in \cals$ lie on the same leaf $L_x$ of $\calf_3$ and
$[\gra]$ is the holonomy equivalence class of all piecewise smooth paths from
$x$ to $y$ lying in $L_x.$ Multiplication in the groupoid $\calg(\cals)$ is
defined on pairs of triples $(x,y,[\gra]),(x',y',[\gra'])$ precisely when
$y=x'$ and then $(x,y,[\gra])\cdot(x',y',[\gra']) =(x,y',[\gra'\gra]).$
Furthermore, the subgroup of triples $(x,x,[\gra])$ with $x$ fixed is
identified with the holonomy group $G(L_x).$ With this structure,
$\calg(\cals)$ is a locally compact topological groupoid [Moo-Sch]. (Actually
$\calg(\cals)$ is a smooth manifold of dimension $4n+6$ but we do not use this
here). We have

\noindent{\sc Proposition \bfol.5.3}: \tensl Let $F:\cals\ra{1.3} \cals'$ be
an isomorphism of 3-Sasakian manifolds. Then $F$ induces an isomorphism
$F_*:\calg(\cals)\ra{1.3} \calg(\cals')$ of topological groupoids. \tenrm

The groupoid $\calg(\cals)$ will be studied in a forthcoming work. For now we
are interested in the unordered list $(\grG_1,\grG_2,\cdots)$ of holonomy
groups in $\calg(\cals)$ up to abstract isomorphism. This list is finite if
$\cals$ is complete and it provides important invariants of a 3-Sasakian
manifold. Since the leaves of the foliation $\calf_3$ are all spherical space
forms, the groups $\grG_i$ are all either subgroups of $Sp(1)$ or all subgroups
of $SO(3),$ depending on whether the Marchiafava-Romani class $\varepsilon$ of
the quaternionic K\"ahler orbifold $\calo$ is $0$ or $1,$ respectively. Notice
that it follows from its definition and \bfol.5.2 above that the class
$\varepsilon$ is an invariant of the 3-Sasakian structure on the manifold
$\cals.$ Indeed, $\varepsilon$ can be identified with a certain secondary
characteristic class of the foliation $\calf_3.$ Thus, the Marchiafava-Romani
class splits the isomorphism classes $\bbs$ of 3-Sasakian manifolds into the
disjoint union $\bbs_0 + \bbs_1$ depending on whether $\varepsilon$ is $0$ of
$1.$  A further rough classification scheme is given by

\noindent{\sc Definition \bfol.5.4}: \tensl $\cals$ is said to be:
\item{(1)} {\it regular} if all the $\grG_i$ are the identity.
\item{(2)} of {\it cyclic type} if all the $\grG_i$ are cyclic.
\item{(3)} of {\it dihedral type} if all the $\grG_i$ are either cyclic or
dihedral or binary dihedral with at least one $\grG_i$ non-Abelian.
\item{(4)} of {\it polyhedral type} if at least one of the $\grG_i$ is one of
the polyhedral groups, tetrahedral, octahedral, or icosahedral (or the
corresponding binary double covers) groups. \tenrm

The definition of regular here coincides with that of Definition \bdef.2.8.
In general 3-Sasakian
dimension $4n+3$ the only known examples of 3-Sasakian manifolds of polyhedral 
or dihedral type are the
spherical space forms $\grG\backslash S^{4n+3}$ and $\grG\backslash
\bbr\bbp^{4n+3},$ where $\grG$ is a binary polyhedral or a binary dihedral
group in the first case
and a polyhedral or a dihedral group in the second. 
The action is that induced by the
diagonal action of $\grG$ on the quaternionic vector space $\bbh^{n+1}.$
However, in dimension 7 there exist 3-Sasakian manifolds of dihedral
or polyhedral type [B-G 3, G-Ni] which are not spherical space forms. 
All other known non-regular 3-Sasakian manifolds are of
cyclic type and are discussed in detail in Section \bex.
The following is essentially due to Salamon:

\noindent{\sc Theorem \bfol.5.5}: \tensl Let $\cals$ be a complete regular
3-Sasakian manifold with $\varepsilon =0.$ Then $\cals \simeq S^{4n+3}$ or
$\bbr\bbp^{4n+3}.$ \tenrm

For more results about regular 3-Sasakian manifolds see Section \btop.4 below.
Next we consider an important infinitesimal rigidity result.
In the regular case this rigidity is a simple consequence 
of the results of LeBrun [Le 2] and Nagatomo [N] (see [G-Sal]).
In the general case it was recently proved by Pedersen and Poon [Pe-Po
2].

\noindent{\sc Theorem \bfol.5.6}: \tensl Complete 3-Sasakian manifolds
are infinitesimally rigid. \tenrm

\noindent{\sc Outline of Proof}: The deformation theory of 3-Sasakian manifolds
is tied to the deformation theory of hypercomplex manifolds studied previously
in [Pe-Po 1]. Let $\cals$ be a complete 3-Sasakian manifold.  Then the compact
manifold $S^1\times \cals$ has a natural hypercomplex structure [B-G-M 2].
Thus, its twistor space $W$ is compact and fibers holomorphically over
$\bbc\bbp^1.$ Moreover, there is a holomorphic foliation on $W$ whose leaves
are elliptic Hopf surfaces, and whose space of leaves is the twistor space
$\calz$ associated to $\cals.$ The geometry of the corresponding deformation
theory is as follows. Deformations $(\cals_t,g_t)$ of the 3-Sasakian
structure $(\cals_0,g_0)$ on $\cals$ correspond to deformations of the
hypercomplex structure on $S^1\times \cals$ of the form $S^1\times \cals_t.$ In
turn these deformations correspond to deformations of the holomorphic fibration
$p: W\ra{1.3} \bbc\bbp^1.$  Thus, there are natural projections:
$$\matrix{&&W&&\cr
          &p\swarrow &&\searrow\Phi &\cr
          &&&&\cr
          &\hskip -25 pt\bbc\bbp^1&&\hskip 25 pt \calz,&\cr} \leqno{\bfol.5.7}$$
where each fiber of $p$ is a divisor in $W$ diffeomorphic to $S^1\times
\cals$ and $\Phi$ is an orbifold submersion whose leaves are elliptic Hopf
surfaces. Now the product map $p\times \Phi:W\ra{1.3} \bbc\bbp^1\times \calz$
is an orbifold submersion whose leaves are elliptic curves. The differential of
$p\times \Phi$ induces the exact sequence of sheaves
$$0\ra{1.5}\calo_W\ra{1.5}\Theta_W\ra{1.5} \Phi^*\Theta_\calz\oplus
p^*\Theta_{\bbc\bbp^1}\ra{1.5}0,$$
where $\calo_W$ denotes the structure sheaf of $W$ and $\Theta$ denotes the
holomorphic tangent sheaf. Then using standard techniques together with the
Kodaira-Baily vanishing theorem and the orbifold version of the Akizuki-Nagano
vanishing theorem, Pedersen and Poon show that the virtual parameter space for
3-Sasakian deformations lies in
$$H^0(\calz,\calo_\calz)\otimes H^1(F,\calo_F)\oplus H^0(\calz,\Theta_\calz)
\otimes H^1(F,\calo_F)$$ $$\oplus H^1(\calz,\Theta_\calz)\otimes H^0(F,\calo_F)
\oplus H^1(W,p^*\Theta_{\bbc\bbp^1}), \leqno{\bfol.5.8}$$
where $F$ is the generic elliptic Hopf surface $S^1\times Sp(1).$ One then
analysis each summand of \bfol.5.8 to show that there are no 3-Sasakian
deformations. For example, possible deformations lying in the last summand
vanish by results of Horikawa, while 3-Sasakian deformations lying in the
second and third summands must preserve the complex contact structure on
$\calz.$ There are no such deformations in the third summand by the
Kodaira-Baily vanishing theorem. Elements in the second summand correspond to
complex contact transformations that are invariant under the $U(1)\times U(1)$
action coming from a discrete quotient of the $\bbc^*$ principal action on
$\call,$ and there are no such elements. Finally, elements of the first summand
correspond to scale changes in the $S^1$ factor of $S^1\times Sp(1)$ and these
hypercomplex deformations do not come from 3-Sasakian ones. \hfill\za

While this theorem says that there is no ``infinitesimal moduli'', there may
well be discrete moduli of 3-Sasakian structures. Indeed, we believe that the
work of Kruggel's [Kru 3] can be used with the aid of a computer to construct
distinct 3-Sasakian structures on the same manifold. See Remark \bex.4.9 below.
\bigskip
\centerline {\bf \bhom. Homogeneous Spaces}
\medskip
In this section we classify Sasakian-Einstein and
$3$-Sasakian homogeneous spaces.  We begin with the Sasakian-Einstein
case.
\smallskip
\noindent
{\bf \bhom.1 Homogeneous Sasakian-Einstein Manifolds}
\smallskip
As a Sasakian vector field $\xi$ is Killing, every Sasakian, and, hence,
Sasakian-Einstein manifold $\cals$ has non-trivial isometries. 
Recall the following well-known terminology. Let $G$ be a complex semi-simple
Lie group. A maximal solvable complex subgroup $B$ is called a Borel subgroup,
and $B$ is unique up to conjugacy.  Any complex subgroup $P$ that contains $B$
is called a parabolic subgroup. Then the homogeneous space $G/P$ is called a
generalized flag manifold. A well-known result of Wang [Ahk] says that every
simply-connected homogeneous K\"ahler manifold is a generalized flag manifold.
 
\noindent{\sc Definition} \bhom.1.1: \tensl A compact
Sasakian-Einstein manifold $\cals$
is called a {homogeneous Sasakian-Einstein} manifold if there is a transitive
group $K$ of isometries on $\cals$ that preserve the Sasakian structure, that
is, if $\phi^k\in \hbox{Diff}~\cals$ corresponds to $k\in K,$ then
$\phi^k_*\xi =\xi.$ (This implies that both $\Phi$ and $\eta$ are also
invariant under the action of $K.$) \tenrm
 
Note that $K$ is a compact Lie group by compactness of $\cals$. The following
is a result of [B-G 2].
 
\noindent{\sc Theorem} \bhom.1.2: \tensl Let $\cals$ be a compact quasi-regular
homogeneous Sasakian-Einstein manifold. Then $\cals$ is an $S^1$-bundle over a
generalized flag manifold $G/P.$ Conversely, given any generalized flag
manifold $G/P$ there is a circle bundle $\pi: \cals \ra{1.3} G/P$ whose total
space $\cals$ is a homogeneous Sasakian-Einstein manifold. \tenrm

\noindent{\sc Proof}: As in Proposition 4.6 of [B-G-M 2], $\cals$ is regular.
By Proposition \bfol.1.2 $\cals$ fibers over a simply-connected Fano variety
$\calz$ with a K\"ahler-Einstein metric of positive scalar curvature. Since the
action of $K$ commutes with $\xi$ it sends fibers to fibers, and thus acts
transitively on $\calz.$ But by Wang's theorem [Akh], $\calz =G/P$ for some
complex semi-simple Lie group $G$ and some parabolic subgroup $P\subset G.$ Now
$K$ preserves the K\"ahler-Einstein structure, and thus the complex structure.
So $K\subset G.$ In fact $G$ is just the complexification of $K$ its maximal
compact subgroup [W].

Conversely, by a theorem of Matsushima [Bes] every $G/P$ admits $K$ invariant
K\"ahler-Einstein metric, where $K$ is the maximal compact subgroup of $G.$
Moreover, there is a subgroup $U\subset K$ such that $G/P=K/U.$ Then by the
Kobayashi construction described in the
previous section there is a circle bundle over $G/P$
whose total space $\cals$ admits a Sasakian-Einstein metric.  By the
construction one easily sees that this metric is homogeneous.  \hfill\za

The following corollary lists all the possible 
$G/P$ in the first three
dimensions:

\noindent{\sc Corollary} \bhom.1.3: \tensl Let $\cals$ be a compact
homogeneous Sasakian-Einstein manifold of dimension
$2n+1$. Then $\cals$ is a circle bundle over
\item{(i)} $\bbc\bbp^1$ when $n=1$,
\item{(ii)} $\bbc\bbp^2$ or $\bbc\bbp^1\times\bbc\bbp^1$ when $n=2$,
\item{(iii)} $\bbc\bbp^3$, $\bbc\bbp^2\times\bbc\bbp^1$,
$\bbc\bbp^1\times\bbc\bbp^1\times\bbc\bbp^1$, the complex flag
$F_{3,2,1}=SU(3)/T^2$,  and the real Grassmannian $Gr_2(\bbr^5)$ when $n=3$.  
\tenrm

\noindent{\sc Remark} \bhom.1.4: Note that $(\cals, g)$ does not have to be
simply-connected. For each $G/P$ and each $k\in\bbz^+$ we get
a homogeneous $S^1$ bundle over $G/P$ with fundamental group
$\pi_1=\bbz_k$. It can be obtained as a discrete
$\bbz_k$-quotient of the unique simply-connected model of such $\cals$.
\smallskip
\noindent
{\bf \bhom.2 Homogeneous 3-Sasakian Manifolds}
\smallskip
Every $3$-Sasakian manifold $(\cals,g)$
has a nontrivial isometry group $I(\cals,g)$ of dimension at least three.  
We first recall some of the
results about $I(\cals,g)$.

\noindent{\sc Definition} \bhom.2.1: \tensl Let $I_0(\cals, g)\subset
I(\cals, g)$ be the subgroup of the isometry group which
preserves the 3-Sasakian structure, that is
if $\phi^k\in \hbox{Diff}~\cals$ corresponds to $k\in I_0(\cals, g)$
then
$\phi^k_*\xi^a =\xi^a,$ for all $a=1,2,3.$ Then $I_0(\cals, g)$  is called
the group of 3-Sasakian isometries and when it acts transitively
on $(\cals, g)$ the space $\cals$ is said to be a 3-Sasakian homogeneous space.
\tenrm

\noindent{\sc Lemma} \bhom.2.2: \tensl Let $({\cal S},g)$ be a
$3$-Sasakian manifold and $X\in \gi$ be a Killing vector field
on ${\cal S}$. Let $\call_X$ denote the Lie derivative with respect to $X.$  
Then the following  conditions are equivalent
$$(i) \ \ {\cal L}_X\Phi^a=0, \ \ a=1,2,3, \quad
(ii) \ \ {\cal L}_X\eta^a=0, \ \ a=1,2,3, \quad 
(iii) \ \ {\cal L}_X\xi^a=0, \ \ a=1,2,3.$$ 
Furthermore, if any (hence, all) of the conditions above is
satisfied, then for any vector field $Y$ on $\cals$ we have
$X\eta^a(Y)=\eta^a([X,Y])$.  \tenrm

The above lemma gives alternative characterizations of
the Lie algebra of $I_0(\cals, g)$ and it easily follows from the
definition and properties of the 3-Sasakian structure.
As its immediate consequence we get the 
following theorem [Tan 1]:

\noindent{\sc Theorem} \bhom.2.3: \tensl Let $(\cals,g)$ be a
complete $3$-Sasakian manifold which is not of constant curvature.
${\Got i}$ and ${\Got i}_0$ denote the Lie algebras of
$I(\cals,g)$ and $I_0(\cals,g)$, respectively. Then as Lie algebras
${\Got i}={\Got i}_0 \oplus \gsp1,$  
where $\gsp1$ is the Lie algebra generated by $\{\xi^1,\xi^2,\xi^3\}$.\tenrm

Notice that any of the first three conditions in Lemma \bhom.2.2 can be used to
describe the Lie subalgebra ${\Got i}_0\subset \gi$.  
Moreover, the equivalence of
conditions (iii) and (i) says that the Lie algebra $\gc(\gsp1)$ of the
centralizer of $Sp(1)$ in $I(\cals,g)$ is precisely $\gi_0$.  Globally, on the
group level we obtain:

\noindent{\sc Proposition} \bhom.2.4: \tensl Let $(\cals,g)$ be a 
complete $3$-Sasakian manifold. Then both the isometry groups $I(\cals,g)$ 
and $I_0(\cals,g)$ are compact.  Furthermore, if $(\cals,g)$ is not 
of constant curvature then either $I(\cals,g)= I_0(\cals,g)\times Sp(1)$
or $I(\cals,g)= I_0(\cals,g)\times SO(3).$   Finally, if $(\cals,g)$ 
does have constant curvature then $I(\cals,g)$ strictly contains either 
$I_0(\cals,g)\times Sp(1)$ or $I_0(\cals,g)\times SO(3)$ as a proper
subgroup and $I_0(\cals,g)$ is the centralizer of $Sp(1)$ or $SO(3)$.  \tenrm

\noindent{\sc Proof}: The first assertion follows from Corollary \bdef.2.6
and a standard result of Myers and Steenrod (cf. [Bes]).  Next, since 
$I_0(\cals,g)$, $Sp(1)$, and $SO(3)$ are all compact, the direct sum on the 
Lie algebra level given in Theorem \bhom.2.3 also gives a direct product of 
Lie groups.  The last assertion follows immediately from lemma \bhom.2.2.
\hfill\za

\noindent{\sc Proposition} \bhom.2.5: \tensl  Let $(\cals,g)$ be a 
$3$-Sasakian homogeneous space.  Then all leaves are diffeomorphic and 
$\cals/\calf_3$ is a quaternionic K\"ahler manifold where the natural 
projection $\pi:\cals\ra{1}\cals/\calf_3$ is a locally trivial Riemannian 
fibration. Furthermore, $I_0(\cals,g)$ acts transitively on the space of 
leaves $\cals/\calf_3.$
\tenrm

\noindent{\sc Proof}: Let $\psi:I_0(\cals,g)\times \cals\ra{1}\cals$ 
denote the action map so that, for each $a\in I_0(\cals,g),$ 
$\psi_a=\psi(a,\cdot)$ is a diffeomorphism of $\cals$ to itself.  
Proposition \bhom.2.4 implies that the isometry group $I(\cals,g)$ contains 
$I_0(\cals,g)\times Sp(1)$ where either $Sp(1)$ acts effectively or its 
$\bbz_2$ quotient $SO(3)\simeq Sp(1)/\bbz_2$ acts effectively.  Since
the Killing vector fields $\xi^a$ for $a=1,2,3$ are both the infinitesimal 
generators of the group $Sp(1)$ and a basis for the vertical distribution 
$\calv$, it follows that $Sp(1)$ acts transitively on each leaf with 
isotropy subgroup of a point some finite subgroup $\grG\subset Sp(1).$  
Now let $p_1$ and $p_2$ be any two points of $\cals$ and let $\call_1$ and
$\call_2$ denote the corresponding leaves through $p_1$ and $p_2$,
respectively.  Since $I_0(\cals,g)$ acts transitively on $\cals$, 
there exists an $a\in I_0(\cals,g)$ such that $\psi_{a}(p_1)=p_2.$  
Now $\psi_{a}$ restricted to $\call_1$ maps $\call_1$ diffeomorphically 
onto its image, and, since the $Sp(1)$ factor acts transitively on each leaf 
and commutes with $I_0(\cals,g)$, the image of $\psi_a$ lies in $\call_2$.  
But the same holds for the inverse map $\psi_{a^{-1}}$ with $\call_1$ and 
$\call_2$ interchanged, so the leaves must be diffeomorphic.  Thus, the 
leaf holonomy is trivial and $\pi:\cals\ra{1}\cals/\calf_3=\calo$ is a locally 
trivial Riemannian fibration.
The fact that the 
space of leaves $\calo$ is a quaternionic K\"ahler manifold now 
follows from Ishihara's theorem \bfol.3.3. Finally, the constructions above 
shows directly that $I_0(\cals,g)$ acts transitively on $\calo.$  
\hfill\za

The following classification theorem is 
now immediate from Proposition \bhom.2.5,
the result of Alekseevsky which states that all homogeneous
quaternionic K\"{a}hler manifolds of positive scalar curvature are symmetric
[Al 2], and Proposition \bdef.2.10:

\noindent{\sc Theorem} \bhom.2.6: \tensl  Let $\cals$ be a 
$3$-Sasakian homogeneous space.  Then $\cals=G/H$ is precisely one of the 
following:

$${Sp(n+1)\over Sp(n)}, ~~~ {Sp(n+1)\over
Sp(n)\!\times\!\bbz_2}, ~~~
{SU(m)\over S\bigl(U(m-2)\!\times\!U(1)\bigr)}, ~~~
{SO(k)\over SO(k-4)\!\times\!Sp(1)},$$ 
$${G_2\over Sp(1)},\qquad {F_4\over Sp(3)},\qquad {E_6\over
SU(6)},\qquad {E_7\over {\rm Spin}(12)},\qquad {E_8\over E_7}.$$ 
Here $n\geq 0$, $Sp(0)$ denotes the trivial group, 
$m\geq 3$, and $k\geq 7.$
Hence, there is one-to-one correspondence
between the simple Lie algebras and the simply-connected $3$-Sasakian
homogeneous manifolds.  \tenrm

Below we give the fundamental diagram $\diamondsuit(G/H)$ for each
3-Sasakian homogeneous space of Theorem \bhom.2.6:

$$\matrix{&&\bbr_+\times G/H&&\cr&
\swarrow&&\searrow&\cr G/H\!\cdot\!U(1)
&&\hskip -15pt\la{4}\hskip -30pt\decdnar{}&& G/H,\cr &
\searrow& &\swarrow&\cr &&G/H\!\cdot\!Sp(1)&&
\cr}\leqno{\bhom.2.7}$$
where $G/H\!\cdot\!Sp(1)$ are precisely the Wolf spaces [Wol].

\noindent{\sc Remark} \bhom.2.8: Note that a homogeneous 3-Sasakian
manifold is necessarily simply-connected with the exception of the
real projective space. This is in sharp contrast with 
the Sasakian-Einstein case.
Also notice that a 3-Sasakian manifold can be Riemannian homogeneous 
(i.e., the full isometry group acts transitively) but not
3-Sasakian homogeneous. This is true for the lens spaces
$\Gamma\backslash S^3$, with $|\Gamma|>2.$
Observe that $\bbz_k\backslash S^3,\  k>2$,
is a homogeneous Sasakian-Einstein manifold but not 3-Sasakian homogeneous.

Theorem \bhom.2.6 does not specify what is the 3-Sasakian metric on the
coset $G/H$. 
In Section \bq.2 we will describe a quotient
construction of the 3-Sasakian homogeneous spaces with $G=SU(n+1)$ and
$G=SO(n+1)$. Here we quote a theorem of 
Bielawski [Bi 1], which gives an 
explicit description of these metrics in all cases.

\noindent{\sc Theorem} \bhom.2.9: \tensl 
Let $\cals=G/H$ be one of the spaces in Theorem \bhom.2.6.
and let ${\Got g}={\Got h}\oplus{\Got m}$ be the 
corresponding decomposition
of the Lie algebras. Then there is a natural decomposition
${\Got m}={\Got s}{\Got p}(1)\oplus{\Got m}^\prime$ and the metric $g$ on
$\cals$ is given in terms of the scalar product on
${\Got m}$
$$|\!|m|\!|^2=-<\sigma,\sigma>-{1\over2}<m^\prime,m^\prime>,$$
where $\sigma\in{\Got s}{\Got p}(1)$, $m^\prime\in{\Got m}^\prime$, and
$<\cdot,\cdot>$ is the Killing form on ${\Got g}$. In particular,
the metric $g$ is not naturally 
reductive
with respect to the homogeneous structure on $\cals$.
\tenrm

\noindent{\sc Remark} \bhom.2.10: In the case when $\cals$
is of constant curvature the canonical metric on $S^{4n+3}$ 
(or $\bbr\bbp^{4n+3}$) is
{\it not} the standard homogeneous metric on the homogeneous space
$Sp(n+1)/Sp(n)$ (or $Sp(n+1)/Sp(n)\times\bbz_2$)
with respect to the reductive decomposition
$\gs\gp(n+1)\simeq \gs\gp(n)+\gm$.  It is, of course, the standard
homogeneous metric with respect to the naturally reductive
decomposition $\go(4n+4)\simeq
\go(4n+3)+\gm.$  
This is quite special to the sphere and orthogonal group. In
general the  $3$-Sasakian homogeneous metrics are not naturally reductive
with respect to any reductive decomposition.
\bigskip
\centerline {\bf \btop. 3-Sasakian Cohomology}
\medskip
In this section we will describe some cohomological
properties of 3-Sasakian manifolds $\cals$. We prove a vanishing
theorem and then derive
a relation between the Betti numbers
of $\cals$ and the Betti numbers of the
associated orbifolds $\calz$ and $\calo$. We conclude with various
implications of these relations in the case $\cals$ is regular.
\smallskip
\noindent
{\bf \btop.1 Sasakian Manifolds and Harmonic Theory}
\smallskip
We start by recalling
some old results about harmonic forms on Sasakian manifolds due to
Tachibana [Tach]. Let $(\cals,g)$ be a compact Sasakian
manifold of dimension $2m+1$ with Sasakian structure
$\{\xi,\eta,\Phi\}$ and let $\Omega^p(\cals)$ be the space of smooth
$p$-forms on $\cals.$ Furthermore, let $\calh^p(\cals)=\{ u\in\Omega^p(\cals):
du=0=d*u\}$ denote the finite-dimensional
space of harmonic $p$-forms. 
By Hodge theory any harmonic form $u$ is necessarily invariant 
under the isometry group $I(\cals,g)$
The tensor $\Phi$ extends to an endomorphism of
$\Omega^p(\cals)$ by setting
$$(\bPhi u)(X_1,X_2,\ldots,X_p)= \sum_{i=1}^p u(X_1,\ldots,\Phi
X_i,\ldots,X_p).\leqno{\btop.1.1}$$
With this notation we have [Tach]

\noindent{\sc Theorem \btop.1.2}:  
\tensl Let $u\in\calh^p(\cals)$, $p\leq m$. Then
$\xi\lfloor u= 0$, and 
$\bPhi(u)\in\calh^p(\cals)$.
\tenrm

\noindent{\sc Proof}: The first statement is easy to prove in the case $p=1$.
Indeed, let $u= \alpha + f\eta$ be a closed invariant 1-form, where
$\alpha(\xi)=0$, and $f$ is a function. The vanishing of the Lie derivative of
$u$ along $\xi$ implies that $0=d(\xi\lfloor u)=df$, so that $f$ is a constant
and $0=d\alpha+fd\eta$. Then
$$0=\int_\cals d(\alpha\wedge (d\eta)^{m-1}\wedge\eta)=-\int_\cals 
f(d\eta)^{m}\wedge\eta.$$
Since $(d\eta)^{m}\wedge\eta$ is a non-zero multiple of the volume form
of $\cals$, we obtain $f=0$ and $\xi\lfloor u=0$. 
The general case of the original proof uses an explicit
computation in local coordinates and we omit it here. The second statement
follows immediately from \btop.1.1 and the fact that $\Phi$ preserves
horizontal subspaces. \hfill\za

Let us define the following 
$\cali:\calh^p(\cals)\rightarrow\calh^p(\cals)$ 
endomorphism for
$p\leq m$:
$$(\cali u)(X_1,\ldots,X_p)=u(\Phi X_1,\ldots, \Phi X_p)\leqno{\btop.1.3}$$
The basic identity \bdef.1.8(i) together with Theorem \btop.1.2
shows that $\cali u$ is a linear combination of $(\bPhi)^ku$ for
$0\leq k\leq p$.
Thus $\cali$ also maps $\calh^p(\cals)$ into itself.
The following proposition 
is now a simple consequence of the 
definition \btop.1.3 and Theorem \btop.1.2 [Bl, Bl-Go]:

\noindent{\sc Proposition \btop.1.4}:
\tensl Let $\cali:\calh^p(\cals)\rightarrow\calh^p(\cals)$ and
$p\leq m$. Then $\cali\circ\cali=(-1)^p$. In particular, when
$p$ is odd, $\cali$ defines an almost complex structure on the vector
space $\calh^p(\cals)$.
\tenrm

\noindent{\sc Corollary \btop.1.5}: 
\tensl Let $(\cals,g)$ be a compact Sasakian manifold of dimension
$2m+1$. Then the Betti numbers $b_p$ for $p$ odd and
$p\leq m$ are even. 
\tenrm

In the case of compact Sasakian-Einstein manifolds this and the
fact that the fundamental group is finite are 
the only known general topological
restrictions on $\cals$. Under some
additional curvature conditions we can get further restrictions.
For example, it is known [Bl] that a compact simply-connected Sasakian manifold
of positive sectional curvature is isometric to a sphere. For other similar
results see [Bl, Go] and references therein.

\smallskip
\noindent
{\bf \btop.2 A Vanishing Theorem}
\smallskip
Now, let $(\cals, g)$ be a compact 3-Sasakian manifold
of dimension $4n+3$ and 3-Sasakian structure $\{\xi^a,\eta^a,\Phi^a\}$.
Throughout this
section we shall suppose that $p\le 2n+1$. Referring to the splitting
of the tangent bundle of $\cals$ into $T\cals=\calv_3\oplus \calh$, we
shall say that a $p$-form 
$u\in\Omega^p(\cals)$ has {\sl bidegree} $(i,p-i)$
if it is a section of the subbundle of $\bigwedge^pT^*\cals$ 
isomorphic to the dual of 
$\bigwedge^i\calv_3\otimes\bigwedge^{p-i}\calh$. In particular, $u$ is called
{\sl 3-horizontal} if it has bidegree $(0,p)$, or equivalently if
$\xi^a\lfloor u=0$ for $a=1,2,3$. An element $\omega\in\Omega^p(\cals)$ 
is called {\sl invariant} if $h^*\omega=\omega$ for all 
$h\in Sp(1)$. In the regular case,
there is a principal $Sp(1)$-bundle $\pi:\cals\rightarrow \calo$, and $\omega$
is both 3-horizontal and invariant if and only if it is the pullback
$\pi^*\hat\omega$ of a form $\hat\omega$ on the quaternionic K\"ahler base
$\calo$. Now the curvature forms $\hi^a$
defined in \bfol.3.4 are horizontal with respect to the foliation $\calf_3.$
The Killing fields $\xi^a$ transform according to the adjoint representation of
$Sp(1)$, and the same is true of the associated triples $\eta^a$, $d\eta^a$,
and $\Phi^a$. For example, if $h\in Sp(1)$, we may write $$h_*\Phi^a=\sum_b
h^{ab}\Phi^b,\quad a=1,2,3,\leqno{\btop.2.1}$$ where $h^{ab}$ are components of
the image of $h$ in $Sp(1)/\bbz_2\cong SO(3)$. The 3-forms 

$$\Eta =
\eta^1\wedge\eta^2\wedge \eta^3,\qquad
\Theta=\sum_a\eta^a\wedge \hi^a=\sum_a\eta^a\wedge d\eta^a+
6\Eta\leqno{\btop.2.2}$$
have respective bidegrees $(3,0),(1,2)$, and are
clearly invariant. Their exterior derivatives are 
$$d\Eta =
\eta^1\wedge\eta^2\wedge\hi^3+\eta^2\wedge\eta^3\wedge\hi^1+
\eta^3\wedge\eta^1\wedge\hi^2,
\qquad d\Theta = \Omega + 2d\Eta,\leqno{\btop.2.3}$$
where the 4-form 
$\Omega$ is defined in section \bfol.3. In fact, $\Omega$ is the canonical
4-form determined by the quaternionic structure of Proposition \bdef.2.4 of the
subbundle $\calh$, and is the pullback of the fundamental 4-form $\hat\Omega$
on the quaternionic K\"ahler orbifold $\calo$ (see section \bfol.2).

Theorem \btop.1.2(i) implies that any harmonic $p$-form with $p\leq2n+1$ on
the compact 3-Sasakian manifold $\cals$ is 3-horizontal. Apply \btop.1.1
so as to obtain 
$\bPhi^a:\calh^p(\cals)\rightarrow\calh^p(\cals),
\quad a=1,2,3,\quad p\leq2n+1,$
and
\btop.1.3 to get 
$$(\cali^au)(X_1,X_2,\ldots,X_p)=u(\Phi^aX_1,
\Phi^aX_2,\ldots,\Phi^aX_p).\leqno{\btop.2.4}$$
Now, using the basic identities of Proposition \bdef.2.4 we can generalize
Proposition \btop.1.4 to get the following result due to Kuo [Kuo]:

\noindent{\sc Proposition \btop.2.5}:
\tensl Let $\cali^a:\calh^p(\cals)\rightarrow\calh^p(\cals)$, $a=1,2,3$, and
$p\leq 2n+1$. Then 
$$\cali^b\circ\cali^a=(-\delta^{ab})^p{\bf I}+\sum_c(\epsilon^{abc})^p\cali^c.
\leqno{\btop.2.6}$$
In particular, when
$p$ is odd, $\{\cali^1,\cali^2,\cali^3\}$ 
defines an almost quaternionic structure on the vector
space $\calh^p(\cals)$.
\tenrm
 
We are now ready to prove the main theorem (Vanishing Theorem)
of this section:

\noindent{\sc Theorem \btop.2.7}: \tensl 
Let $u\in\calh^p(\cals)$, $p\leq 2n+1$. 
\item{(i)} If $p$
is odd then $u\equiv0$.
\item{(ii)} If $p$ is even then $\cali^au=u$ for
$a=1,2,3$.
\tenrm
 
\noindent{\sc Proof}: Let $u\in\calh^p(\cals)$. We shall in fact show that 
$\cali^1u=\cali^2u$
irrespective of whether $p$ is even or odd; the result then follows
from the identities \btop.2.6 and symmetry between the indices
1,2,3. By \btop.2.1, we may choose an isometry $h\in Sp(1)$ so that
$h_*\Phi^1=\Phi^2$. Both $u$ and $\cali^1u$ are harmonic, so $h^*u=u$
and 
$$(\cali^1u)(X_1,\ldots,X_p)=(h_*(\cali^1u))(X_1,\ldots,X_p)
=u((h_*\Phi^1)(X_1),\ldots,(h_*\Phi^1)(X_p))=$$
$$=u(\Phi^2X_1,\ldots,\Phi^2X_p)= (\cali^2u)(X_1,\ldots,X_p).$$
\hfill\za

\noindent{\sc Corollary \btop.2.8}: \tensl
Let $(\cals,g)$ be a compact 3-Sasakian manifold of dimension
$4n+3$. Then the odd Betti numbers $b_{2k+1}$ of
$\cals$ are all zero for $0\leq k\leq n$.
\tenrm
   
We should point out that Corollary \btop.2.8 does not apply to compact
Sasakian or even Sasakian-Einstein manifolds. In [B-G 2] the authors construct
examples of Sasakian-Einstein manifolds with certain non-vanishing odd Betti
numbers within the range given in Corollary \btop.2.8. For example, in
dimension 7 there are circle bundles over Fermat hypersurfaces in
$\bbc\bbp^3,$ as well as circle bundles over certain complete intersections
that admit Sasakian-Einstein structures and have $b_3\neq 0.$ These are the
only known examples of Sasakian-Einstein manifolds which cannot admit any
3-Sasakian structure.

\smallskip
\noindent
{\bf \btop.3 3-Sasakian Cohomology As Primitive Cohomology}
\smallskip
We are going to consider connection between the cohomology 
of $\cals$ and that of $\calz$ and $\calo$. We will use the vanishing theorem
and orbifold Gysin sequence arguments for the diagram of orbifold
bundles of $\diamondsuit(\cals)$:
$$\matrix{\cals&&\ra{2}&&
\calz.&\cr&&&&&\cr\decdnar{}&
&\swarrow&&&\cr&&&&&\cr
\calo&&&&&\cr}\leqno{\btop.3.1}$$

\vskip -8pt

\noindent{\sc Proposition \btop.3.2}: \tensl Let ${\cals}$ be a compact 
3-Sasakian manifold 
of dimension $4n+3$ and
$\calz=\cals/S^1$ be the twistor space. Then
$b_{p}(\cals)=b_{p}(\calz)-b_{p-2}(\calz),$
for $p\leq 2n+1.$ 
In particular, all odd Betti numbers of $\calz$ vanish.
\tenrm

\noindent{\sc Proof}: The result follows form the rational Gysin sequence
applied to the orbifold fibration $S^1\rightarrow
\cals\rightarrow\calz$.  First, note that the
bundle $S^1\longrightarrow \cals\longrightarrow\calz$ is a circle $V$-bundle
over a compact K\"ahler-Einstein orbifold $\calz$. 
As explained in Section \bfol, up to a possible $\bbz_2$ cover,
$\cals$ is
the total space of the unit circle bundle in the dual of the contact line
V-bundle on $\calz$, and the
K\"ahler-Einstein metric of $\calz$ arises in accordance with
the orbifold version of the Kobayashi's theorem [Kob, B-G 1] . 
It follows that the connecting
homomorphism $\delta$ is given by wedging with a non-zero multiple of
the K\"ahler form of $\calz$. When $\calz$ is smooth 
this is well-known to be injective so
long as $p\leq 2n+2$. However, the Lefschetz decomposition is equally true
for compact orbifolds and the result still holds in this more general
situation [B-G 1].
The Gysin sequence therefore reduces to a series
of short exact sequences up to and including $H^{2n+1}(\cals)$, and the
proposition follows. 
\hfill\za

\noindent{\sc Proposition \btop.3.3}: \tensl Let ${\cals}$ be a compact
3-Sasakian manifold 
of dimension $4n+3$ and let
$\calo=\cals/\calf_3$. Then
$b_{2p}(\cals)=b_{2p}(\calo)-b_{2p-4}(\calo),$ for $p\leq 2n+1.$
\tenrm   
 
\noindent{\sc Proof}: The result follows form the Gysin sequence applied to
the orbifold fibration $\call\rightarrow \cals\rightarrow\calo$. 
Since the principal orbit of the $Sp(1)$ action (or generic leaf $\call$)
is either $S^3$ or $SO(3)$ the usual Gysin sequence argument applies
as well in this situation (see the Appendix). We have
$$\cdot\cdot\cdot\rightarrow H^{i}(\cals,\bbq)\rightarrow 
H^{i-3}(\calo,\bbq){\buildrel \delta \over \rightarrow} 
H^{i+1}(\calo,\bbq)\rightarrow  
H^{i+1}(\cals,\bbq)\rightarrow
H^{i-2}(\cals,\bbq)\rightarrow\cdot\cdot\cdot$$  
and the statement of the proposition follows easily
from the vanishing
of the odd Betti numbers of $\cals$.\hfill\za

Recall that the vector space of
primitive harmonic $p$-forms 
$\calh_0^{p}(\calz,\bbq)$ 
of the orbifold $\calz$  
is isomorphic to the cokernel of the injective
mapping
$L_\calz:\calh^{p-2}(\calz)\hookrightarrow \calh^p(\calz),\quad p\leq 2n$
defined by wedging with the K\"ahler 2-form. We define
the primitive Betti numbers $b^0_p(\calz)$ of $\calz$ as the
dimension of $\calh_0^{p}(\calz)$. Proposition \btop.3.2 says that
the primitive Betti numbers of $\calz$ are the usual Betti numbers of
$\cals$ and it follows from the fact that
for, $0\leq r\leq 2n+1$, an $r$-form on
$\cals$ is harmonic if and only if it is the lift of a primitive harmonic form
on $\calz$ [B-G 1].
Similarly, the vector space of
primitive harmonic $p$-forms
$\calh_0^{p}(\calo,\bbq)$
of the orbifold $\calo$
is isomorphic to the cokernel of the injective
mapping
$L_\calo:\calh^{p-4}(\calo)\hookrightarrow \calh^p(\calo),\quad p\leq 2n+2$
defined by wedging with the quaternionic K\"ahler 4-form
$\Omega$. The injectivity of this mapping is well-known
in the smooth case [Bon, Fuj, Kra] and it extends to the
orbifold case. We define
the primitive Betti numbers $b^0_p(\calo)$ of $\calo$ as the
dimension of $\calh_0^{p}(\calo)$. Proposition \btop.3.3 says that
the primitive Betti numbers of $\calo$ are the usual Betti numbers of
$\cals$. Again, Proposition \btop.3.3 is a consequence of the fact
that an $r$-form on
$\cals$ is harmonic if and only if it is the lift of a primitive harmonic form
on $\calo$, $0\leq r\leq 2n+1$.
\smallskip
\noindent
{\bf \btop.4 Regular 3-Sasakian Cohomology, Finiteness, and Rigidity}
\smallskip
In this Section we shall assume that $\cals$ is regular and, hence,
both $\calz$ and $\calo$ are smooth. In this instance, using the results of the
previous section, one can easily translate all the results
about strong rigidity of positive quaternion K\"ahler manifolds 
[Le 1, Le-Sal, Sal 3]
(see the chapter in this volume on Quaternionic K\"ahler Manifolds by S.
Salamon) to compact regular 3-Sasakian manifolds. In particular, we get

\noindent{\sc Proposition \btop.4.1}:  \tensl
Let $\cals$ be a compact regular 3-Sasakian manifold of dimension $4n+3$.
Then
$\pi_1(\cals)=0$ unless $\cals=\bbr\bbp^{4n+3}$ and 
$$\pi_2(\cals)=
\cases{
\bbz~&\hbox{iff $\cals=SU(n+2)/S(U(n)\times U(1))$,}\cr
finite &\hbox{otherwise.}\cr}$$
Furthermore, up to isometries, for 
each $n\geq1$ there are only finitely many regular 3-Sasakian
manifolds $\cals$.
\tenrm

\noindent{\sc Proof}: Using the long exact homotopy sequence for the vertical
map in \btop.3.1, this follows from the strong rigidity theorem of LeBrun and
Salamon [Le-Sal,Le 1] for positive quaternionic K\"ahler manifolds, and
Salamon's theorem that a positive quaternionic K\"ahler manifold with vanishing
Marchiafava-Romani class must be $\bbh\bbp^n.$ \hfill\za

\tenrm

$$
\beginpicture
\setcoordinatesystem units <1pt, 1pt>
\setlinear
\plot -95 83  -95 -88 /
\plot -94 83  -94 -88 /
\plot -60 83  -60 -88 /
\plot 220 83  220 -88 /
\plot 219 83  219 -88 /
\plot -95 83  220 83 /
\plot -95 82  220 82 /
\put {$n$} [l] at -85 75 
\put {Relation on Betti numbers or coefficients thereof} [l] at -30 75 
\plot -95 68  220 68 /
\plot -95 67  220 67 /
\put {$2$} [l] at -85 60 
\put {$b_2=b_4$} [l] at -45 60 
\plot -95 53  220 53 /
\put {$3$} [l] at -85 46 
\put {$b_2=b_6$} [l] at -45 46 
\plot -95 39  220 39 /
\put {$4$} [l] at -85 32 
\put {$2b_2+b_4 = b_6+2b_8$} [l] at -45 32 
\plot -95 25  220 25 /
\put {$5$} [l] at -85 18 
\put {$5b_2+4b_4 = 4b_8+5b_{10}$} [l] at -45 18 
\plot -95 11  220 11 /
\put {$6$} [l] at -85 04 
\put {$5b_2+5b_4+2b_6 = 2b_8+5b_{10}+5b_{12}$} [l] at -45 04 
\plot -95 -03  220 -03 /
\put {$7$} [l] at -85 -10 
\put {$7b_2+8b_4+5b_6 =
5b_{10}+8b_{12}+7b_{14}$} [l] at -45 -10 
\plot -95 -17  220 -17 /
\put {$8$} [l] at -85 -24 
\put {$28b_2+35b_4+27b_6+10b_8 =
10b_{10}+27b_{12}+35b_{14}+28b_{16}$} [l] at -45 -24 
\plot -95 -31  220 -31 /
\put {$9$} [l] at -85 -38 
\put {$12b_2+16b_4+14b_6+8b_8 =
8b_{12}+14b_{14}+16b_{16}+12b_{18}$} [l] at -45 -38 
\plot -95 -45  220 -45 /
\plot -95 -45.25  220 -45.25 /
\plot -95 -44.75  220 -44.75 /
\put {$10$} [l] at -85 -52 
\put {$15,21,20,14,5$} [l] at -45 -52 
\plot -95 -59  220 -59 /
\put {$16$} [l] at -85 -66 
\put {$40, 65, 77, 78, 70, 55, 35, 12$} [l] at -45 -66 
\plot -95 -73  220 -73 /
\put {$28$} [l] at -85 -80 
\put {$126, 225, 299, 350, 380, 391, 385, 364, 330,
285, 231, 170, 104,35$} [l] at -45 -80 
\plot -95 -87  220 -87 /
\plot -95 -88  220 -88 /
\put {{\bf Table 1:} Betti number relations in lower dimensions} [l] 
at -60 -110
\endpicture $$

\noindent{\sc Proposition \btop.4.2}:  \tensl
The Betti numbers of a regular compact 3-Sasakian manifold $\cals$ of
dimension $4n+3$ satisfy
\item{(i)}  $b_2\leq1$, with equality iff $\cals=SU(l+2)/S(U(l)\times U(1))$,
\item{(ii)} $\displaystyle\sum_{k=1}^nk(n+1-k)(n+1-2k)b_{2k}=0$.
\tenrm
 
\noindent{\sc Proof}:
(i) follows from Proposition \btop.4.1 and (ii) for 
Salamon's relation on Betti numbers of $\calo$ via
Theorem \btop.3.3.
\hfill\za
 
\noindent
The following is a 3-Sasakian version of a theorem
of Salamon [G-Sal]:
 
\noindent{\sc Proposition \btop.4.3}:  \tensl
Let $\cals$ be a regular compact 3-Sasakian manifold of
dimension $4n+3$. If
$n=3,4$ and $b_4=0$, then $\cals$ is either
a sphere $S^{4n+3}$
or a real projective space $\bbr\bbp^{4n+3}.$
\tenrm
 
The linear Betti number relations in Proposition \btop.4.2(ii) exhibit
an interesting symmetry of the coefficients which, for lower
values of $n$, are listed in Table 1.

One can compute the 
Poincar\'e polynomials of all known regular 3-Sasakian manifolds, that
is 3-Sasakian homogeneous space of
Theorem \bhom.2.6. We get [G-Sal]
\bigskip

\noindent{\sc Proposition \btop.4.4}:  \tensl
The Poincar\'e polynomials of the homogeneous
3-Sasakian manifolds are as given in Table 2.
\tenrm

$$
\beginpicture
\setcoordinatesystem units <1pt, 1pt>
\setlinear
\plot -155 91  -155 -111 /
\plot -154 91  -154 -111 /
\plot -95 91  -95 -111 /
\plot 15 91  15 -111 /
\plot 220 91  220 -111 /
\plot 219 91  219 -111 /

\plot -155 91  220 91 /
\plot -155 90  220 90 /
\put {$G$} [l] at -130 80
\put {$H$} [l] at -70 80
\put {$P(G/H,t)$} [l] at 40 80
\plot -155 71  220 72 /
\plot -155 70  220 71 /
\put {$SU(n+2)$} [l] at -150 60
\put {$SU(n-1)\times_{\bbz_n}T^1$} [l] at -90 60
\put {$\sum_{i=0}^n(t^{2i}+t^{4n+3-2i})$} [l] at 20 60
\plot -155 50  220 50 /
\put {$SO(2k+3)$} [l] at -150 40
\put {$SO(2k-1)\times SU(2)$} [l] at -90 40
\put {$\sum_{i=0}^{k-1}(t^{4i}+t^{8k-1-4i})$} [l] at 20 40
\plot -155 30  220 30 /
\put {$Sp(n+1)$} [l] at -150 20
\put {$Sp(n)$} [l] at -90 20
\put {$1+t^{4n+3}$} [l] at 20 20
\plot -155 10  220 10 /
\put {$SO(2l+4)$} [l] at -150 0
\put {$SO(2l)\times SU(2)$} [l] at -90 0
\put {$t^{2l}+t^{6l+3}+\sum_{i=0}^l(t^{4i}+t^{8l+3-4i})$} [l] at 20 0
\plot -155 -10  220 -10 /
\put {$E_6$} [l] at -150 -20
\put {$SU(6)$} [l] at -90 -20
\put {$1+t^6+t^8+t^{12}+t^{14}+t^{20}+\cdots$} [l] at 20 -20
\plot -155 -30  220 -30 /
\put {$E_7$} [l] at -150 -40
\put {Spin(12)} [l] at -90 -40
\put {$1+t^8+t^{12}+t^{16}+t^{20}+t^{24}+t^{32}+\cdots$} [l] at 20 -40
 
\plot -155 -50  220 -50 /
\put {$E_8$} [l] at -150 -60
\put {$E_7$} [l] at -90 -60
\put {$1+t^{12}+t^{20}+t^{24}+t^{32}+t^{36}+t^{44}+t^{56}+\cdots$}
[l] at 20 -60
 
\plot -155 -70  220 -70 /
\put {$F_4$} [l] at -150 -80
\put {$Sp(3)$} [l] at -90 -80
\put {$1+t^8+t^{23}+t^{31}$} [l] at 20 -80
 
\plot -155 -90  220 -90 /
\put {$G_2$} [l] at -150 -100
\put {$SU(2)$} [l] at -90 -100
\put {$1+t^{11}$} [l] at 20 -100
\plot -155 -110  220 -110 /
\plot -155 -111  220 -111 /
\put {{\bf Table 2:} Betti numbers of 3-Sasakian homogeneous
spaces} [l]
at -90 -130
\endpicture
$$

We conclude this section with a translation of two well-known classification
results for positive quaternionic K\"ahler manifolds.

\noindent{\sc Theorem \btop.4.5}: \tensl Let $(\cals,g)$ be a compact
regular 3-Sasakian manifold of dimension $4n+3$. If $n<3$ then
then $\cals=G/H$ is homogeneous, and hence one of the spaces listed
in Theorem \bhom.2.6.
\tenrm

The $n=0$ case is trivial and it was an observation made by
Tanno [Tan 2]. The $n=1$ case is based on [Hit 1, Fr-Kur] 
and it was first observed
in [Fr-Kat 2, B-G-F-K]. The $n=2$ case is based on [Po-Sal] and was stated in
[B-G-M 1].
\settabs 3\columns
\vfil\eject
\bigskip
\centerline {\bf \bg2. Killing Spinors and $G_2$-Structures}
\medskip
In this section we discuss some additional properties of Sasakian
and Sasakian-Einstein manifolds which are connected with spin
structure and eigenvalues of the Dirac operator
\smallskip
\noindent{\bf \bg2.1 Killing Spinors}
\smallskip
\noindent{\sc Definition \bg2.1.1}: \tensl
Let $(M,g)$ be a complete $n$-dimensional Riemannian spin manifold, and let
$S(M)$ be the spin bundle of $M$ and $\psi$ a smooth section of
$S(M)$.
We say that
$\psi$ is a {\it Killing spinor} if
$$\nabla_X\psi=\alpha\!\cdot\!X\!\cdot\!\psi,\qquad\forall X\in\Gamma(TM),
\leqno{\bg2.1.2}$$
where $\nabla$ is the Levi-Civita connection of $g$ and $X\!\cdot\!\psi$
denotes the Clifford product of $X$ and $\psi$. We say that
$\psi$ is {\it imaginary} when $\alpha\in {\rm Im}(\bbc^*)$, $\psi$ is 
{\it parallel} if $\alpha=0$ and $\psi$ is {\it real} if $\alpha\in
{\rm Re}(\bbc^*)$. 
\tenrm

From the point of view of Einstein geometry
the importance of Killing spinors is an immediate consequence of the
following theorem of Friedrich [Fr]:

\noindent{\sc Theorem \bg2.1.2}: \tensl
Let $(M,g)$ be an $n$-dimensional; complete Riemannian spin manifold with
a Killing spinor. Then $M$ is Einstein with Einstein constant
$\lambda=4(n-1)\alpha^2.$ In particular, when $\alpha\in {\rm Re}(\bbc^*)$,
$M$ is compact of positive scalar curvature.
\tenrm

On the other hand,
Friedrich showed that if $M$ is a compact spin manifold
of positive scalar curvature and
$R_0$ is the minimum of the scalar curvature, 
then for all eigenvalues $\beta$ of
the Dirac operator $D$ one has $\beta^2\geq {1\over4}{nR_0\over n-1}$ [Fr 1].
If the equality holds than it follows that the corresponding eigenspinor
must be a Killing spinor with $\alpha=\pm{1\over2}[{R_0\over n(n-1)}]^{1/2}$. 
We have the following important property of manifolds with Killing spinors
[B-G-F-K]:

\noindent{\sc Theorem \bg2.1.3}: \tensl
Let $(M^n,g)$ be a connected Riemannian spin manifold admitting 
a non-trivial Killing spinor with $\alpha\not=0$. Then $(M,g)$ is 
locally irreducible. Furthermore, if $M$ is locally symmetric, or
$n\leq4$, then $M$ is a space of 
constant sectional curvature equal $4\alpha^2$.
\tenrm

From now on we will be interested only in the case of real Killing spinors.
It was Friedrich and Kath [Fr-Kat 1-2] who first noticed that
in some low odd dimensions the existence of real Killing spinors leads
naturally to the existence of Sasakian-Einstein or 3-Sasakian structures.
Later, the problem found a simple classification in terms of
the holonomy of the associated metric cone $C(M)$ [B\"ar]. First,
we have the following definition:

\noindent{\sc Definition \bg2.1.4}: \tensl
We say that $M$ is of type
$(p,q)$ if it carries exactly $p$ linearly independent real Killing spinors 
with $\alpha>0$ and exactly $q$ linearly independent 
real Killing spinors with $\alpha<0$, or vice versa.
\tenrm

For, example, the standard sphere $S^n$ is of type
$(2^{[n/2]},2^{[n/2]})$. B\"ar shows that when $M$ admits a real
Killing spinor then the cone 
$(C(M),\bar{g})$ has a parallel spinor. In particular,
$C(M)$ is always Ricci-flat and, when $M$ is simply-connected,  then
only a few holonomy groups ${\rm Hol}(\bar{g})$ are possible [Wan 3]:

\noindent{\sc Theorem \bg2.1.5}: \tensl
Let $(M^n,g)$ be a simply-connected Riemannian spin manifold admitting
a non-trivial Killing spinor and let
${\rm Hol}(\bar{g})$ be the holonomy group 
of the metric cone $(C(M),\bar{g})$. Then there are only the following
6 possibilities for the triple $\bigl(n, {\rm Hol}(\bar{g}), (p,q)\bigr)$:
\bigskip
\+\ \ \ \ (1)\ \ $n$ arbitrary, &${\rm Hol}(\bar{g})={\rm id}$, 
&$(p,q)=(2^{[n/2]},2^{[n/2]})$,\cr
\+\ \ \ \ (2)\ \ $n=2m+1, m$ even, 
&${\rm Hol}(\bar{g})=SU(m+1)$, &$(p,q)=(1,1)$,\cr
\+\ \ \ \ (3)\ \ $n=4m+3$,  &${\rm Hol}(\bar{g})=SU(2m+2)$, &$(p,q)=(2,0)$,\cr
\+\ \ \ \ (4)\ \ $n=4m+3$, &${\rm Hol}(\bar{g})=Sp(m+1)$, &$(p,q)=(m+2,0)$,\cr
\+\ \ \ \ (5)\ \ $n=7$, &${\rm Hol}(\bar{g})={\rm Spin}(7)$, &$(p,q)=(1,0)$,\cr 
\+\ \ \ \ (6)\ \ $n=6$, &${\rm Hol}(\bar{g})=G_2$, &$(p,q)=(1,1)$.\cr
\tenrm   

The first case is special as $M$ is the $n$-dimensional round sphere.
Since $M$ is assumed to be simply-connected,
in the next two cases, by Proposition \bdef.1.9, $M$ must be
Sasakian-Einstein. In the case (4), by Definition \bdef.2.1, 
$M$ is 3-Sasakian. Specifically, we get the following theorem [B\"ar]:

\noindent{\sc Theorem \bg2.1.6}: \tensl
Let $(M^n,g)$ be a complete simply-connected Riemannian spin manifold admitting
a non-trivial Killing spinor with $\alpha>0$ or $\alpha<0$. If
$n=2m+1$, $m\geq2$ even, then there are two possibilities:
\item{(i)} $(M,g)=(S^n,g_{can})$,
\item{(ii)} $(M,g)$ is of type $(1,1)$ and it is a Sasakian-Einstein manifold.

Conversely, if $(M,g)$ is a complete simply-connected Sasakian-Einstein
manifold of dimension $4m+1$, then $M$ carries Killing spinors with $\alpha>0$
and $\alpha<0$.  \tenrm

\noindent{\sc Remark \bg2.1.7}:
Note that in the converse statement we do not need to assume that
$M$ is spin. When $\pi_1(M)=0$ this is automatic by Corollary \bdef.1.11.
When $\pi_1(M)\not=0$ then the `if' part of Theorem  \bg2.1.6 can be 
generalized and we still get two possibilities:
(i) either $M$ is
a spin spherical space form, or (ii) it is 
of type $(1,1)$ with a Sasakian-Einstein structure and ${\rm Hol}(\bar{g})=
SU(m+1)$ [Wan 3].

\noindent{\sc Theorem \bg2.1.8}: \tensl
Let $(M^n,g)$ be a complete simply-connected 
Riemannian spin manifold admitting 
a non-trivial Killing spinor with $\alpha>0$ or $\alpha<0$. If 
$n=4m+3$, $m\geq2$, then there are three possibilities:    
\item{(i)} $(M,g)=(S^n,g_{can})$,   
\item{(ii)} $(M,g)$ is a Sasakian-Einstein manifolds of type $(2,0)$, but
$(M,g)$ is not 3-Sasakian,
\item{(iii)} $(M,g)$ is of type $(m+2,0)$ and it is 3-Sasakian. 

Conversely, if $(M,g)$ is a complete simply-connected 3-Sasakian manifold, 
of dimension $4m+3$ which is not of constant curvature, 
then $M$ carries $(m+2)$ linearly independent
Killing spinors with $\alpha>0$. If $(M,g)$ is a complete simply-connected
Sasakian-Einstein manifold of dimension $4m+3$ which is not 3-Sasakian
then $M$ carries $2$ linearly independent
Killing spinors with $\alpha>0$.
\tenrm   

\noindent{\sc Remark \bg2.1.9}: 
Note that in Theorem \bg2.1.8(ii) we are not excluding the possibility
of $M$ having another 3-Sasakian structure with a different metric $g'$.
We are only saying that the holonomy group ${\rm Hol}(\bar{g})=
SU(2m+2)$ rather than $Sp(m+1)\subset SU(2m+2)$, which, by definition,
means that $g$ cannot be 3-Sasakian. 
However, we are not aware of any such example.
We have excluded
${\rm dim}(M)=7$ because in this case we have one
more possibility due to Theorem \bg2.1.5 and we want to discuss the
associated geometry in more detail later. 
Again, one can generalize 
Theorems \bg2.1.8 to $\pi_1(M)\not=0$. For the full list of
possible holonomy groups ${\rm Hol}(\bar{g})$ see [Wan 2]. 
The coresponding $M$ are then only locally Sasakian-Einstein or locally
3-Sasakian [Or-Pi, Pi]. The problem of the existence of Killing spinors
on a Sasakian-Einstein or 3-Sasakian manifold with 
$\pi_1(M)\not=0$ is, however, more subtle. 

\noindent{\sc Corollary \bg2.1.10}: \tensl
Let $(\cals,g)$ be a compact Sasakian-Einstein 
manifold of dimension $2m+1$. Then
$\cals$ is locally symmetric if and only if $\cals$ is of constant
curvature. Moreover, $(\cals,g)$ is locally irreducible as a
Riemannian manifold.
\tenrm

\noindent{\sc Proof}: If necessary, go to the universal cover
$\tilde\cals$. This is a compact simply-connected Sasakian-Einstein manifold;
hence, it admits a non-trivial Killing spinor by Theorems \bg2.1.6 and
\bg2.1.8.  The statement then follows from the Theorem \bg2.1.3.\hfill\za

\noindent{\sc Corollary} \bg2.1.11: \tensl Let $(\cals, g)$ be a compact
Sasakian-Einstein manifold of dimension
$2m+1$. Then
${\rm Hol}(g)=SO(2m+1)$.
\tenrm

\noindent{\sc Proof}: Let us consider universal cover
$\tilde{\cals}$. This is a compact simply-connected Sasakian-Einstein
manifold; hence, it admits a non-trivial Killing spinor. By the previous 
corollary, it can be symmetric if
only if it is isomorphic to a space of constant curvature, that is, a sphere.
Then $\cals$ is a spherical space form and ${\rm Hol}(g)=SO(2m+1)$.
Assume $\cals$ is not locally symmetric. By Corollary \bg2.1.10 $\cals$ is
locally Riemannian irreducible, so for dimensional reasons and Berger's 
famous classification theorem [Ber], 
the only possibilities for the restricted
holonomy group ${\rm Hol}^0(g)$ are $SO(2m+1)$ and $G_2$ in dimension 7. But
$G_2$ holonomy implies Ricci-flat and, hence, not Sasakian-Einstein. Hence, the
restricted holonomy group ${\rm Hol}^0(g)=SO(2m+1)$. Since $\cals$ is
orientable this coincides with the holonomy group ${\rm Hol}(g)$.  \hfill\za
\smallskip 
\noindent{\bf \bg2.2 $G_2$-Structures}
\smallskip
Recall, that geometrically $G_2$ is defined to be the Lie group
acting on $\bbr^7$ and preserving the 3-form
$$\varphi=\alpha_1\wedge\alpha_2\wedge\alpha_3+
\alpha_1\wedge(\alpha_4\wedge\alpha_5-
\alpha_6\wedge\alpha_7)+
\alpha_2\wedge(\alpha_4\wedge\alpha_6-
\alpha_7\wedge\alpha_5)+
\alpha_3\wedge(\alpha_4\wedge\alpha_7-
\alpha_5\wedge\alpha_6),\leqno{\bg2.2.1}$$
where $\{\alpha_i\}_{i=1}^7$ is a fixed orthonormal basis of the dual of
$\bbr^7$. A $G_2$ structure on a 7-manifold $M$ is, by definition, a reduction
of the structure group of the tangent bundle to $G_2$. This is equivalent to
the existence of a global 3-form $\varphi\in\Omega^3(M)$
which may be written locally as \bg2.2.1.  Such a 3-form defines
an associated Riemannian metric, an orientation class, and a spinor field of
constant length. The following terminology is due to Gray [Gra 2]:

\noindent{\sc Definition \bg2.2.2}: \tensl
Let $(M,g)$ be a complete $7$-dimensional Riemannian manifold. 
We say that that $(M,g)$ has  {\it weak holonomy} $G_2$ if there exist
a global 3-form $\varphi\in\Omega^3(M)$ which locally can be written in terms
of a local orthonormal basis as in \bg2.2.1, and $d\varphi=c\star \varphi$,
where $\star$ is the Hodge star operator associated to $g$ and $c$ is a
constant whose sign is fixed by an orientation convention.  \tenrm

The equation $d\varphi=c\star \varphi$ 
implies that $\varphi$ is `nearly parallel' in the sense that only a
1-dimensional component of $\nabla\varphi$ is different from zero [Fe-Gra].
Thus, a weak holonomy $G_2$ structure is sometimes called a {\it nearly
parallel $G_2$ structure}. The case of $c=0$ is somewhat special. In
particular, it is known [Sal 4] that the condition $d\varphi =0=d\star\varphi$
is equivalent to the condition that $\varphi$ be parallel, i.e.,
$\nabla\varphi=0$ which is equivalent to the condition that the metric $g$ has
holonomy group contained in $G_2.$ For a discussion of this very interesting
and very difficult case, see the article by D. Joyce in this volume. The
following theorem provides the connection with the previous discussion on
Killing spinors [B\"ar]

\noindent{\sc Theorem \bg2.2.3}: \tensl
Let $(M,g)$ be a complete $7$-dimensional Riemannian manifold with
weak holonomy $G_2$. Then the holonomy group ${\rm Hol}(\bar{g})$
of the metric cone $(C(M),\bar{g})$ is contained in ${\rm Spin}(7)$.
In particular, $C(M)$ is Ricci-flat and 
$M$ is Einstein with positive Einstein constant
$\lambda=6$.
\tenrm

\noindent{\sc Remark \bg2.2.4}:
The sphere $S^7$ with its constant curvature metric is isometric to
the isotropy irreducible space $Spin(7)/G_2$. The fact that $G_2$
leaves invariant (up to constants) a unique 3-form and a unique 4-form
on $\bbr^7$ implies immediately that this space has weak holonomy $G_2$.

\noindent{\sc Definition \bg2.2.5}: \tensl
Let $(M,g)$ be a complete $7$-dimensional Riemannian manifold.
We say that $g$ is a {\it proper $G_2$-metric}
if ${\rm Hol}(\bar{g})={\rm Spin}(7)$.
\tenrm

\noindent{\sc Theorem \bg2.2.6}: \tensl
Let $(M^7,g)$ be a complete simply-connected Riemannian spin manifold of
dimension 7 admitting
a non-trivial Killing spinor with $\alpha>0$ or $\alpha<0$. Then
there are four possibilities:
\item{(i)} $(M,g)$ is of type $(1,0)$ and it is
a proper $G_2$-manifold,
\item{(ii)} $(M,g)$ is of type $(2,0)$ and it is 
a Sasakian-Einstein manifold, but
$(M,g)$ is not 3-Sasakian,
\item{(iii)} $(M,g)$ is of type $(3,0)$ and it is 3-Sasakian,
\item{(iv)} $(M,g)=(S^7,g_{can})$ and is of type $(8,8)$.
 
Conversly, if $(M,g)$ is a compact simply-connected 
proper $G_2$-manifold then it carries a Killing spinor with
$\alpha>0$. If $(M,g)$ is a compact simply-connected
Sasakian-Einstein 7-manifold which is not 3-Sasakian
then $M$ carries $2$ linearly independent
Killing spinors with $\alpha>0$. Finally, if $(M,g)$ is a
3-Sasakian 7-manifold,
which is not of constant curvature,
then $M$ carries $3$ linearly independent
Killing spinors with $\alpha>0$. 
\tenrm   

\noindent{\sc Remark \bg2.2.7}
The four possibilities of the Theorem \bg2.2.6 correspond to the
sequence of inclusions
$${\rm Spin}(7)
\supset SU(4)\supset Sp(2)\supset \{{\rm id}\}.$$
All of the corresponding cases are examples of weak holonomy $G_2$
metrics. If we exclude the trivial case when the associated cone is
flat, we have three types of the weak holonomy $G_2$ geometries.
Following [F-K-M-S] we use the number of linearly independent Killing spinors
to classify the types of weak holonomy $G_2$ geometries. We call these
type I, II, and III corresponding to cases (i), (ii), and (iii) of Theorem
\bg2.2.6, respectively.

\noindent{\sc Remark \bg2.2.8} In the case $\pi_1(M)\not=0$, then
$M$ is either a spin spherical space form or
${\rm Hol}(\bar{g})$ equals to $SU(4)$, $SU(4)\bowtie\bbz_2$,
$Sp(2)$, or ${\rm Spin}(7)$ of type
$(2,0), (1,0), (3,0), (1,0)$, respectively. Hence, we have just one more
possible  geometry for $M$ [Wan 3]. Note that in the case
${\rm Hol}(\bar{g})=SU(4)\bowtie\bbz_2$, the cone $C(M)$ is not
K\"ahler so that $M$ cannot be Sasakian, but it is locally so.

Recall that $S^7$, regarded
as the space $Sp(2)/Sp(1)$ and fibering over $S^4$, admits a
`squashed' Einstein metric which does not have constant curvature.
This metric also has weak holonomy $G_2$ since the associated cone
metric has holonomy equal to $Spin(7)$ and therefore $S^7$
with this metric is a proper $G_2$-manifold. We can generalize this
example to get [G-Sal, F-K-M-S]:
 
\noindent{\sc Theorem \bg2.2.9}: \tensl Let $(\cals,g)$ be a 7-dimensional
3-Sasakian manifold. Then the metric $g$ has weak holonomy $G_2.$ Moreover, the
second Einstein metric $g'$ given by Theorem \bfol.4.2 has weak holonomy $G_2.$
In fact $g'$ is a proper $G_2$ metric. \tenrm

\noindent{\sc Proof}: For the second Einstein metric $g'$ we have three
mutually orthonormal 1-forms
$\alpha^1=\sqrt{t}\eta^1,\quad\alpha^2=\sqrt{t}\eta^2,\quad
\alpha^3=\sqrt{t}\eta^3,$ where $t$ is the parameter of the canonical variation
discussed in section \bfol.4. Let $\{\alpha^4,\alpha^5,\alpha^6,\alpha^7\}$ be
local 1-forms spanning the annihilator of $\calv_3$ in $T^*\cals$ such that
$$\hi^1= 2(\alpha^4\wedge\alpha^5-\alpha^6\wedge\alpha^7),$$
$$\hi^2= 2(\alpha^4\wedge\alpha^6-\alpha^7\wedge\alpha^5),$$
$$\hi^3= 2(\alpha^4\wedge\alpha^7-\alpha^5\wedge\alpha^6).$$
Then the set $\{\gra^1,\cdots \gra^7\}$ forms a local orthonormal coframe for
the metric $g'.$  In terms of the 3-forms $\Eta$ and $\Theta$ of \btop.2.2 we
have $\varphi={1\over2}\sqrt{t}\Theta+\sqrt{t}^3\Eta.$ One easily sees that
this is of the type of Equation \bg2.2.1 and, therefore, defines a
compatible $G_2$-structure.  Moreover, a straightforward computation gives
$$d\varphi ={1\over2}\sqrt{t}\Omega +
\sqrt{t}(t+1)d\Eta,\qquad \star\varphi = -{1\over2}td\Eta - {1\over24}\Omega.$$
Thus, $d\varphi=c\star\varphi$ is solved with $\sqrt{t}=1/\sqrt5$, and
$c=-12/\sqrt5$. So $g'$ has weak holonomy $G_2.$ That $g'$ is a proper $G_2$
metric is due to [F-K-M-S]. The idea is to use Theorem \bg2.2.6. Looking at the
four possibilities given in that theorem, we see that it suffices to show that
$g'$ is not Sasakian-Einstein. The details are in [F-K-M-S]. \hfill\za

\noindent{\sc Examples \bg2.2.10}: 3-Sasakian 7-manifolds are plentiful and
examples will be discussed in next section. These give, by Theorem \bg2.2.9,
many examples of type I and type III geometries. 
Examples of simply-connected
type I geometries that do not arise via 
Theorem \bg2.2.9 are the homogeneous Aloff-Wallach spaces
$N_{k,l}=SU(3)/T^1_{k,l}$, with ${\rm gcd}(k,l)=1$ and $(k,l)\not=(1,1)$
[C-M-Sw, B-G-F-K] together with
the homogeneous real Stiefel manifold $SO(5)/SO(3)$ [Bry]. 
All the known type II geometries are the 3 homogeneous examples from the list of
Corollary \bhom.1.3(iii) (not 3-Sasakian) and the inhomogeneous 
simply-connected circle
bundles over $P_k\times\bbc\bbp^1$, where $P_k$ is the del Pezzo
surface with $2<k<9$ [F-K-M-S]. Actually,
$N_{1,1}$ has three Einstein metrics. One is 3-Sasakian and is denoted by
$\cals(1,1,1)$ in section \bex.4 below. The second is the proper $G_2$ metric
of Theorem \bg2.9, while the third Einstein metric also has weak holonomy $G_2$ 
most likely of type I but we could not positively exclude
type II as a possibility [C-M-Sw]. 
\bigskip
\centerline {\bf \bq. The Quotient Construction}
\medskip
In this section we give a general $3$-Sasakian reduction procedure
which constructs new $3$-Sasakian manifolds from a given $3$-Sasakian
manifold $\cals$ with a non-trivial $3$-Sasakian isometry 
group $I_0(\cals,g)$ [B-G-M 2].
Actually, this is a reduction that is 
associated with a quadruple of spaces of the fundamental
diagram $\diamondsuit(\cals)$. At the level of the hyperk\"ahler cone
$C(\cals)$ the reduction was discovered by 
Lindstr\"om and Ro\v cek [L-R] in the context 
of supersymmetric $\sigma$-model and later rigorously described by
Hitchin et al. in [H-K-L-R]. In the case of the quaternionic
K\"ahler base $\calo$ the reduction was discovered by the second
author and H.B. Lawson [G, G-L]. The lift of the quaternionic 
K\"ahler quotient to the twistor space $\calz$ was described
by Hitchin [Hit 2]. In this section we restrict  ourselves to describing the
general procedure of reduction together with the homogeneous case arising from
reduction by a circle group, as well as a brief discription of the singular
case. The large class of 3-Sasakian toric manifolds obtained by reduction is
relagated to a separate section, namely section \bex.
It should also be understood that every 3-Sasakian
reduction gives as well a reduction procedure for each of the spaces of the
fundamental diagram $\diamondsuit(\cals).$

\smallskip
\noindent
{\bf \bq.1. The 3-Sasakian Moment Map}
\smallskip
Let $(\cals,g)$ be a $3$-Sasakian 
manifold with a nontrivial group $I_0(\cals,g)$ of $3$-Sasakian 
isometries.  By the Definition \bdef.2.1, $C(\cals)=\cals \times \bbr^+$ 
is a hyperk\"ahler manifold with respect to 
the cone metric $\bar{g}.$  The isometry group 
$I_0(\cals,g)$ extends to a group $I_0(C(\cals),\bar{g})\cong 
I_0(\cals,g)$ of isometries on $C(\cals)$ by defining each element to act 
trivially on $\bbr^+.$  Furthermore, it follows easily from the definition 
of the complex structures $I^a$ given in 
equation \bdef.1.7 that these isometries 
$I_0(C(\cals),\bar{g})$ are hyperk\"ahler; that is, they preserve the hyperk\"ahler 
structure on $C(\cals).$  Recall [H-K-L-R] shows that any subgroup 
$G\subset I_0(M,\bar{g})$ gives rise to a hyperk\"ahler moment map 
$\mu:M \ra{1} \gg^*\otimes \bbr^3,$ 
where $\gg$ denotes the Lie algebra of $G$ and $\gg^*$ is its dual.  
Thus, we can define a $3$-Sasakian moment map 
$$\mu_{\cals} :\cals ~\ra{1.5}~ \gg^*\otimes \bbr^3 \leqno{\bq.1.1}$$
by restriction $\mu_{\cals} =\mu\!\mid\!\cals.$  
We denote the components of $\mu_{\cals}$ with respect 
to the standard basis of $\bbr^3,$ which we have identified with the 
imaginary quaternions, by $\mu^a_{\cals}.$  
Recall that ordinarily moment maps determined by Abelian 
group actions (in particular, those associated to $1$-parameter groups) are 
only specified up to an arbitrary constant.  This is not the case for 
$3$-Sasakian moment maps since we require that the group $Sp(1)$ generated 
by the Sasakian vector fields $\xi^a$ acts on the level sets of 
$\mu_{\cals}.$  However, we shall see that $3$-Sasakian moment maps are 
given by a particularly simple expression.  

\noindent{\sc Proposition} \bq.1.2: \tensl Let $(\cals,g)$ be a
$3$-Sasakian manifold with a connected 
compact Lie group $G$ acting on $\cals$ by
3-Sasakian isometries.  Let $\tau$ be an element of the Lie algebra $\gg$ of 
$G$ and let $X^{\tau}$ denote the corresponding infinitesimal isometry.  
Then there is a unique $3$-Sasakian moment map $\mu_{\cals}$ such that the 
zero set $\mu^{-1}_{\cals}(0)$ is invariant under the group $Sp(1)$ 
generated by the vector fields $\xi^a$.  This moment map is given by
$$<\mu^a_{\cals},\tau> ~=~ {1\over 2}\eta^a(X^{\tau}). \leqno{\bq.1.3}$$
Furthermore, the zero set $\mu_{\cals}^{-1}(0)$ is $G$ invariant. \tenrm

\noindent{\sc Proof}: Using the Definition \bdef.2.1 we can define the 
$2$-forms $\gro^a_{\cals}$ on $\cals$ as the restriction of the 
hyperk\"ahler $2$-forms $\gro^a.$  Then any $3$-Sasakian moment map 
$\mu^a_{\cals}(\tau)$ determined by $\tau\in\gg$ satisfies
$2d\mu^a_{\cals}(\tau) = 2X^{\tau}\rfloor \gro^a_{\cals} 
= -X^{\tau}\rfloor d\eta^a.$
As $X^{\tau}$ is a $3$-Sasakian infinitesimal isometry, Lemma \bhom.2.2
implies that $2<\mu^a_{\cals},\tau>$ differs from $\eta^a(X^{\tau})$ by a
constant depending on $a$ and $\tau.$ One then uses the invariance of the zero
set $\mu_{\cals}^{-1}(0)$ to show that these constants must vanish. See [B-G-M
2] for details. \hfill\za

Henceforth by the $3$-Sasakian moment map, we shall mean the moment map 
$\mu_{\cals}$ determined in Proposition \bq.1.2.  Hence, the 
Definition \bdef.2.1
and Proposition \bq.1.2 imply 

\noindent{\sc Theorem \bq.1.5}: \tensl Let $(\cals,g)$ be a
$3$-Sasakian manifold with a
connected compact Lie group $G$ acting on $\cals$ smoothly and properly by
$3$-Sasakian isometries.  Let $\mu_{\cals}$ be the corresponding $3$-Sasakian
moment map and assume both that $0$ is a regular value of $\mu_{\cals}$ and
that $G$ acts freely on the submanifold $\mu^{-1}_{\cals}(0)$.  Furthermore,
let $\gri:\mu^{-1}_{\cals}(0) \ra{1.5} \cals$ and
$\pi:\mu^{-1}_{\cals}(0) \ra{1.5} \mu^{-1}_{\cals}(0)/G$
denote the corresponding embedding and submersion.  Then
$(\cals/\!\!/\!\!/G=\mu^{-1}_{\cals}(0)/G,\check{g})$
is a smooth $3$-Sasakian manifold of dimension
$4(n-\hbox{dim}~\gg)+3$ with metric $\check{g}$ and characteristic
vector fields $\check{\xi}^a$ determined uniquely by the two conditions
$\gri^*g = \pi^*\check{g}$ and
$\pi_*(\xi^a\!\mid\!\mu^{-1}_{\cals}(0)) = \check{\xi}^a.$
\tenrm

We conclude this part with the following fact concerning 
$3$-Sasakian isometries whose proof can be found in [B-G-M2].

\noindent{\sc Proposition} \bq.1.6:  \tensl  Assume that the hypothesis of
Theorem \bq.1.5 holds. In addition assume that $(\cals,g)$ is complete 
and hence compact.  Let $C(G)\subset I_0(\cals,g)$ denote the 
centralizer of $G$ in $I_0(\cals,g)$ and let $C_0(G)$ denote the
subgroup of $C(G)$ given by the connected component of the identity. 
Then $C_0(G)$ acts on the 
submanifold $\mu_{\cals}^{-1}(0)$ as isometries with respect to the 
restricted metric $\gri^*g$ and the $3$-Sasakian isometry group 
$I_0(\cals/\!\!/\!\!/G,\check{g})$ of the quotient
$(\cals/\!\!/\!\!/G,\check{g})$ determined in Theorem \bq.1.5 contains an
isomorphic copy of $C_0(G).$  Furthermore, if $C_0(G)$ acts transitively on
$\cals/\!\!/\!\!/G$, then $\cals/\!\!/\!\!/G$ is 
a $3$-Sasakian homogeneous space.
\tenrm

It should be mentioned that it is 
not required that the isometry group $I_0(\cals/\!\!/\!\!/G,\check{g})$ acts
effectively.

\smallskip
\noindent
{\bf \bq.2 Regular Quotients And Classical Homogeneous Metrics}
\smallskip
We now apply the reduction procedure given in Theorem \bq.1.4 to the
round unit sphere $S^{4n+3}$ to explicitly construct the Riemannian metrics
for the $3$-Sasakian homogeneous manifolds arising from the simple classical
Lie algebras.  These metrics are precisely the ones associated to the three
infinite families appearing in Theorem \bhom.2.6. The quotient
construction applied to $\diamondsuit(\cals^{4n+3})$ explicitly describes
all metrics in the fundamental diagrams $\diamondsuit(G/H)$, where 
$G$ is either the special unitary $SU(n+1)$ or the orthogonal group
$SO(n+1)$. 
To carry out this reduction we must set some conventions.
We describe the unit sphere $S^{4n+3}$ by
its embedding in flat space and we represent an element
${\bf u} = (u_1,\cdots, u_{n+1}) \in \bbh^{n+1}$
as a column vector.  The quaternionic components of this vector are denoted
by ${\bf u}^0$ for the real component and by ${\bf u}^a$ for the three
imaginary components so that we can write
$\bfu=\bfu^0+i\bfu^1+j\bfu^2+k\bfu^3$ using the quaternionic units 
$\{i,j,k\}.$ We also define quaternionic conjugate 
$\bar\bfu=\bfu^0-i\bfu^1-j\bfu^2-k\bfu^3$. 

Now, the infinitesimal
generators of the subgroup $Sp(1)\subset \bbh^*$ 
acting by the right multiplication on $\bfu$ are the
defining vector fields $\xi^a$ for the Sasakian $3$-structure. 
These vector fields are given by
$$\xi^a_r ~=~ {\bf u}^0\cdot {\partial \over \partial {\bf u}^a}
- {\bf u}^a\cdot {\partial \over \partial {\bf u}^0} -
\gre^{abc}{\bf u}^b\cdot {\partial \over
\partial {\bf u}^c}, \leqno{\bq.2.1}$$
where the dot indicates sum over the vector components $u_i$ and the
subscript $r$ means that these vector fields are the generators of the
right action.
 
We will first  consider $G=U(1)$ acting on the sphere $S^{4n+3}$ as follows
$$\varphi_t(\bfu)=\tau\bfu,\qquad \tau=e^{2\pi it},\qquad \bfu\in S^{4n+3}.
\leqno{\bq.2.2}$$
Note that this action is actually free on $S^{4n+3}$ and hence it will be
automatically free on the level set of the moment map. To compute
the moment map
we identify the imaginary quaternions $\bbr^3$ with the Lie
algebra $\gsp1$ in equation \bq.1.1 and the Lie algebra of
$U(1)$ with $\bbr$, 
so the moment map is $\mu_{\cals}:S^{4n+3} ~\ra{1.5}~
\bbr\otimes \gsp1$  and it can easily be computed
$$\mu_{\cals}(\bfu)=\sum_{\alpha=1}^{n+1}\bar{u}_\alpha i u_\alpha. 
\leqno{\bq.2.3}$$
One can easily
identify the zero-level set of the moment map with the
Stiefel manifold of complex 2-frames in $\bbc^{n+1},$ and the following
proposition is then an immediate consequence of Theorem \bq.1.5. 
 
\noindent{\sc Proposition} \bq.2.4: \tensl Let $N=\mu_{\cals}^{-1}(0)$ and
$\iota: N\hookrightarrow  S^{4n+3}$ be the inclusion. Then 
$\iota$ is an embedding and $(N, \iota^*g_{can})$ is
the complex homogeneous Stiefel 
manifold $V_{2,n+1}^\bbc=SU(n+1)/SU(n-1)$
of 2-frames in $\bbc^{n+1}$. Hence, the 3-Sasakian quotient
$S^{4n+3}/\!\!/\!\!/U(1)=V_{2,n+1}^\bbc/U(1)=
SU(n+1)/S(U(n-1)\times U(1))$ with the 3-Sasakian metric 
$\check{g}$ given by inclusion $\iota$ and submersion
$\pi:N ~\ra{1.5}~ N/U(1)$, i.e., $\iota^*g_{can}=\pi^*\check{g}$.
\tenrm

\noindent{\sc Remark} \bq.2.5: A similar construction can be carried
out for the $Sp(1)$-action on $S^{4n+3}$ defined by the left
multiplication of $\bfu$ by a unit quaternion $\sigma$, i.e.,
$$\varphi_\sigma(\bfu)=\sigma\bfu,\qquad \sigma\bar{\sigma}=1,
\qquad \bfu\in S^{4n+3}.
\leqno{\bq.2.6}$$
This action is free on $S^{4n+3}$ and the zero-level set
of the corresponding moment map can be easily identified with the
real Stiefel manifold 
$V^\bbr_{4,n+1}\simeq SO(n+1)/SO(n-3)$
of 4-frames in $\bbr^{n+1}$ with $n\geq 4.$
Hence, the reduced
space $S^{4n+3}/\!\!/\!\!/Sp(1)={SO(n+1)\over SO(n-3)\times Sp(1)}$.
For the more detailed and uniform description of the geometry of these
two quotients see [B-G-M 2].
\smallskip
\noindent 
{\bf \bq.3 The Structure of Singular Quotients}
\smallskip
In this section we will describe a more general situation, when
the zero-level set of the 3-Sasakian moment map \bq.1.1 is not
necessarily smooth and the group action on the level set is
not necessarily locally free.

Let $G$ be a Lie group acting smoothly and properly on a manifold
$\cals$ and let $H\subset G$ be a subgroup. 
Using standard notation we will denote
by $\cals_{H}\subset\cals$ the set of points 
in $\cals$ where the stability group is exactly equal to $H$ and by
$\cals_{(H)}\subset\cals$ the set of points with stabilizer conjugate
to $H$ in $G$. It follows than that the normalizer $N(H)$ of $H$ in $G$
acts freely on $\cals_{(H)}$. 
Then we have the following theorem due to Dancer and
Swann [D-Sw]:

\noindent{\sc Theorem} \bq.3.1: \tensl 
Let $(\cals,g)$ be a
$3$-Sasakian manifold with a
connected compact Lie group $G$ acting on $\cals$ smoothly and properly by
$3$-Sasakian isometries.  Let $\mu_{\cals}$ be the corresponding $3$-Sasakian
moment map. Then the quotient $\mu_{\cals}^{-1}(0)/G$ is a union of the
smooth, 3-Sasakian manifolds $(\cals_{(H)}\cap \mu_{\cals}^{-1}(0))/G$,
where $(H)$ runs over the conjugacy classes of stabilizers of points in
$\cals$. 
\tenrm

Quite often $\cals_{(H)}$ does not meet the zero locus of
the moment map. Then the stratum 
$(\cals_{(H)}\cap \mu_{\cals}^{-1}(0))/G$ is empty.

\noindent{\sc Example} \bq.3.2: We start with the
3-Sasakian sphere $S^{4n+3}$ in the notation of the previous section.
But now we consider a different circle action $U(1)$, namely
$$\varphi_t^{p,q;m}(\bfu)=(\tau^pu_1,\ldots,\tau^p u_m,\tau^q u_{m+1},\ldots,
\tau^qu_{n+1}),\qquad \tau=e^{2\pi it},\qquad \bfu\in S^{4n+3},$$
where $0\leq m\leq n+1$ and $p,q\in\bbz$.
Let $\cals(p,q;m)=S^{4n+3}/\!\!/\!\!/U(1)$ be the quotient. 
\item{(i)}
First, let $p,q$ be relatively prime positive integers bigger than 1 and
$2\leq m\leq n-1$. Then, the stratified
manifold $\cals(p,q;m)$ consists of 3 strata. The stratum of the
highest dimension corresponding to
$H=\{{\rm id}\}$ is an open incomplete 3-Sasakian manifold. The two strata
of lower dimension are easily seen to be the homogeneous spaces:
one with $H=\bbz_q$ is the homogeneous 3-Sasakian space of
$SU(n+1-m)$ and the one with $H=\bbz_p$ is the homogeneous 3-Sasakian space of
$SU(m)$. In this case,  $\cals(p,q;m)$ is actually a compact 3-Sasakian
orbifold and the stratification of Theorem \bq.3.1 coincides with the
orbifold stratification. 
\item{(ii)}
Consider $\cals(0,p;m)$, where
$p>1$ and $2\leq m\leq n-1$. There are two strata now: the
stratum of the highest dimension corresponds to
$H=\bbz_p$ and the second stratum is just the sphere $S^{4m-1}$
with $H=U(1)$.
The space $\cals(0,p;m)$ is not an orbifold but, as pointed out in [D-Sw],
it does have a length space structure.
\item{(iii)} Consider $\cals(0,1;n)$.
Here $H$ is either $U(1)$ or trivial but the
set $S^{4n+3}_{{\rm id}}$ does not meet the zero locus of the moment map.
Hence, there is only one stratum and $\cals(0,1;n)=S^{4n-1}$.
\item{(iv)} Finally, consider $\cals(0,1;n-1)$. The stability group 
$H$ is either $U(1)$ or trivial. The stratum corresponding to
$H=U(1)$ is the sphere $S^{4n-5}$. We leave it as an exercise to the reader
to show that $\cals(0,1;n-1)$ is an orbifold and that it can be identified
with $S^{4n-1}/\bbz_2$, where $(w_1,....,w_n)\in S^{4n-1}$, where
$\bbz_2$ acts on the last quaternionic coordinate by multiplication by
$\pm1$.
\bigskip
\centerline {\bf \bex. Toric 3-Sasakian Manifolds}
\medskip
In this section we shall describe the quotient construction of large
families of 3-Sasakian manifolds $\cals(\Omega)$. They all have the
property that $I_0(\cals(\Omega), g(\Omega))\supset T^m$, where 
${\rm dim}(\cals(\Omega))=4m-1$, and following the ideas of 
[Bi-D] we shall call such 3-Sasakian manifolds {\it toric} (See \bex.6.1 for a
precise definition).  We also describe some interesting geometric and
topological properties of such spaces. Up until now all known examples of
3-Sasakian manifolds are either homogeneous or toric or discrete quotients of
them.
\smallskip
\noindent {\bf \bex.1 Toral Reductions of Spheres}
\smallskip
Using the notation of the previous section we start with
the unit $(4n+3)$-dimensional sphere embedded in the quaternionic vector space
$\bbh^{n+1}.$  The subgroup of the full isometry group $O(4n+3)$ that preserves
the quaternionic structure is $Sp(n+1)\!\cdot\!Sp(1)$ acting by
$$\varphi_{\bba,\sigma}(\bfu)=
\bba\bfu\sigma^{-1},\leqno{\bex.1.1}$$
where $\bba\in Sp(n+1)$ is the
quaternionic $(n+1)\times (n+1)$ matrix of the quaternionic
representation of $Sp(n+1)$, 
and $\sigma\in Sp(1)$ is a unit quaternion. As 
the diagonal $\bbz_2$ acts trivially this is indeed an
$Sp(n+1)\!\cdot\!Sp(1)$ action.
The group $Sp(n+1)$ is the subgroup of $Sp(n+1)\!\cdot\!Sp(1)$ which preserves
the 3-Sasakian structure on $S^{4n+3},$ so we have $I_0(S^{4n+3},
g_{can})=Sp(n+1)$.  We shall consider the maximal torus $T^{n+1}\subset
Sp(n+1)$ and its subgroups.  Every quaternionic representation of a $k$-torus
$T^k$ on $\bbh^{n+1}$ can be described by an exact sequence
$0\ra{1.3}T^k\fract{f_\Omega}{\ra{1.3}} T^{n+1}\ra{1.3} T^{n+1-k}\ra{1.3}0.$
The monomorphism $f_\grO$ can be represented by the matrix
$$f_\Omega(\tau_1,\ldots,\tau_k)=\pmatrix
{\displaystyle{\prod_{i=1}^k\tau_i^{a^i_1}}&\dots&0\cr
\vdots&\ddots&\vdots\cr
0&\dots&\displaystyle{\prod_{i=1}^k\tau_i^{a^i_{n+1}}}\cr},\leqno{\bex.1.2}$$
where $(\tau_1,..,\tau_k)\in S^1\times\cdots\times S^1=T^k$ are
the complex coordinates on $T^k,$ and $a^i_j\in \bbz$  are
the coefficients of 
a $k\times (n+1)$ integral {\it weight} matrix
$\Omega=(a^l_\alpha)_{\alpha=1,...,n+1}^{l=1,...,k}\in\calm_{k,n+1}(\bbz).$

Let $\{e_l\}_{l=1}^k$ denote the standard basis for 
$\gt_k^*\simeq\bbr^k.$ Then
the $3$-Sasakian moment map $\mu_\Omega:S^{4n+3}\ra{1.3} \gt_k^*\otimes
\bbr^3$  of the $k$-torus action defined by
$\varphi_{(\tau_1,\ldots,\tau_k)}(\bfu)=f_\Omega(\tau_1,\ldots,\tau_k)\bfu,$
is given by
$\mu_\Omega
=\sum_l\mu^l_\Omega e_l$ with
$$\mu^l_\Omega(\bfu)=\sum_{\gra}\bar{u}_\gra ia^l_\gra u_\gra.
\leqno{\bex.1.3}$$
Let us further denote the triple 
$(T^k, f_\Omega,\varphi_{(\tau_1,\ldots,\tau_k)})$ by $T^k(\Omega)$. 

\noindent{\sc Definition} \bex.1.4: \tensl
$N(\Omega)=\mu_\Omega^{-1}(0)$ and $\cals(\Omega)=
S^{4n+3}/\!\!/\!\!/T^k(\Omega)=N(\Omega)/T^k(\Omega)$.
\tenrm

Let $S^{4n+3}_H$ denote all the points on the sphere where the
stability group $H\subset T^k$ is exactly $H$. Because $T^k$ is
Abelian $S^{4n+3}_H=S^{4n+3}_{(H)}$. Furthermore,
let $K_H=T^k/H$ and denote by $\cals(\Omega;H)=
S^{4n+3}_H\cap N(\Omega)/K_H$.
Following Theorem \bq.3.1 we have

\noindent{\sc Proposition} \bex.1.5: \tensl The quotient
$\cals(\Omega)=\bigcup_H\cals(\Omega;H)$ is
a disjoint union of  3-Sasakian manifolds, where each stratum 
$\cals(\Omega;H)$ is smooth.
\tenrm

We will be interested in the case
when $\cals(\Omega)$ is a compact orbifold (all stability groups
$H$ for which $S^{4n+3}_H\cap N(\Omega)$ are non-empty
are discrete) or a compact smooth manifold (there is only
one stratum).
Necessary and sufficient conditions for this to happen
can be expressed in terms of properties of the matrix $\Omega.$ First observe
that, without loss of generality, we can assume that
the rank of $\grO$ equals $k$. Otherwise, one simply has an action
of a torus of lower dimension and the whole problem reduces to considering
another weight matrix $\Omega$ with fewer rows.

We introduce the following terminology:
Consider the ${n\choose k}$ minor determinants
$$\Delta_{\alpha_1...\alpha_k}={\rm det}\pmatrix{
a_{\alpha_1}^1&\dots&a_{\alpha_k}^1\cr \vdots&&\vdots\cr
a_{\alpha_1}^k&\dots&a_{\alpha_k}^k\cr} \leqno{\bex.1.6}$$ 
obtained by deleting
$n+1-k$ columns of $\Omega.$ 

\noindent{\sc Definition} \bex.1.7: \tensl Let $\Omega\in
{\cal M}_{k,n+1}(\bbz)$ be the weight matrix.
\item{(i)} If 
$\Delta_{\alpha_1\cdots\cdots\alpha_k}\not=0,
\forall\ \ \ 1\leq\alpha_1<\cdots
<\alpha_{k}\leq n+1$, then we say that $\Omega$ is {\it non-degenerate}.

Suppose $\Omega$ is non-degenerate and let $g$ be the $k$th determinantal
divisor, i.e., the $gcd$ of all the $k$ by $k$ minor determinants
$\Delta_{\alpha_1\cdots\cdots\alpha_k}.$ Then $\grO$ is said to be {\it
admissible} if in addition we have
\item{(ii)} 
${\rm gcd}(\Delta_{\alpha_2\cdots\alpha_{k+1}},...,
\Delta_{\alpha_1\cdots\hat\alpha_s\cdots\alpha_{k+1}},...,
\Delta_{\alpha_1\cdots\alpha_{k}})=g$
for all sequences of length (k+1) such that
$1\leq\alpha_1<\cdots<\alpha_s<\cdots< \alpha_{k+1}\leq n+1.$
\tenrm

\smallskip
\noindent{\bf \bex.2 Equivalence Problem and Admissibility}
\smallskip
Before we show how these properties of
the matrix $\Omega$ impact on the geometry of the
quotient $\cals(\Omega)$ we
need to discuss the notion of the equivalence of
$T^k$-actions on $S^{4n+3}$ and
obtain a normal form for admissible weight matrices. We are free to change
bases of the Lie algebra $\gt_k.$ This can be done by the group of unimodular
matrices $GL(k,\bbz).$ Moreover, if we fix a maximal torus $T^{n+1}$ 
of $Sp(n+1),$
its normalizer, the Weyl group 
$\calw(Sp(n+1))\simeq \grS_{n+1}\bowtie (\bbz_2)^{n+1},$
preserves the $3$-Sasakian structure on $S^{4n+3}$ and intertwines the $T^k$
actions.  Thus, there is an induced 
action of $GL(k,\bbz)\times \calw(Sp(n+1))$
on the set of weight matrices $\calm_{k,n+1}(\bbz).$ 
The group $GL(k,\bbz)$
acts on $\calm_{k,n+1}(\bbz)$ by matrix multiplication from the left, and
the Weyl group $\calw(Sp(n+1))$ 
acts by permutation and overall sign changes of
the columns.  Actually we want a slightly stronger notion of equivalence than
that described above. If the $i$th row of $\grO$ has a gcd $d_i$ greater than
one, then by reparameterizing the 
one-parameter subgroup $\grt'_i=\grt_i^{d_i}$
we obtain $\grt_i^{a^j_{\gra}}=(\grt'_i)^{b^i_{\gra}}$, where
$\gcd\{b^i_{\gra}\}_{\gra}=1.$ So the 
action obtained by using the matrix whose
$i$th row is divided by its gcd $d_i$ 
is the same as the original action. The integers $d_i$ all
divide the $k$th determinantal divisor $g.$ 
We say that a non-degenerate matrix $\grO$
is in {\it reduced} form (or
simply {\it reduced}) if $g=1.$ The following easy lemma says that among
non-degenerate matrices it is sufficient
to consider matrices in a reduced form.

\noindent{\sc Lemma} \bex.2.1: \tensl Every 
non-degenerate weight matrix $\grO$ is 
equivalent to a matrix in a reduced form.
\tenrm

\noindent Henceforth, we shall only consider matrices in a reduced form. 

\noindent{\sc Definition} \bex.2.2: \tensl  
Let $\cala_{k,n+1}(\bbz)\subset \calm_{k,n+1}(\bbz)$
denote the subset of reduced 
admissible matrices. This subset is invariant under the
action of $GL(k,\bbz)\times \calw(Sp(n+1)),$ so the set
$\cala_{k,n+1}(\bbz)/GL(k,\bbz)\times \calw(Sp(n+1))$ of equivalence classes
$[\Omega]$ is well defined. We let $\cale_{k,n+1}(\bbz)\subset
\cala_{k,n+1}(\bbz)$ denote a fundamental domain for the action. \tenrm

Our interest in $\cala_{k,n+1}(\bbz)$ is the following:

\noindent{\sc Theorem} \bex.2.3:  \tensl Let $\cals(\grO)$
be the quotient space of definition \bex.1.6. Then
\item{(i)} if $\grO$ is non-degenerate, $\cals(\grO)$ is an orbifold.
\item{(ii)} If $\grO$ is degenerate, 
then either $\cals(\grO)$ is a singular stratified space which is 
{\it not} an orbifold or        
it is an orbifold obtained by reduction of a lower dimensional sphere
$S^{4n-4r-1}$ by a torus $T^{k-r}(\grO')$ 
or a finite quotient of such, where        
$1\leq r\leq k$ and $\grO'$ is non-degenerate.  
(When $r=k$ the quotient is the      
sphere $S^{4n-4k-1}).$
\item{(iii)} Assuming that $\grO$ is 
non-degenerate $\cals(\grO)$ is a smooth        
manifold if and only if $\grO$ is admissible.
\tenrm

One can easily see that the non-degeneracy of $\grO$ is not
necessary for the quotient space $\cals(\Omega)$ to be smooth or
a compact orbifold (see Example \bq.3.3(iv)). However,
Theorem \bex.2.5(ii) shows that then we can reformulate the whole problem
in terms of another quotient and a new non-degenerate weight matrix
$\Omega^\prime$ and can be found in [B-G-M 7]. 
Theorem \bex.2.5(iii) shows then the importance
of admissible matrices in the construction and it easily follows
from the fact that non-degeneracy implies that at most $n-k$ quaternionic
coordinates $u_j$ can simultaneously vanish on $N(\Omega)$ [B-G-M-R 1]. 

\noindent{\sc Remark} \bex.2.4: Our discussion shows clearly
that, if $\Omega,\Omega'\in\cala_{k,n+1}(\bbz)$ such that
$[\Omega]=[\Omega']$ then the quotients
$\cals(\Omega)\simeq\cals(\Omega')$ are equivalent as 3-Sasakian
manifolds. We believe that the converse of this is also true, though
we will establish it later only in certain cases.
\smallskip 
\noindent
{\bf \bex.3.3 Combinatorics and Admissibility}
\smallskip
In general Theorem \bex.2.3 is not yet an existence theorem, since
$\cala_{k,n+1}(\bbz)$ could be empty. Indeed, for many pairs $(k,n)$
this is the case and we shall demonstrate this next.

Let $\grO\in \cala_{k,n+1}(\bbz).$ Since $\grO$ is reduced there is a
$k$ by
$k$ minor determinant that is odd. 
By permuting columns if necessary this minor
can be taken to be the first $k$ columns. Now consider the mod 2 reduction
$\calm_{k,n+1}(\bbz)\ra{1.3} \calm_{k,n+1}(\bbz_2).$ We have the
following commutative diagram
$$\matrix{GL(k,\bbz)\times \calm_{k,n+1}(\bbz) &\ra{1.6} &
\calm_{k,n+1}(\bbz)\cr
\decdnar{} && \decdnar{}\cr
GL(k,\bbz_2)\times \calm_{k,n+1}(\bbz_2) &\ra{1.6} &
\calm_{k,n+1}(\bbz_2).}\leqno{\bex.3.1}$$
Let $\tgrO\in \cala_{k,n+1}(\bbz_2)$ 
denote the mod 2 reduction of $\grO\in \cala_{k,n+1}(\bbz).$ 
Since the first $k$ by $k$ minor
determinant of $\grO$ is odd, the mod 2 
reduction of this minor in $\tgrO$ is        
invertible.  Thus, we can use the 
$GL(k,\bbz_2)$ action to put $\tgrO$ in the
form
$$\tgrO =\pmatrix{1&0&\dots&0&a_{k+1}^1&\dots &a_{n+1}^1\cr
0&1&\dots&0&a_{k+1}^2&\dots &a_{n+1}^2\cr
\vdots&\vdots&\ddots&\vdots&\vdots&\cdots&\vdots\cr
0&0&\cdots&1&a_{k+1}^k&\dots &a_{n+1}^k\cr}\leqno{\bex.3.2}$$
with $a^i_j\in \bbz_2.$

\noindent{\sc Lemma} \bex.3.3: \tensl The set $\cala_{k,n+1}(\bbz)$ is
empty for $n>k+1$ and $k>4.$ \tenrm

\noindent{\sc Proof}: The second admissibility condition is equivalent to the
condition that every $k$ by $k+1$ submatrix of $\tgrO$ has rank $k.$ By
considering $k-1$ of the first $k$ columns and $2$ of last $n+1-k$ columns,
this condition implies
$(a^j_l,a^j_m)\neq (0,0)$
for all $j=1,\cdots,k,$ and $k+1\leq l<m\leq n+1.$ Similarly, by considering
$k-2$ of the first $k$ columns and $3$ of last $n+1-k$ columns \bex.3.2
implies
$$\pmatrix{a^i_l&a^i_m&a^i_r\cr a^j_l&a^j_m&a^j_r}\neq\pmatrix{1&1&1\cr
1&1&1}, \qquad \pmatrix{a^i_l&a^i_m&a^i_r\cr
a^j_l&a^j_m&a^j_r}\neq\pmatrix{0&1&1\cr 0&1&1}, \leqno{\bex.3.4}$$
where the last inequality is understood to be up to column permutation.
Hence, it follows that, up to column and row permutations, that any four
triples of the last $n-k$ columns of an admissible $\tgrO$ must have the form
$$\pmatrix{1&1&1\cr 0&1&1\cr 1&0&1\cr 1&1&0}. \leqno{\bex.3.5}$$
So we see that we
cannot add another row without violating the above conditions.  
It follows that
$k\leq 4.$ \hfill\za
 
Similar analysis shows that
 
\noindent{\sc Lemma} \bex.3.6: \tensl 
The set $\cala_{k,n+1}(\bbz)$ is empty if
$k>1$ and $n-k\geq 4.$ \tenrm

\noindent{\sc Remark} \bex.3.7: In view of the above lemmas and the fact
that in the remainder of this section we will be  
interested only in the smooth and compact quotients we
are left with the following possibilities:
\item{(i)} Trivial case of $n=k$. 
Then there are many admissible matrices
$\Omega$ but ${\rm dim}(\cals(\Omega))=3$ and it follows
that $\cals(\Omega)=S^3/\bbz_p$, where $p=p(\Omega)$ depends on
$\Omega$. This case is of little interest.
\item{(ii)} Bi-quotient 
geometry with $k=1$ and $n>1$ arbitrary. Here $\Omega$ is
just a row vector $\bfp$. The admissibility condition means
that the entries are non-zero and pairwise relatively prime.
The quotient $\cals(\bfp)$ turns out to be a bi-quotient of the
unitary group $U(n+1)$ and we shall discuss its geometry and
topology in the next subsection.
\item{(iii)} The most interesting,
7-dimensional case of $k=n-1$. Here one easily sees that there are many
admissible matrices and we analyze the geometry and topology of the
quotients in a separate subsection.
\item{(iv)} ``Special" quotients: $(k,n)=\{
(2,4), (2,5), (3,5), (3,6), (4,6), (4,7)\}$. These quotients
are 11- or 15-dimensional and we give examples of
admissible weight matrices in each case. We shall show also that they provide
counterexamples to certain Betti number relations that are satisfied in the
regular case [G-Sal].
\smallskip
\noindent{\bf \bex.4 3-Sasakian Structures on Bi-Quotients}
\smallskip
When $k=1$ we have $\Omega=\bfp=(p_1,...,p_{n+1})$ and we shall write 
$\cals(\Omega)=\cals(\bfp)$, $N(\Omega)=N(\bfp)$, 
$f_\Omega=f_\bfp$, and $\tau_1=\tau$. 
The quotients $\cals(\bfp)$ are generalizations 
of the homogeneous examples discussed in Section \bq.2. We get
$$\cala_{1,n+1}(\bbz)=\{\bfp\in(\bbz)^{n+1}\ \ | \ \ 
p_i\not=0 \ \forall i=1,...,n+1
\ {\rm and}\ 
{\rm gcd}(p_i,p_j)=1\qquad \forall i\not=j\},$$
$$\cale_{1,n+1}(\bbz)=\{\bfp\in\bbz^{n+1}\ \ | \ \
0<p_1\leq \cdots\leq p_{n+1} \ {\rm and} \ \ 
{\rm gcd}(p_i,p_j)=1\qquad \forall i\not=j\}.$$

\noindent Note that $\cale_{1,n+1}(\bbz)$ can be identified with a certain
integral lattice in the positive Weyl chamber in $\gt^*_{n+1}.$

First, by studying the geometry of the
foliations in the diagram $\diamondsuit(\cals(\bfp))$ [B-G-M 6] one can solve
the equivalence problem in this case. We get [B-G-M 3]:
 
\noindent{\sc Proposition \bex.4.1}: \tensl Let $n\geq 2$ and
$\bfp,\bfq\in\cala_{1,n+1}(\bbz)$ so the quotients
$\cals(\bfp)$ and $\cals(\bfq)$ are smooth manifolds. Then
$\cals(\bfp)\simeq\cals(\bfq)$ are 3-Sasakian equivalent if and only if
$[\bfp]=[\bfq].$
\tenrm

It is easy to see that for $\bfp\in\cala_{1,n+1}(\bbz)$
the zero locus
of the moment map $N(\bfp)$ is always diffeomorphic to the 
Stiefel manifold
$V^\bbc_{2,n+1}$ of complex 2-frames in $\bbc^{n+1}$. Hence, the quotient
$\cals(\bfp)=V^\bbc_{2,n+1}/S^1$. We first
observe that one can
identify $V^\bbc_{2,n+1}$ with the homogeneous space $U(n+1)/U(n-1)$. Using
this identification we have 

\noindent{\sc Proposition} \bex.4.2: \tensl  
For each $\bfp\in\cale_{1,n+1}(\bbz)$, there is an equivalence
$\cals(\bfp)\simeq U(1)_\bfp\backslash U(n+1)/U(n-1)$ as smooth
$U(1)_\bfp\times U(n-1)$-spaces, where the action of $U(1)_\bfp\times U(n-1)
~\subset~ U(n+1)_L\times U(n+1)_R$ is given by the formula
$$\varphi^\bfp_{\tau,\bbb}(\bbw)=
f_\bfp(\tau)
\bbw\pmatrix{\bbi_2&\bbo\cr \bbo&\bbb\cr}.\leqno{\bex.4.3}$$
Here $\bbw\in U(n+1)$ and $(\tau,\bbb)\in S^1\times U(n-1).$
\tenrm

Note that the identification 
$\cals(\bfp)\simeq U(1)_\bfp\backslash U(n+1)/U(n-1)$
is only true after assuming that
all the weights are positive, as the right-hand side is not invariant under
such sign changes. Proposition \bex.4.2 shows that, in a way, the
quotients $\cals(\bfp)$ can be though of as 
``bi-quotient deformation" of the homogeneous
model $\cals({\bf 1})$.
Now let 
$\iota_\bfp: N(\bfp)\hookrightarrow S^{4n+3}$ be the inclusion
and $\pi_\bfp:N(\bfp)
\rightarrow \cals(\bfp)$ be the Riemannian submersion of the moment map. 
Then the metric $g(\bfp)$ is the unique metric on $\cals(\bfp)$ that satisfies
$\iota^*_\bfp g_{can}=\pi^*_\bfp g(\bfp).$
Using the geometry of the inclusion $\iota_\bfp$ one can show the
following [B-G-M 3,6]

\noindent{\sc Theorem \bex.4.4}: \tensl Let $I_0(\cals(\bfp),g(\bfp))$ 
be the group of
$3$-Sasakian isometries of $\bigl(\cals(\bfp),g(\bfp)\bigr)$
and let $k$ be
the number of $1$'s in $\bfp$. Then the connected component of $I_0$
is $S(U(k)\times U(1)^{n+1-k})$, where we define $U(0)=\{e\}.$ Thus,
the connected component of the isometry group is the product
$S(U(k)\times U(1)^{n+1-k})\times SO(3)$
if the sums $p_i+p_j$ are even for all $1\leq i,j\leq n+1,$ and
$S(U(k)\times U(1)^{n+1-k})\times Sp(1)$
otherwise.  \tenrm
 
In the case that $\bfp$ has no repeated $1$'s, 
the cohomogeneity can easily be
determined, viz. [B-G-M 3]
 
\noindent{\sc Corollary \bex.4.5} : \tensl If the number of $1$'s in
$\bfp$ is $0$ or $1$ then the dimension of the principal orbit
in $\cals(\bfp)$ equals $n+3$ and the cohomogeneity of
$g(\bfp)$ is $3n-4.$
In particular, the 7-dimensional $\cals(\bfp)$ the family 
$(\cals(\bfp), g(\bfp))$ contains metrics of cohomogeneity 0,1, and 2.
\tenrm

Combining Proposition \bex.4.2 with techniques developed by Eschenburg [Esch
1-2] in the study of certain $7$-dimensional bi-quotients of $SU(3)$ one can
compute the integral cohomology ring of $\cals(\bfp)$ [B-G-M 2]:
 
\noindent{\sc Theorem \bex.4.6}: \tensl Let $n\geq 2$ 
$\bfp\in\cale_{1,n+1}(\bbz)$. Then, as rings,
$$H^*\bigl(\cals(\bfp),\bbz\bigr) ~\cong~
\Biggl({\bbz[b_2]\over [b_2^{n+1}=0]} \otimes
E[f_{2n+1}]\Biggr) / \calr(\bfp).$$
Here the subscripts on $b_2$ and $f_{2n+1}$ denote the cohomological
dimension of each generator. Furthermore,
the relations $\calr(\bfp)$ are generated by
$\grs_{n}(\bfp)b_2^{n}=0$ and $f_{2n+1}b_2^{n}=0$, where
$\grs_{n}(\bfp)=\sum_{j=1}^{n+1} p_1\cdots \hat{p_j}\cdots p_{n+1}$
is the $n^{th}$ elementary symmetric polynomial in the entries of
$\bfp.$
\tenrm
 
Notice that Theorem \bex.4.6 shows that
$H^{2n}(\cals(\bfp);\bbz) ~=~ \bbz_{\grs_{n}(\bfp)}$
and hence has the following corollary.
 
\noindent{\sc Corollary \bex.4.7}:
\tensl  The quotients $(\cals(\bfp),g(\bfp))$
give infinitely many homotopy inequivalent simply-connected compact
inhomogeneous 3-Sasakian manifolds in dimension
$4n-1$ for every $n\geq 2.$ In fact, there are infinite families that are not
homotopy equivalent to any homogeneous space. \tenrm

\noindent{\sc Remark \bex.4.8}: Corollary \bex.4.7 shows that the finiteness
results for regular 3-Sasakian manifolds discussed in Section \btop.4
fail for non-regular 3-Sasakian manifolds. Moreover, combining our results with
a well-known finiteness theorem of Anderson [An] we have

\noindent{\sc Corollary \bex.4.9}: \tensl For each $n\geq 2$ there are
infinitely many $3$-Sasakian $4n-1$-manifolds with arbitrarily small
injectivity radii. \tenrm

When $n=2$ the spaces $\cals(\bfp)=\cals(p_1,p_2,p_3)$ give a subfamily of
the more general bi-quotients of $U(3)$ studied by Eschenburg [Esch 1-2]. This
large collection of spaces contains not only our 3-Sasakian subfamily, but also
the well-known Aloff-Wallach spaces [Al-Wa] which are of much interest since
they admit Einstein metrics of positive sectional curvature [Wan 1].  These two
subfamilies intersect at the homogeneous 3-Sasakian manifold $\cals(1,1,1),$
that is $\cals(1,1,1)$ is diffeomorphic to the Aloff-Wallach space $N_{1,1}$
mentioned in \bg2.2.10.  Then following Eschenburg [Esch 1] we can make use of
the Cheeger $\grr^*$-topology on the space of Riemannian manifolds to show the
existence of an infinite number of 3-Sasakian manifolds that admit metrics of
positive sectional curvature. More precisely [B-G-M 2],

\noindent{\sc Corollary \bex.4.10}: \tensl For all sufficiently large odd
positive integers $c$, the 3-Sasakian manifolds $\cals(c,c+1,c+2)$ admits a
metric of positive sectional curvature. \tenrm

In the next subsection we give a result in the opposite direction.  We shall
exhibit an infinite family of 3-Sasakian manifolds that cannot admit any metric
whose sectional curvature is bounded from below by a fixed arbitrary negative
number.

We end this subsection with a discussion of topological and differential
invariants of the 7-manifolds $\cals(p_1,p_2,p_3).$ Homotopy invariants for
Eschenburg space have been worked out independently by Kruggel [Kru 1,2] and
Milgram [Mil]. The homeomorphism and diffeomorphism classification was first
done for a certain subclass of Eschenburg spaces which include some of the
$\cals(p_1,p_2,p_3)$ by Astey, Micha, and Pastor [A-M-P]. Later Kruggel [Kru
3] obtained the diffeomorphism and homeomorphism classification of all
Eschenburg's bi-quotients by computing the Kreck-Stolz invariants [K-S 1].
This, in principle, gives a complete differential topological description of
the 7-dimensional family $\cals(p_1,p_2,p_3).$ Using this classification
together with the help of a computer program, one would expect to find examples
$\cals(\bfp)$ and $\cals(\bfq)$ with $[\bfp]\not=[\bfq]$ such that the
quotients  are homeomorphic, but not diffeomorphic, as well as examples that
are diffeomorphic, but not 3-Sasakian equivalent.  The later would show that a
smooth 7-manifold can admit more than one inequivalent 3-Sasakian structure.
In the case of the former, such exotic structures are known to exist for the
family of Aloff-Wallach spaces [K-S 2], but the examples involve large integers
and were obtained with help of a computer program.  The analysis of the above
mentioned invariants for our family $\cals(p_1,p_2,p_3)$ proves even harder due
to the positivity of the weights. For a fixed $\sigma_2=p_1p_2+p_2p_3+p_3p_1$
there are only finitely many positive integer solution
$\bfp\in\cale_{1,3}(\bbz)$. 
\smallskip
\noindent{\bf \bex.5 7-dimensional Toric 3-Sasakian Manifolds} 
\smallskip
In this case we can easily see that there are many examples of
admissible weight matrices $\Omega$. The simplest family of examples is given
by matrices of the form
$$\grO =\pmatrix{1&0&\dots&0&a_1&b_1\cr
0&1&\dots&0&a_2&b_2\cr
\vdots&\vdots&\ddots&\vdots&\vdots&\vdots\cr
0&0&\cdots&1&a_{k}&b_k\cr},\leqno{\bex.5.1}$$
for which we have 

\noindent{\sc Proposition \bex.5.2}: \tensl 
Let $k$ be a positive integer, and let $\grO\in\calm_{k,k+2}(\bbz)$ be as in
\bex.5.1. Then $\grO\in\cala_{k,k+2}(\bbz)$ if and only if
$(\bfa,\bfb)\in (\bbz^*)^k \oplus (\bbz^*)^k$ and
the components
$(a^i,b^i)$ are pairs of
relatively prime integers for $i=1,\cdots,k$ such that if
for some pair $i,j$ $a^i=\pm a^j$ or $b^i=\pm b^j$ then we must have $b^i\neq
\pm b^j$ or $a^i\neq \pm a^j,$ respectively. 
\tenrm

Proposition \bex.5.2 shows that 
$\cala_{k,k+2}(\bbz)$ is never empty and we have many examples of compact smooth
7-dimensional quotients $\cals(\Omega)$ for arbitrary $k>1$. Some of these
examples were first mentioned in [B-G-M 1] and the idea of the quotient
is based on the result of [G-Ni]. As we
shall not present here the complete solution to the equivalence problem, we
shall further assume that $\Omega\in\cala_{k,k+2}(\bbz)$ is arbitrary and shall
determine some important topological properties of the quotients
$\cals(\Omega)$.  More explicitly,

\noindent{\sc Theorem} \bex.5.3: \tensl Let $\grO\in\cala_{k,k+2}(\bbz)$ 
Then $\pi_1(\cals(\grO))=0$ and $\pi_2(\cals(\grO))=\bbz^k$.
\tenrm

Because of Corollary \btop.2.8 and Poincare duality, Theorem \bex.5.3
completely determines the rational homology of the 3-Sasakian 7-manifolds
$\cals(\grO).$ The proof given below is a compilation with some simplifications
of the proofs in [B-G-M-R 1, B-G-M 8], while some of the more tedious details
are left to those references.

\noindent{\sc Proof}:  First note that the groups $T^{k+2}\times
Sp(1)$ and $T^{2}\times Sp(1)$ act as isometry groups on $N(\grO)$ and
$\cals(\grO),$ respectively. Let us define the following quotient spaces:
$$Q(\grO)= N(\grO)/T^{k+2}\times Sp(1)\qquad B(\grO)= N(\grO)/Sp(1).$$ 
We have the following commutative diagram
$$\matrix{N(\grO)&\ra{2.8}& B(\grO)\cr
          \decdnar{} &&\decdnar{}\cr
          \cals(\grO)&\ra{2.8}& Q(\grO).\cr}\leqno{\bex.5.4}$$
The top horizontal arrow and the left vertical arrow are principal bundles with
fibers $Sp(1)$ and $T^k,$ respectively. The remaining arrows are not
fibrations. The right vertical arrow has generic fibers $T^{k+2},$ while the
lower horizontal arrow has generic fibers $T^{2}\!\cdot\!Sp(1)$ homeomorphic
either to $T^{2}\times \bbr\bbp^3$ or $T^{2}\times S^3$ depending on $\grO.$
The dimension of the orbit space $Q(\grO)$ is $2.$ 

The difficulty is in proving that both $N(\grO)$ and $B(\grO)$ are 2-connected.
Once this is accomplished the result follows by applying the long exact
homotopy sequence to the left vertical arrow in diagram \bex.5.4.

\noindent{\sc Lemma} \bex.5.5: \tensl Both $N(\grO)$ and $B(\grO)$ are
$2$-connected. \tenrm

\noindent{\sc Proof}: The idea is to construct a stratification giving a Leray
spectral sequence whose differentials can be analyzed. Let us define the
following subsets of $N(\grO):$ (Recall that, in this case, at most one
quaternionic coordinate can vanish.)
$$\eqalign{N_0(\grO)&=\{\bfu\in N(\grO)|~u_\gra =0~\hbox{for some}~ \gra
=1,\cdots,k+2\}, \cr
N_1(\grO)&=\{\bfu\in N(\grO)|~\hbox{for all}~ \gra
=1,\cdots,k+2,~ u_{\gra}\neq 0~\hbox{and} \cr
&\phantom{=\{} \hbox{there is a pair $(u_\gra,u_\grb)$ that
lies on the same} \cr
&\phantom{=\{} \hbox{complex line in $\bbh$} \},\cr
N_2(\grO)&=\{\bfu\in N(\grO)|~\hbox{for all}~ \gra
=1,\cdots,k+2,~ u_{\gra}\neq 0~\hbox{and} \cr
&\phantom{=\{} \hbox{ no pair $(u_\gra,u_\grb)$ lies on the
same complex line } \cr
&\phantom{=\{} \hbox{ in $\bbh$} \}.} \leqno{\bex.5.6}$$ 
Clearly, $N(\grO)=N_0(\grO)\sqcup N_1(\grO)\sqcup N_2(\grO)$ and $N_2(\grO)$ is
a dense open submanifold of $N(\grO).$ This stratification is compatible with
the diagram \bex.5.4 and induces corresponding stratifications
$$B(\grO)=B_0(\grO)\sqcup B_1(\grO)\sqcup B_2(\grO)\quad 
Q(\grO)=Q_0(\grO)\sqcup Q_1(\grO)\sqcup Q_2(\grO).\leqno{\bex.5.7}$$
The $B_i(\grO)$ fiber over the $Q_i(\grO)$ whose fibers are tori $T^{k+i}.$
The strata are labeled by the dimension of the cells in the resulting CW
decomposition of $Q(\grO).$ Using known results about cohomogeneity 2 actions
[Bre] one can easily prove:

\noindent{\sc Lemma} \bex.5.8: \tensl 
\item{(i)} The orbit space $Q(\grO)$ is homeomorphic to the closed
disc $\bar{D^2},$ and the subset of singular orbits  $Q_1(\grO)\sqcup
Q_0(\grO)$ is homeomorphic to the boundary $\partial\bar{D^2}\simeq S^1.$
\item{(ii)} $Q_2(\grO)$ is homeomorphic to the open disc $D^2.$
\item{(iii)} $Q_1(\grO)$ is homeomorphic to the disjoint union of $k+2$ copies
of the open unit interval.
\item{(iv)} $Q_0(\grO)$ is a set of $k+2$ points. \tenrm

Next one can easily show that $\pi_1(B(\grO))$ is Abelian; hence,
$\pi_1(B(\grO))=H_1(B(\grO)).$ Now we claim that $H_1(B(\grO))=H_2(B(\grO))=0,$
and since the fibers of the top horizontal arrow of \bex.5.4 are $S^3$'s this
together with Hurewicz will prove Lemma \bex.5.5. To prove this claim we define
$Y_0 = Q_0(\grO), ~ Y_1 = Q_0(\grO)\cup Q_1(\grO),$ and $Y_2=Q(\grO).$ Then, we
filter $B(\grO)$ by $X_i= \pi^{-1}(Y_i)$ to obtain the increasing filtration
$$  X_0 = B_0(\grO), ~~ X_1 = B_0(\grO)\cup B_1(\grO), 
\hbox{~~and~~} X_2=B(\grO).$$

The Leray spectral sequence associated to this filtration has $E^1$
term given by
$$  E^1_{s,t} ~\cong~  H_{s+t}(X_t,X_{t-1};\bbz) $$
with differential
$ d_1:  H_{s+t}(X_t,X_{t-1};\bbz) ~\ra{1.5}~ 
H_{s+t-1}(X_{t-1},X_{t-2};\bbz),$
where we use the convention that $X_{-1}=\emptyset.$

To compute these $E^1$ terms notice that all the pairs $(X_t,X_{t-1})$
are relative manifolds so that one can apply the Alexander-Poincar\'e
duality theorem.  Hence, by \bex.5.7
$$\eqalign{  H_s(X_0;\bbz) &~\cong~ H_s(\sqcup_{k+2}T^k;\bbz); \cr
H_s(X_1,X_0;\bbz) &~\cong~ H^{k+2-s}(\sqcup_{k+2}T^{k+1};\bbz); \cr 
H_s(X_2,X_1;\bbz) &~\cong~ H^{k+4-s}(T^{k+2};\bbz), \cr}$$ 
where $\sqcup_jT^l$ means the disjoint union of $j$ copies of $T^l.$
Hence, the $E^1_{s,t}$ term of the spectral sequence is described by the
diagram
$$
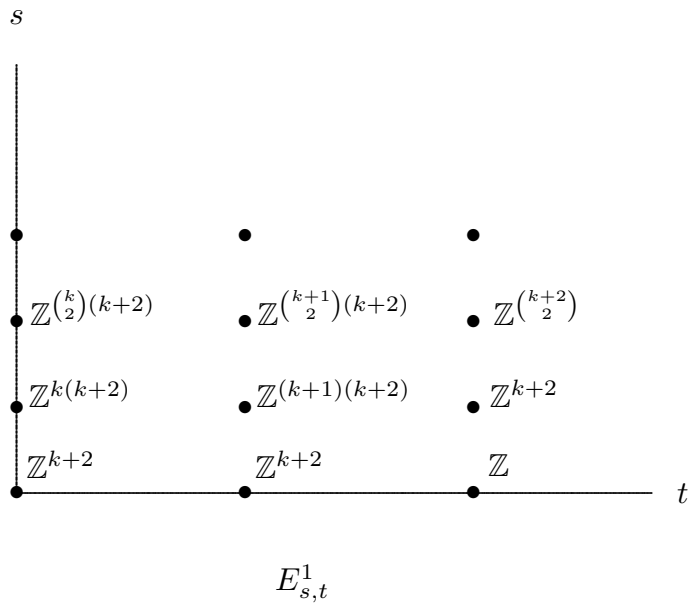
\beginpicture
\setcoordinatesystem units <2pt,1.5pt>
\setlinear
\put {\hbox to .2pt{\hfill}} [t] at 20 65    

\plot 0 90  0 0  100 0 /

\multiput {$\bullet$} at 0 0  0 18  0 36  0 54  /
\multiput {$\bullet$} at 36 0  36 18 36 36  36 54 /
\multiput {$\bullet$} at 72 0  72 18 72 36 72 54 /
                                                            
\multiput {$\bbz^{k+2}$} at 7 6  43 6  80 21 /
\multiput {$\bbz^{(k+1)(k+2)}$} at 50 21 / 
\multiput {$\bbz$} at  76 6  / 
\multiput {$\bbz^{k(k+2)}$} at  10 21 /
\multiput {$\bbz^{{k+1\choose 2}(k+2)}$} at  50 39  /      
\multiput {$\bbz^{k+2\choose2}$} at  82 39   /      
\multiput {$\bbz^{{k\choose 2}(k+2)}$} at  12 39   /      

\put {$t$} at 105 0 
\put {$s$} at  0 100
\put {\hbox{$E^1_{s,t}$}} [t] at  45 -15 
\endpicture
$$

The computation of the differentials is fairly tedious and we refer the reader
to [B-G-M 8] for details. Suffice it to say here that after making certain
choices Lemma \bex.5.8 can be used to represent $Q(\grO)$ topologically as a
polygon
$$
\beginpicture
\setcoordinatesystem units <1pt, 1pt>
\setlinear
\plot -35 54  0 72  62.35 36  62.35 -36  0 -72  -62.35 -36  -62.35 -8 /
\multiput {$\bullet$} at   0 72  62.35 36  62.35 -36  0 -72  -62.35
-36 /
\put {$v_k$} [b] at 0 75
\put {$v_1$} [t] at 0 -75
\put {$.$} [r] at -36 53.5
\put {$.$} [r] at -37 53
\put {$v_{k+1}$} [l] at 66 -36
\put {$v_2$} [r] at -66 -36
\put {$v_{k+2}$} [l] at 66 36
\put {$\vdots$} [rb] at -61 0
\put {$e_{k+1}$} [lt] at 35 -54
\put {$\grs_2$} [r] at 0 0
\put {$e_{k+2}$} [l] at 66, 0
\put {$e_1$} [rt] at -35 -54
\put {$e_k$} [lb] at 35 54
\put {\hbox{Diagram \bex.5.9}} [t] at 0 -95 \endpicture $$ 

The $d_1$ differential can then be computed and the result is that the
$E^2_{s,t}$ term has zeros for $(s,t)=(1,0),(2,0),(1,1),(1,2),(0,1),(0,2).$
Then $E^2_{s,t}=E^\infty_{s,t}$ which converges to $H_{s+t}(B(\grO),\bbz),$ so
this proves Lemma \bex.5.5 and hence, Theorem \bex.5.3. \hfill\za

By further analysis of the differentials it should be possible to determine the
torsion in $H_3(\cals(\grO),\bbz).$ This should be given in terms of symmetric
functions of the invariants $|\grD_{\gra_1,\cdots,\gra_k}|.$

\noindent{\sc Remark \bex.5.10}: It was pointed out to the authors by Karsten
Grove that if one takes the metric geometry into account, the internal angles
in Diagram \bex.5.9 are all less than $90$ degrees. This indicates the presence
of hyperbolic geometry.

Now, using Theorem \bex.5.3 with Propositions \btop.3.2, \btop.3.3 gives:

\noindent{\sc Proposition} \bex.5.11: \tensl Let
$\grO\in\cala_{k,k+2}(\bbz)$ so that $\cals(\Omega)$ is a smooth manifold.
Let $\calz(\Omega)$ and $\calo(\Omega)$ be the associated twistor space and
quaternionic K\"ahler orbifold, respectively.  Then we have
$$b_2(\cals(\Omega))=b_2(\calo(\Omega))=b_2(\calz(\Omega))-1=k.$$
\tenrm

This shows that inequality $b_2\leq1$ in Proposition \btop.4.2 does not hold
for non-regular 3-Sasakian manifolds. Finally we give several interesting
corollaries of our work.

\noindent{\sc Corollary} \bex.5.12: \tensl There exists a simply-connected
3-Sasakian 7-manifold for every rational homology type allowed by Corollary
\btop.2.8. \tenrm

Our next corollary follows from the results of this section and remarkable
theorem of Gromov [Gro]:

\noindent{\sc Corollary \bex.5.13}: \tensl For any non-positive real number
$\grk$ there are infinitely many 3-Sasakian 7-manifolds which do not admit
metrics whose sectional curvatures are all greater than or equal to $\grk.$
\tenrm

For such an infinite family of 3-Sasakian 7-manifolds, the appearance of
negative curvature is foretold by Remark \bex.5.10. Corollary \bex.5.13 can be
contrasted with Corollary \bex.4.7.

\noindent{\sc Corollary \bex.5.14}| \tensl There exist $7$-manifolds with
arbitrary second Betti number having metrics of weak holonomy $G_2.$ \tenrm

Of course, Corollary \bex.5.13 also applies to these weak holonomy $G_2$
metrics.

\noindent{\sc Corollary \bex.5.15}: \tensl There exist $\bbq$-factorial
contact Fano $3$-folds $X$ with $b_2(X)=l$ for any positive integer $l.$ \tenrm

This corollary should be contrasted to the smooth case, where Mori and Mukai
have proven that $b_2\leq 10$ [Mo-Mu]:

\noindent{\sc Corollary \bex.5.16}: \tensl If the second Betti number
$b_2(\cals(\Omega))=k>3,$ the $3$-Sasakian manifolds $\cals(\Omega)$
are not homotopy equivalent to any homogeneous space. \tenrm

This corollary can be compared to Corollary \bex.4.7. Finally we have

\noindent{\sc Corollary \bex.5.17}: \tensl There exist compact,
$T^2$-symmetric, self dual Einstein orbifolds of positive scalar curvature with
arbitrary second Betti number.  \tenrm

Again this should be contrasted to the smooth case where we must have $b_2\leq
1.$ 
\smallskip
\noindent{\bf \bex.6 Higher Dimensional Toric 3-Sasakian Manifolds}
\smallskip
We begin with the definition of a toric 3-Sasakian manifolds which is
motivated by the hyperk\"ahler case [Bi-D].

\noindent{\sc Definition} \bex.6.1: \tensl A 3-Sasakian manifold (orbifold)
of dimension $4m-1$ is said to be a {\it toric 3-Sasakian} manifold (orbifold)
if it admits an effective action of a $m$-torus $T^m$ that preserves the
3-Sasakian structure. \tenrm

The importance of toric 3-Sasakian manifolds is underlined by the following
recent Delzant-type theorem of Bielawski:

\noindent{\sc Theorem} \bex.6.2 [Bi 3]: \tensl Let $\cals$ be a toric
3-Sasakian manifold of dimension $4n-1.$ Then $\cals$ is isomorphic as a
3-Sasakian $T^n$-manifold to a 3-Sasakian quotient of a sphere by a torus, that
is to a $\cals(\grO)$ for some $\grO.$ \tenrm

This theorem includes the degenerate case when the quotient is a sphere or a
discrete quotient of such. The Betti numbers of a 3-Sasakian orbifold obtained
by a toral quotient of a sphere were computed by Bielawski [Bi 2] using
different techniques than the ones employed in Section \bex.5:

\noindent{\sc Theorem} \bex.6.3 [Bi 2]: \tensl 
Let $\Omega\in\calm_{k,n+1}(\bbz)$ be non-degenerate so that
$\cals(\Omega)$ is a compact 3-Sasakian orbifold of dimension
$4(n-k)+3$. Then we have 
$$b_{2i}=\pmatrix{k+i-1\cr k\cr},\qquad i< n+1-k.\leqno{\bex.6.7}$$
Furthermore, the Betti number constraints of Proposition \btop.4.2(ii)
can hold for $\cals(\Omega)$ if and only if $k=1$.
\tenrm

Combining Theorems \bex.6.2 and \bex.6.3 with Lemmas \bex.3.3 and \bex.3.6
which give obstructions to smoothness gives the somewhat surprising result
[B-G-M 7],

\noindent{\sc Theorem} \bex.6.4: \tensl Let $\cals$ be a toric 3-Sasakian
manifold.
\item{(i)} If the dimension of $\cals$ is $19$ or greater, then $b_2(\cals)\leq
1.$
\item{(ii)} If the dimension of $\cals$ is $11$ or $15$, then $b_2(\cals)\leq
4.$ 
\item{(iii)} If $b_2(\cals)>4,$ then the dimension of $\cals$ is $7.$
\tenrm

A corollary due to Bielawski [Bi 3] is:

\noindent{\sc Corollary} \bex.6.5: \tensl Let $\cals$ be a regular toric
3-Sasakian manifold. Then $\cals$ is one of the 3-Sasakian homogeneous spaces
$S^{4n-1},\bbr\bbp^{4n-1}$ or ${SU(n)\over S\bigl(U(n-2)\!\times\!U(1)\bigr)}.$
\tenrm

Next we give an explicit construction of toric 3-Sasakian manifolds not
eliminated by Theorem \bex.6.4.  It is enough to show that
$\cala_{4,8}$ and $\cala_{4,7}$ are not empty 
as the rest follow by deletion of rows
of the corresponding $\Omega\in\cala_{4,*}$. We shall present two three
parameter families of solutions, namely        
$$A_1= \pmatrix{1&1&1\cr 2&1&1+2^l\cr 1&16&1+2^m\cr -1&3&2c}, \qquad
A_2= \pmatrix{1&1&1&2\cr 2&1&1+2^{l'}&-1\cr 1&16&1+2^{2n}&3\cr -1&3&2c'&-1},
\leqno{\bex.6.6}$$
where $l,l',m,n\in \bbz^+,$ and $c,c'\in \bbz.$

With the aid of MAPLE symbolic manipulation program, we find

 \noindent{\sc Lemma} \bex.6.7 [B-G-M 7]: \tensl Let $\grD =2(31c+6 +19\cdot
2^{l-1}-7\cdot 2^{m-1}).$ 
\item{(i)} $\Omega_1=\pmatrix{\bbi_4& A_1\cr}$
is admissible if and only if $c\neq 0$ and is not
divisible by $3,$ and $\grD\neq 0$ and is not divisible by $7,19$ nor $31.$
\item{(ii)} $\Omega_2=\pmatrix{\bbi_4& A_2\cr}$
is admissible if and only if $c$ and all minor determinants
of $A_2$ are non-vanishing, and $c'\not\equiv 0 \pmod{3},$ $l'\not\equiv 0
\pmod{4},$ $c'\not\equiv 5 \pmod{7},$ and $11,19,37,71$ do not
divide $\det A=19\cdot 2^{2n}-63-148c'-11\cdot 2^{l'},$ and the following
conditions hold: 
$$\eqalign{\gcd(3,4c'+2^{l'}+1,2c'-2^{l'}-1)&~=~1\cr
           \gcd(7,2^{2n+1}-2^{l'}+1,3\cdot 2^{l'}+2^{2n}+4)&~=~1\cr
           \gcd(19,2^{2n}-2^{l'+4}-15,3\cdot 2^{l'}+2^{2n}+4)&~=~1\cr
           \gcd(25,32c'-3\cdot 2^{2n}-3,6c'+2^{2n}+1)&~=~1.\cr}$$  \tenrm

The conditions in this proposition guarantee that the quotient spaces denoted
by $\cals(c,l,m)$ and $\cals(c',l',n)$ are smooth manifolds of dimension $11$
and $15,$ respectively.

It is routine to verify that the three parameter
infinite family given by
$$ c ~\equiv~ 14  \pmod{21},\qquad
             l ~\not\equiv~ 1  \pmod{5},\qquad
             m ~\not\equiv~ \gra(c)  \pmod{18},\leqno{\bex.6.8}$$
where $2^{\gra(c)}=22(31c+6)\pmod{18}$
satisfies the conditions in (i) of Lemma \bex.6.3. This gives examples in
dimension $11.$ 
(Notice that as $2$ is a primitive root of $19$ the equation
defining $\gra(c)$ has a unique solution $\pmod{18}$ for each value of $c.$)
Similarly, it is straightforward to verify that the
infinite family given by
$$c' ~=~ 2, \qquad
             l' ~=~ 1, \qquad
             n ~\equiv~ 21 \pmod{90},\leqno{\bex.6.9}$$
satisfies the conditions (ii) of Lemma \bex.6.7. We have arrived at:

\noindent{\sc Theorem} \bex.6.10 [B-G-M 7]: \tensl  There exist toric
3-Sasakian manifolds $\cals$ of dimensions $11$ and $15$ with
$b_2(\cals)=2,3,4.$ Consequently, the Betti number relations of Proposition
\btop.4.2 do not hold generally. More explicitly there are compact
11-dimensional 3-Sasakian manifolds for which $b_2\not=b_4$, and compact
15-dimensional 3-Sasakian manifolds for which $b_2\not=b_6.$  \tenrm
\bigskip
\centerline {\bf \bc. Open Problems and Questions}
\medskip
We conclude this section with a short list of interesting problems.
Some minor questions do appear in the text 
but these are usually of  more technical ones. There, quite often,
we simply could not provide a complete answer only because of the time
constraint imposed by the fact that this chapter is a part of a collection of
articles. Here we try to concentrate on, what we believe, are more fundamental
questions.

\noindent{\sc Problem} \bc.1: \tensl
Classify all compact simply-connected Sasakian-Einstein manifolds in dimension 5. 
\tenrm 

All the known examples are regular and regular spaces were
classified in [B-G-F-K]. Can one find irregular examples?
Consider the connected sum $\cals_k=S^5{\#}^k(S^2\times S^3)$. Now, 
$\cals_k$ admits a Sasakian-Einstein metric for $k=0,1,3,4,5,6,7,8$.
How about $k=2$ and $k>8$? In this case $\cals_k$ would necessarily
have to be a Seifert fibered space with the space of leaves
a positive scalar curvature K\"ahler-Einstein orbifold (and not
a smooth manifold) $X_k$ with
$b_2(X_k)=k$. If one could construct such structures for each remaining $k$,
by the result of Smale, every compact simply-connected  5-manifold with no
2-torsion
would admit an Einstein metric of positive scalar curvature. The same problem
in any dimension $2m+1$, $m>2$
appears to be much more involved as it would necessarily
have to include the classification
of 3-Sasakian 7-manifolds [B-G 2].

\noindent{\sc Problem} \bc.2: \tensl
Classify all compact simply-connected 3-Sasakian manifolds in dimension 7.
\tenrm

Again, regular examples were classified in [B-G-M 1, Fr-Kat 2].
This appears to be a difficult problem. Its solution would amount to
a classification of good self-dual and Einstein orbifold of positive 
scalar curvature which, in smooth case, was done in [Hi 1, Fr-Kur].
Certainly, a more modest, partial
classification could be in reach. In particular, in terms of Definition
\bfol.5.4 one easily sees that all toric 3-Sasakian manifolds are regular or of
cyclic type. Is the converse true? That is:

\noindent{\sc Question} \bc.3: \tensl Is every 3-Sasakian 7-manifold of cyclic
type toric (which includes discrete quotients of a spheres as a degenerate
case)? \tenrm

In terms of the classification by symmetries one can ask:

\noindent{\sc Question} \bc.4: \tensl 
Is every compact 3-Sasakian 7-manifold of cohomogeneity 
$\leq2$ toric? 
\tenrm

\noindent{\sc Question} \bc.5: \tensl Let $(\cals,g)$ be a 
simply-connected 3-Sasakian 7-manifold. Can $g$ be of
maximal cohomogeneity 4?
\tenrm   

We are not aware of any such examples. All toric examples are of
cohomogeneity 0,1,2 and some 
new construction of [B-G 2] gives 3-Sasakian
7-manifolds of cohomogeneity 3.

Concerning topology and Problem \bc.2, we can ask the following questions:

\noindent{\sc Question} \bc.6: \tensl Let $(\cals,g)$ be a compact
simply-connected 3-Sasakian 7-manifold. Can $\cals$ be topologically
a product? 
\tenrm

If so then $\cals$ must be $S^2\times S^5$. In the Sasakian-Einstein case it
is known that such a splitting can occur. The simplest example is
$S^2\times S^3$ which has a Sasakian-Einstein structure [Tan 4]. Of course, the
above problem and questions have versions in higher dimension. More generally,

\noindent{\sc Questions} \bc.7: \tensl Other than the vanishing of the odd
Betti numbers up to the middle dimension and the finiteness of the fundamental
group, what more can be said about the topology of a compact 3-Sasakian
manifold? For instance, is $H_2(\cals,\bbz)$ always torsion free? Are there
further restrictions on the fundamental group? \tenrm

Specifically in higher dimensions we ask:

\noindent{\sc Question} \bc.8: \tensl Are there 3-Sasakian manifolds $\cals$ of
dimension $19$ or greater with $b_2(\cals)> 1?$
\tenrm

From a differentiable topological viewpoint we can ask:

\noindent{\sc Questions} \bc.9: \tensl Let $(\cals,g)$, $(\cals', g')$ 
be two compact 
simply-connected 3-Sasakian 7-mani- folds which are
not 3-Sasakian equivalent. Can $\cals$ be diffeomorphic (homeomorphic)
to $\cals'$? In particular, is there a non-standard 3-Sasakian structure
on $S^7$? Can one have 3-Sasakian structures on exotic 7-spheres?
\tenrm    

As pointed out in the Remark \bex.4.11, we expect the positive answer to
the first question. But the problem of existence of other 3-Sasakian structures
on $S^7$ or exotic spheres lacks even the slightest hint, one way or the other.
In general, due to the local rigidity, the moduli space of
inequivalent 3-Sasakian structures must be discrete. So we
have

\noindent{\sc Question} \bc.10: \tensl Is the moduli space
always finite, or can a 3-Sasakian manifold admit
infinitely many inequivalent 3-Sasakian structures?
\tenrm

Concerning related geometries, we have

\noindent{\sc Problem} \bc.11: \tensl Classify all compact
simply-connected proper $G_2$-manifolds.
\tenrm   

This appears to be more involved than Problem \bc.2 because
of Theorem \bg2.2.9. On the other hand, perhaps the $G_2$-structure can be
investigated without reference to the 3-Sasakian geometry. It could happen 
that Problem \bc.10 might admit a simpler solution and become the right
approach to Problem \bc.2. Maybe even one could try to classify all
weak holonomy $G_2$-manifolds. At the moment we do not even know if the
converse of the Theorem \bg2.2.9 is true, that is if a proper $G_2$-manifold
always admits a metric which is 3-Sasakian. This is unlikely though and one
could start by looking for possible proper $G_2$-manifolds with
$b_3\not=0$.

Last but not least, we turn to the regular case. All regular
3-Sasakian manifolds in dimension 7 and 11 are known as explained in
Section \btop.5. Any classification in higher dimensions would
translate into the classification of positive quaternionic K\"ahler manifolds.
Below we give the 3-Sasakian version of the conjecture that all compact
positive quaternionic K\"ahler spaces are symmetric:

\noindent{\sc Conjecture} \bc.12: \tensl Let $(\cals,g)$ be a compact
regular 3-Sasakian manifold of dimension $4n+3$. Then
$\cals$ is homogeneous.
\tenrm

This is simply theorem \btop.4.5 without $n<3$ in the hypothesis.
One might hope that 3-Sasakian geometry would provide some new
input in the regular case. So far we have mostly used results about
positive quaternionic K\"ahler manifolds to describe properties of
regular 3-Sasakian manifolds. But Section \btop \ does give some indication
that 3-Sasakian geometry can be used, at the very least, 
to give new proves of known
theorems.

\noindent{\sc Remark} \bc.13: \tensl Sasakian-Einstein, 3-Sasakian, and
proper $G_2$-manifolds in the AdS/CFT Correspondence.
\tenrm

Very recently Sasakian-Einstein geometry has emerged quite naturally
in the conformal field theory and string theory. In particular,
Klebanov and Witten [K-W] cnosidered $\cals = S^3\times S^2$ in the
context of superconformal field theory dual to the string
theory on $Ad S_5\times \cals$.
Their article originates in an influential
result of Maldacena [Mal] who noticed that large $N$
limit of certain conformal field theories in $d$ dimensions can be described
in terms of supergravity (and string theory) on a product of $(d+1)$-dimensional
anti-de-Sitter $AdS_{d+1}$ space with a compact manifold $M$. The idea was later
examined by Witten who proposed a precise correspondence between
conformal field theory observables and those of supergravity [Wi].
It turns out, and this observation has recently been made by Figueroa [Fi],
that $M$ necessarily has real Killing spinors and
the number of them determines the number of supersymmetries
preserved. Depending on the dimension and the amount of supersymmetry,
the following geometries are possible: spherical in any dimension,
Sasakian-Einstein
in dimension $2k+1$, 3-Sasakian in dimension $4k+3$, 7-manifolds with
weak $G_2$-holonomy, and 6-dimensional nearly K\"ahler manifolds [A-F-H-B].
The case when ${\rm dim}(M)=5,7$ seems to be of particular interest.
For other results concerning Sasakian and 3-Sasakian manifolds in
supersymmetric field theories see [M-P, O-T, G, G-R, C$^2$-D-F$^2$-T].
\tenrm
\bigskip

\def\tU{\tilde{U}}
\def\tx{\tilde{x}}

\centerline {\bf Appendix: Fundamentals of Orbifolds}
\medskip

The notion of orbifold was introduced under the name V-manifold 
by Satake [Sat 1]
in 1956, and subsequently he developed Riemannian geometry on 
V-manifolds [Sat 2]
ending with a proof of the Gauss-Bonnet theorem for V-manifolds.
Contemporaneously, Baily introduced complex V-manifolds and generalized both
the Hodge decomposition theorem [Bai 1], and Kodaira's projective embedding
theorem [Bai 2] to V-manifolds. Somewhat later in the late 1970's and early
1980's Kawasaki generalized various index theorems [Kaw 1-3] to the
category of V-manifolds. It was about this time that Thurston [Thu]
rediscovered the concept of V-manifold, under the name of orbifold, in his
study of the geometry of 3-manifolds, and defined the orbifold fundamental
group $\pi_1^{orb}.$  By now orbifold has become the accepted term for these
objects and we shall follow suit.  However, we do use the name V-bundle for
fibre bundles in this category.

Orbifolds arise naturally as spaces of leaves of Riemannian foliations with
compact leaves, and we are particularly interested in this point of view.
Conversely, every orbifold can be realized in this way. In fact, given an
orbifold $\calo,$ we can construct on it the V-bundle of orthonormal frames
whose total space $P$ is a smooth manifold with a locally free action of the
orthogonal group $O(n)$ such that $\calo =P/O(n).$ Thus, every orbifold can be
realized as the quotient space by a locally free action of a Lie group. We are
not certain of the history of this connection, but it was surely well
understood by Haefliger [Hae] in 1982 who developed the basic techniques for
studying the topology of orbifolds.

\noindent{\sc Definition} \borb.1: \tensl A smooth {\it orbifold} (or {\it
V-manifold}) is a second countable Hausdorff space $X$ together
with a family $\{U_i\}_{\i\in I}$ of open sets that satisfy:
\item{\rm i)} $\{U_i\}_{\i\in I}$ is an open cover of $X$ that is closed under
finite intersections.
\item{\rm ii)} For each $i\in I$ a {\it local uniformizing system} consisting
of a triple $\{\tU_i,\grG_i,\varphi_i\}$, where $\tU_i$ is connected open subset
of $\bbr^n$ containing the origin, $\grG_i$ is a finite group of
diffeomorphisms acting effectively and properly on $\tU_i,$ and
$\varphi_i:\tU_i\ra{1.3} U_i$ is a continuous map onto $U_i$ such that
$\varphi_i\circ \grg=\varphi_i$ for all $\grg\in\grG_i$ and the induced natural
map of $\tU_i/\grG_i$ onto $U_i$ is a homeomorphism. The finite group $\grG_i$
is called a {\it local uniformizing group}.
\item{\rm iii)} Given $\tx_i\in \tU_i$ and $\tx_j\in \tU_j$ such that
$\varphi_i(\tx_i)=\varphi_j(\tx_j),$ there is a diffeomorphism
$g_{ji}:\tV_i\ra{1.3} \tV_j$ from a neighborhood $\tV_i\subset \tU_i$ of
$\tx_i$ onto a neighborhood $\tV_j\subset \tU_j$ of $\tx_j$ such that
$\varphi_i =\varphi_j\circ g_{ji}.$ \tenrm

\noindent{\sc Remarks} \borb.2: 1) We can always take the finite subgroups
$\grG_i$ to be subgroups of the orthogonal group $O(n)$ and in the orientable
case $SO(n).$ 2) Condition iii) implies that for each $\grg_i\in \grG_i$ there
exists a unique $\grg_j\in \grG_j$ such that $g_{ji}\circ \grg_i =\grg_j\circ
g_{ji}.$ 3) One can define the notion of equivalence of families of open sets,
any such family of open sets is contained in a unique maximal family satisfying
the required properties. 4) The standard notions of smooth maps between
orbifolds, and isomorphism classes of orbifolds, etc.  can then be given in an
analogous manner to manifolds (see [Sat 1-2, Bai 1-2]).  We leave this to the
reader to fill in.  Notice that in particular a diffeomorphism between
orbifolds gives a homeomorphism of the underlying topological spaces.
Similarly, a {\it complex orbifold} can be defined by making the obvious
changes. 

An alternative definition of orbifold given by Haefliger [Hae] can be obtained
as follows: Let $G_{\grG,g}$ denote the groupoid of germs of diffeomorphisms
generated by the germs of elements in $\grG_i$ and the germs of the
diffeomorphisms $g_{ji}$ described above. Let $\tU =\sqcup_i \tU_i$ denote the
disjoint union of the $\tU_i.$ Then $x,y\in \tU$ are equivalent if there is a
germ $\grg \in G_{\grG,g}$ such that $y=\grg(x).$ The quotient space $X
=\tU/G_{\grG,g}$ defines an orbifold (actually an isomorphism class of
orbifolds). In the case that an orbifold $X$ is given as the space of leaves of
a foliation $\calf$ on a smooth manifold, the groupoid $G_{\grG,g}$ is the
transverse holonomy groupoid of $\calf.$

The following result relating to foliations, which is given in Molino, is
fundamental to our work:

\noindent{\sc Theorem} \borb.3: [Mol: Proposition 3.7] \tensl 
Let $(M,\calf,g)$ be a Riemannian foliation of codimension $q$ with 
compact leaves and bundle-like metric $g$.
Then the space of leaves $M/\calf$ admits the structure of a $q$-dimensional
orbifold such that the natural projection $\pi:M\ra{1} M/\calf$ is an
orbifold submersion. \tenrm

Let $X$ be an orbifold and choose a local uniformizing system
$\{U,\grG,\varphi\}.$ Let $x\in X$ be any point, and let $p\in
\varphi^{-1}(x),$ then up to conjugacy the isotropy subgroup $\grG_p\subset
\grG$ depends only on $x,$ and accordingly we shall denote this isotropy
subgroup by $\grG_x.$ A point of $X$ whose isotropy subgroups $\grG_x\neq
\hbox{id}$ is called a {\it singular} point.  Those points with
$\grG_x=\hbox{id}$ are called {\it regular} points. The subset of regular
points is an open dense subset of $X.$ The isotropy groups give a natural
stratification of $X$ by saying that two points lie in the same stratum if
their isotropy subgroups are conjugate. Thus, the dense open subset of regular
points forms the principal stratum. In the case that $X$ is the space of leaves
of a foliation, the isotropy subgroup $\grG_x$ is precisely the leaf holonomy
group of the leaf $x.$ An orbifold $X$ is a smooth manifold or in the complex
analytic category a complex manifold if and only if $\grG_x=\hbox{id}$ for all
$x\in X.$ In this case we can take $\grG=\hbox{id}$ and $\varphi=\hbox{id},$
and the definition of an orbifold reduces to the usual definition of a smooth
manifold.  

Many of the usual differential geometric concepts that hold for smooth or
complex analytic manifolds also hold in the orbifold category,  in particular
the important notion of a fiber bundle.

\noindent{\sc Definition} \borb.4: \tensl A {\it V-bundle} over an orbifold
$X$ consists of a bundle $B_{\tU}$ over $\tU$ for each local uniformizing
system $\{\tU_i,\grG_i,\varphi_i \}$ with Lie group $G$ and fiber $F$
(independent of $\tU_i$) together with a homomorphism
$h_{\tU_i}:\grG_i\ra{1.3} G$ satisfying:
\item{\rm i)} If $b$ lies in the fiber over $\tx_i\in \tU_i$ then for each
$\grg\in \grG_i,$ $bh_{\tU_i}(\grg)$ lies in the fiber over $\grg^{-1}\tx_i.$ 
\item{\rm ii)} If $g_{ji}:\tU_i \ra{1.5}\tU_j$ is a diffeomorphism onto an open
set, then there is a bundle map $g_{ij}^*:B_{U_j}|g_{ji}(\tU_i)\ra{1.3}
B_{\tU_i}$ satisfying the condition that if $\grg\in \grG_i,$ and $\grg'\in
\grG_j$ is the unique element such that $g_{ji}\circ \grg=\grg'\circ g_{ji},$
then $h_{\tU_i}(\grg)\circ g_{ji}^*=g_{ji}^*\circ h_{\tU_j}(\grg'),$ and if
$g_{kj}:\tU_j\ra{1.3} \tU_k$ is another such diffeomorphism then $(g_{kj}\circ
g_{ji})^*=g_{ji}^*\circ g_{kj}^*.$ 

\noindent If the fiber $F$ is a vector space and $G$ acts on $F$ as linear
transformations of $F$, then the V-bundle is called a {\it vector}
V-bundle. Similarly, if $F$ is the Lie group $G$ with its right action, then
the V-bundle is called a {\it principal} V-bundle.  \tenrm

The {\it total space} of a V-bundle over $X$ is an orbifold $E$ with local
uniformizing systems $\{B_{\tU_i},\grG_i^*,\varphi_i^*\}.$ By choosing the
local uniformizing neighborhoods of $X$ small enough, we can always take
$B_{\tU_i}$ to be the product $\tU_i\times F$ which we shall heretofore assume.
There is an action of the local uniformizing group $\grG_i$ on $\tU_i\times F$
given by sending $(\tx_i,b)\in \tU_i\times F$ to
$(\grg^{-1}\tx_i,bh_{\tU_i}(\grg)),$ so the local uniformizing groups
$\grG_i^*$ can be taken to be subgroups of $\grG_i.$ We are particularly
interested in the case of a principal bundle. In the case the fibre is the Lie
group $G,$ so the image $h_{\tU_i}(\grG_i^*)$ acts freely on $F.$ Thus the
total space $P$ of a principal V-bundle will be smooth if and only if
$h_{\tU_i}$ is injective for all $i.$

\noindent{\sc Remarks} \borb.5: We shall often denote a V-bundle by the
standard notation $\pi:P\ra{1.3} X$ and think of this as an ``orbifold
fibration''. It must be understood, however, that an orbifold fibration is not
a fibration in the usual sense. Shortly, we shall show that it is a fibration
rationally.   Again the standard notions of smooth maps between
V-bundles, and isomorphism classes of V-bundles can be given in the usual
manner. We let this description to the reader. An {\it absolute} V-bundle
resembles a bundle in the ordinary sense, and corresponds to being able to take
$h_{\tU} =\hbox{id},$ for all local uniformizing neighborhoods $\tU.$ In
particular, the trivial V-bundle $X\times F$ is absolute. Another important
notion introduced by Kawasaki [Kaw 2] is that of proper. A V-bundle $E$ is said
to be {\it proper} if the local uniformizing groups $\grG_i^*$ of $E$ act
effectively on $X$ when viewed as subgroups of the local uniformizing groups
$\grG_i$ on $X.$ Any V-bundle with smooth total space is clearly proper. The
Kawasaki index theorems such as his Riemann-Roch Theorem used in section
\bfol.2 require the V-bundles to be proper.

Since an orbifold fibration is not a fibration in the usual sense, the usual
techniques in topology for fibrations do not apply directly. However, Haefliger
[Hae] has defined orbifold homology, cohomology, and homotopy groups which do
have an analogue in the standard theory. Let $X$ be an orbifold of dimension
$n$ and let $P$ denote the bundle of orthonormal frames on $X.$ It is a smooth
manifold on which the orthogonal group $O(n)$ acts locally freely with the
quotient $X.$ Let $EO(n)\ra{1.3} BO(n)$ denote the universal $O(n)$ bundle.
Consider the diagonal action of $O(n)$ on $EO(n)\times P$ and denote the
quotient by $BX.$ Now there is a natural projection $p:BX\ra{1.3} X$ with
generic fiber the contractible space $EO(n),$ and Haefliger defines the
orbifold cohomology, homology, and homotopy groups by
$$H^i_{orb}(X,\bbz)=H^i(BX,\bbz), \quad H^{orb}_i(X,\bbz)=H_i(BX,\bbz), \quad
\pi_i^{orb}(X)=\pi_i(BX). \leqno{\borb.6}$$ 
This definition of $\pi_1^{orb}$ is equivalent to Thurston's better known
definition [Thu] in terms of orbifold deck transformations, and when $X$ is a
smooth manifold these orbifold groups coincide with the usual groups. Moreover,
we have

\noindent{\sc Proposition} \borb.7 [Hae]: \tensl The map $p:BX\ra{1.3} X$
induces an isomorphism $H^i_{orb}(\cals,\bbz)\otimes \bbq \simeq
H^i(\cals,\bbz)\otimes \bbq.$ \tenrm

Now with this in hand for the orbifold category, the circle V-bundles over
$\cals$ are classified [H-S] by $H^2_{orb}(\cals,\bbz).$ Of course, rationally
there is no difference by Proposition \borb.7. The rational Gysin sequence for
orbifold sphere bundles whose generic fibres are spheres also holds.
Haefliger's theory also applies to the following situation. Let $G$ be a
compact Lie group acting locally freely on an orbifold $Y$ with quotient $X.$
This gives rise to a fibration $EO(n)\times G\ra{1.3} BY\ra{1.3} BX,$ which
induces the long exact homotopy sequence
$$\cdots \ra{1.5}\pi_i(G)\ra{1.5}\pi_i^{orb}(Y)\ra{1.5}\pi_i^{orb}(X)\ra{1.5}
\pi_{i-1}(G)\ra{1.5}\cdots. \leqno{\borb.8}$$
This was used by Haefliger and Salem [H-S] in their study of torus actions on
orbifolds.

We are particularly interested in the case of circle V-bundles. Using the
exponential exact sequence one sees as the usual case that $H^2_{orb}(X,\bbz)$
classifies equivalence classes of circle V-bundles over an orbifold $X.$
Furthermore, in [H-S] it is shown that $H^2(X,\bbz)$ classifies circle
V-bundles up to local equivalence. This gives a monomorphism
$H^2(X,\bbz)\ra{1.3} H^2_{orb}(X,\bbz)$ which is an isomorphism rationally.

In [B-G 1] we introduced the set $\hbox{Pic}^{orb}(X)$ of equivalence classes
holomorphic line V-bundles over a complex orbifold $X$ and one easily sees
[B-G 1]:

\noindent{\sc Lemma} \borb.9: \tensl $\hbox{Pic}^{orb}(X)$ forms an Abelian
group. Furthermore, there is a monomorphism $\hbox{Pic}(X)\ra{1.3}
\hbox{Pic}^{orb}(X)$ which is an isomorphism rationally.   \tenrm

The notion of sections of bundles works just as well in the orbifold category.

\noindent{\sc Definition} \borb.10: \tensl Let $E$ be a V-bundle over an
orbifold $X.$ Then a {\it section} $\grs$ of $E$ over the open set $V\subset X$
is a section $\grs_U$ of the bundle $B_U$ for each local uniformizing system
$\{U,\grG,\varphi\}\in \calf_V$ such that for any $x\in U$ we have
\item{(i)} For each $\grg\in \grG$ $\grs_U(\grg^{-1}x)=h_U(\grg)\grs_U(x).$
\item{(ii)} If $\grl:\{U,\grG,\varphi\}\ra{1.5}\{U',\grG',\varphi'\}$ is an
injection, then $\grl^*\grs_{U'}(\grl(x))=\grs_U(x).$ \tenrm

\noindent If each of the local sections $\grs_U$ is continuous, smooth,
holomorphic, etc., we say that $\grs$ is continuous, smooth, holomorphic, etc.,
respectively. Given local sections $\grs_U$ of a vector V-bundle we can
always construct $\grG$-invariant local sections by ``averaging over the
group'', i.e.,  we define
$\grs_U^I={1\over |\grG|}\sum_{\grg\in \grG}\grs_U\circ
\grg.$
A similar procedure holds for product structures.  For example, if $L$ is a
holomorphic line V-bundle on $X,$ and if $\grs$ is a holomorphic section, we
can construct local invariant sections $\grs_U^I$ of $L^{|\grG|}$ by taking
products, viz.,
$\grs_U^I={1\over |\grG|}\prod_{\grg\in \grG}\grs_U\circ
\grg.$

The standard notions of tangent bundle, cotangent bundle, and all the
associated tensor bundles all have V-bundle analogues [Bai 1, Sat 1-2]. In
particular, if $V$ is an open subset of $\varphi(U)$ then the integral of an
$n$-form (measurable) $\grs$ is defined by
$\int_V\grs = {1\over |\grG|}\int_{\varphi^{-1}(V)}\grs_U.$
All of the standard integration techniques, such as Stokes' theorem, hold on
V-manifolds.

Riemannian metrics also exist by the standard partition of unity argument, and
we shall always work with $\grG$-invariant metrics. Moreover, all the standard
differential geometric objects involving curvature and metric concepts, such as
the Ricci tensor, Hodge star operator, etc., hold equally well. On a complex
orbifold there is a $\grG$-invariant tensor field $J$ of type $(1,1)$ which
describes the complex structure on the tangent V-bundle $TX.$ The almost
complex structure $J$ gives rise in the usual way to the V-bundles $A^{r,s}$
of differential forms of type $(r,s).$ The standard concepts of Hermitian and
K\"ahler metrics hold equally well on V-manifolds, and all the special
identities involving K\"ahler, Einstein, or K\"ahler-Einstein geometry hold. In
particular, the standard Weizenb\"ock formulas hold.

Finally, there is associated to every compact orbifold $X$ an integer $m_0$
called the {\it order} of $X$ and defined to be the least common multiple of
the orders of the local uniformizing  groups. 

\bigskip
\hfuzz=4pt
\medskip
\centerline{\bf Bibliography}
\medskip
\font\ninesl=cmsl9
\font\bsc=cmcsc10 at 10truept
\parskip=1.5truept
\baselineskip=11truept
\ninerm 
\item{[A-F-H-S]} {\bsc B. S. Acharya, J. M. Figueroa-O'Farrill, C. M. Hull, and
B. Spence}, {\ninesl Branes at Conical Singularities and Holography},
preprint, August 1998; hep-th/9808014.
\item{[Akh]} {\bsc D. N. Akhiezer}, {\ninesl Homogeneous complex manifolds}, in
Enc. Math. Sci. vol 10, {\ninesl Several Complex Variables IV}, S. G. Gindikin
and G. M. Khenkin (Eds)., Springer-Verlag, New York, 1990.
\item{[Al 1]} {\bsc D. V. Alekseevski}, {\ninesl 
Riemannian manifolds with exceptional holonomy groups}, Functional
Anal. Appl. 2 (1968), 106-114.
\item{[Al 2]} {\bsc D. V. Alekseevski}, {\ninesl Classification of
quaternionic spaces with solvable group of motions}, Math.
USSR-Izv. 9 (1975), 297-339.
\item{[Al-Wa]} {\bsc S. Aloff and N. Wallach}, {\ninesl
An infinite family of distinct 7-manifolds 
admitting positively curved Riemannian structures},
Bull. Amer. Math. Soc. 81 (1975), 93-97.
\item{[A-M-P]} {\bsc L. Astey, E. Micha, and G. Pastor}, {\ninesl Homeomorphism
and diffeomorphism types of Eschenberg spaces}, Differential Geom. Appl. 
7 (1997), 41-50.
\item{[An]} {\bsc M. Anderson}, {\ninesl Convergence and rigidity of manifolds
under Ricci curvature bounds}, Invent. Math. 102 (1990), 429-445.
\item{[Bai 1]} {\bsc W. L. Baily}, {\ninesl The decomposition theorem for
V-manifolds}, Amer. J. Math. 78 (1956), 862-888.
\item{[Bai 2]} {\bsc W. L. Baily}, {\ninesl On the imbedding of V-manifolds in
projective space}, Amer. J. Math. 79 (1957), 403-430.
\item{[B\"ar]} {\bsc C. B\"ar}, {\ninesl Real Killing spinors and holonomy},
Comm. Math. Phys. 154 (1993), 509-521.
\item{[BeB\`er]} {\bsc L. Berard B\`ergery}, {\ninesl
Vari\'et\'es quaternionnienns}, Notes d'unde conf\'rence \`a la table
ronde ``Varie\'et\'es d'Einstein", Espalion (1997); (unpublished).
\item{[Bes]} {\bsc A. L. Besse}, {\ninesl Einstein Manifolds}, 
Springer-Verlag, New York (1987). 
\item{[Ber]} {\bsc M. Berger}, {\ninesl Sur les groupes d'holonomie
des vari\'et\'es \`a connexion et des vari\'et\'es riemanniennes},
Bull. Soc. Math. France 83 (1955), 279-330.
\item{[B-F-G-K]} {\bsc H. Baum, T. Friedrich, R. Grunewald, and I. Kath}, 
{\ninesl Twistors and Killing Spinors on Riemannian Manifolds},
Teubner-Texte f\"ur Mathematik, vol. 124, Teubner, Stuttgart, Leipzig, 1991.
\item{[B-G 1]} {\bsc C. P. Boyer and  K. Galicki}, {\ninesl
The twistor space of a 3-Sasakian manifolds},
Int. J. Math. 8 (1997), 31-60.
\item{[B-G 2]} {\bsc C. P. Boyer and  K. Galicki}, {\ninesl On
Sasakian-Einstein geometry}, UNM preprint, September 1998.
\item{[B-G 3]} {\bsc C. P. Boyer and  K. Galicki}, {\ninesl
Polygons, gravitons, and Einstein manifolds}, in preparation.
\item{[B-G-M 1]} {\bsc C. P. Boyer, K. Galicki, and B. M. Mann}, {\ninesl
Quaternionic reduction and Einstein manifolds}, Comm.
Anal. Geom. 1 (1993), 1-51. 
\item{[B-G-M 2]} {\bsc C. P. Boyer, K. Galicki, and B. M. Mann}, {\ninesl
The geometry and topology of 3-Sasakian manifolds}, 
J. reine angew. Math. 455 (1994), 183-220.
\item{[B-G-M 3]} {\bsc C. P. Boyer, K. Galicki, and B. M. Mann}, {\ninesl
On strongly inhomogeneous Einstein manifolds},
Bull. London Math. Soc. 28 (1996), 401-408.
\item{[B-G-M 4]} {\bsc C. P. Boyer, K. Galicki, and B. M. Mann}, {\ninesl
New examples of inhomogeneous Einstein manifolds of
positive scalar curvature}, Math. Res. Lett. 1 (1994), 115-121.
\item{[B-G-M 5]} {\bsc C. P. Boyer, K. Galicki, and B. M. Mann}, {\ninesl
3-Sasakian manifolds}, Proc. Japan Acad. vol. 69, Ser. A (1993),
335-340.
\item{[B-G-M 6]} {\bsc C. P. Boyer, K. Galicki, and B. M. Mann}, {\ninesl
Hypercomplex structures on Stiefel manifolds}, Ann. Global Anal. Geom. 14
(1996), 81-105.
\item{[B-G-M 7]} {\bsc C. P. Boyer, K. Galicki, and B. M. Mann}, {\ninesl
A note on smooth toral reductions of spheres},
Manuscripta Math. 95 (1998), 321-344.
\item{[B-G-M 8]} {\bsc C. P. Boyer, K. Galicki, and B. M. Mann}, {\ninesl
Hypercomplex structures from 3-Sasakian structures}, J. reine angew. Math.,
501 (1998), 115-141.
\item{[B-G-M-R 1]} {\bsc C. P. Boyer, K. Galicki, B. M. Mann, and E. Rees},
{\ninesl Compact 3-Sasakian 7-manifolds with arbitrary second Betti number},
Invent. Math. 131 (1998), 321-344.
\item{[B-G-M-R 2]} {\bsc C. P. Boyer, K. Galicki, B. M. Mann, and E. Rees},
{\ninesl Einstein manifolds of positive scalar curvature with arbitrary second
Betti number}, Balkan J. Geom. Appl. 1 no. 2 (1996), 1-8.
\item{[Bi 1]} {\bsc R. Bielawski},
{\ninesl On the hyperk\"ahler metrics associated
to singularities of nilpotent varieties}, Ann. Global Anal. Geom.
14 (1996), 177-191.
\item{[Bi 2]} {\bsc R. Bielawski},
{\ninesl Betti numbers of 3-Sasakian quotients of
spheres by tori}, 
Bull. London
Math. Soc. 29 (1997), 731-736.
\item{[Bi 3]} {\bsc R. Bielawski}, 
{\ninesl Complete $T^n$-invariant hyperk\"ahler $4n$-manifolds}, preprint MPI
(1998).
\item{[Bi-D]} {\bsc R. Bielawski and A. Dancer}, {\ninesl
The geometry and topology of
toric hyperk\"ahler manifolds}, to appear in Comm. Anal. Geom. (1998).
\item{[Bl]} {\bsc D. E. Blair}, {\ninesl Contact Manifolds in
Riemannian Geometry}, Lecture Notes in Mathematics 509, Springer-Verlag,
New Yrok 1976.
\item{[Bl-Go]} {\bsc D. E. Blair and S. I. Goldberg}, {\ninesl 
Topology of almost contact manifolds}, J. Differential Geom. 1 (1967), 347-354.
\item{[Bon]} {\bsc E. Bonan}, {\ninesl Sur les G-structures de type
quaternionien}, Cah. Top. Geom. Differ. 9 (1967), 389-461.
\item{[Bre]} {\bsc G.E. Bredon}, {\ninesl Introduction to Compact
Transformation Groups}, Academic Press, New York (1972).
\item{[Bry]} {\bsc R. Bryant}, {\ninesl Metrics with exceptional holonomy},
Ann. Math. 126 (1987), 525-576.
\item{[C$^2$-D-F$^2$-T]} 
{\bsc L. Castellani, A. Ceresole, R. D'Auria, S. Ferrara,
P. Fr\'e, and M. Trigiante}, {\ninesl $G/H$ $M$-branes and
$AdS_{p+2}$ Geometries}, preprint, March 1998; hep-th/9803039.
\item{[C-M-Sw]} {\bsc F. M. Cabrera, M. D. Monar, and A. F. Swann},
{\ninesl Classification of $G_2$-structures}, J. London Math. Soc. 53 (1996),
407-416.
\item{[D-Sw]} {\bsc A. Dancer and A. Swann}, {\ninesl
The geometry of singular quaternionic K\"ahler
quotients}, Int. J. Math. 8 (1997), 595-610.
\item{[Esch 1]} {\bsc J. H. Eschenburg}, {\ninesl New examples of manifolds
with strictly positive curvature}, Invent. Math. 66 (1982), 469-480.
\item{[Esch 2]} {\bsc J. H. Eschenburg}, {\ninesl Cohomology of biquotients},
Manuscripta Math. 75 (1992), 151-166.
\item{[Fe-Gra]} {\bsc M. Fern\'andez and A. Gray}, {\ninesl
Riemannian manifolds with
structure group $G_2$}, Ann. Mat. Pura Appl. 32 (1982), 19-45.
\item{[Fi]} {\bsc J. M. Figueroa-O'Farrill}, {\ninesl Near-Horizon
Geometries of Supersymmetric Branes}, preprint, July 1998; hep-th/9807149.
\item{[F-K-M-S]} {\bsc T. Friedrich, I. Kath, A. Moroianu, and
U. Semmelmann}, {\ninesl On nearly parallel $G_2$-structures},
J. Geom. Phys. 23 (1997), 259-286.
\item{[Fr]} {\bsc T. Friedrich}, {\ninesl
der erste Eigenwert des Dirac-Operators einer kompakten Riemannschen
Mannigfaltigkeiten nichtnegativer Skalarkr\"ummung}, Math. Nach. 97 (1980),
117-146.
\item{[Fra]} {\bsc A. Franc}, {\ninesl Spin structures and Killing spinors on
lens spaces}, J. Geom. Phys. 4 (3) (1987), 277-287.
\item{[Fr-Kat 1]} {\bsc T. Friedrich and I. Kath}, {\ninesl Compact
five-dimensional Riemannian manifolds with parallel spinors},
Math. Nachr. 147 (1990), 161-165.
\item{[Fr-Kat 2]} {\bsc T. Friedrich and I. Kath}, {\ninesl Compact 
seven-dimensional
manifolds with Killing spinors}, Comm. Math. Phys. 133 (1990), 543-561.
\item{[Fr-Kur]} {\bsc T. Friedrich and Kurke}, {\ninesl
Compact four-dimensional self-dual Einstein manifolds with
positive scalar curvature}, Math. Nach. 106 (1982), 271-299.
\item{[Fuj]} {\bsc A. Fujiki}, {\ninesl
On the de$\,$Rham cohomology group of a compact K\"ahler
symplectic manifold}, in Algebraic Geometry, Sendai, 1985 (Advanced
Studies in Pure Mathematics 10), ed. T. Oda, North Holland (1987).
\item{[G]} {\bsc K. Galicki}, {\ninesl
Geometry of the Scalar Coupling in $N=2$ Supergravity Models},
Class. Quan. Grav. 9(1) (1992), 27-40.
\item{[Gei]} {\bsc H. Geiges}, {\ninesl Normal contact structures
on 3-manifolds}, T\^ohoku Math. J. 49 (1997), 415-422.
\item{[G-L]} {\bsc K. Galicki and B. H. Lawson, Jr.}, {\ninesl
Quaternionic reduction and quaternionic orbifolds}, Math. Ann. 282 (1988),
1-21.
\item{[G-Ni]} {\bsc K. Galicki and T. Nitta}, {\ninesl
Nonzero scalar curvature generalizations of the ALE
instantons},
J. Math. Phys. 33
(1992), 1765-1771.
\item{[G-R]} {\bsc G. W. Gibbons and P. Rychenkova}, {\ninesl
Cones, tri-Sasakian structures and
superconformal invariance}, preprint, September 1998; hep-th/9809158.
\item{[Go]} {\bsc S. I. Goldberg}, {\ninesl Nonegatively curved
contact manifolds}, Proc. Amer. Math. Soc. 96 (1986), 651-656.
\item{[Gra 1]} {\bsc A. Gray}, {\ninesl A note on manifolds whose holonomy
group is a subgroup of $Sp(n)\!\cdot\! Sp(1)$}, Mich. Math. J. 16 (1965),
125-128.
\item{[Gra 2]} {\bsc A. Gray}, {\ninesl Weak holonomy groups}, Math. Z.
123 (1971), 290-300.
\item{[Gro]} {\bsc M. Gromov}, {\ninesl Curvature, diameter and Betti numbers},
Comment. Math. Helvetici 56 (1981) 179-195.
\item{[G-Sal]} {\bsc K. Galicki and S. Salamon}, {\ninesl On
Betti numbers of 3-Sasakian manifolds},  Geom. Ded. 63 (1996), 45-68.
\item{[Hae]} {\bsc A. Haefliger}, {\ninesl Groupoides d'holonomie et
classifiants}, Ast\'erisque 116 (1984), 70-97.
\item{[Hit 1]} {\bsc N. J. Hitchin}, {\ninesl K\"ahlerian twistor
spaces}, Proc. Lond. Math. Soc. 43 (1981), 133-150.
\item{[Hit 2]} {\bsc N. J. Hitchin}, unpublished.
\item{[H-K-L-R]} {\bsc N. J. Hitchin, A. Karlhede, U. Lindstr\"{o}m and
M. Ro\v cek}, {\ninesl Hyperk\"ahler metrics and supersymmetry},
Comm. Math. Phys. 108 (1987), 535-589.
\item{[H-S]} {\bsc A.  Haefliger and E. Salem}, {\ninesl Actions of tori on
orbifolds}, Ann. Global Anal. Geom. 9 (1991), 37-59.
\item{[I-Kon]} {\bsc  S. Ishihara and M. Konishi}, {\ninesl  Fibered
Riemannian spaces with Sasakian 3-structure}, Differential
Geometry, in honor of K. Yano, Kinokuniya, Tokyo (1972), 179-194.
\item{[Ish 1]} {\bsc  S. Ishihara}, {\ninesl  Quaternion K\"ahlerian
manifolds}, J. Differential Geom. 9 (1974), 483-500.
\item{[Ish 2]} {\bsc  S. Ishihara}, {\ninesl  Quaternion K\"ahlerian
manifolds and fibered Riemannian spaces with Sasakian 3-structure},
Kodai Math. Sem. Rep. 25 (1973), 321-329.
\item{[Kas]} {\bsc  T. Kashiwada}, {\ninesl  A note on a Riemannian
space with Sasakian 3-structure}, Nat. Sci. Reps. Ochanomizu Univ.
22 (1971), 1-2.
\item{[Kaw 1]} {\bsc T. Kawasaki}, {\ninesl The Riemann-Roch theorem for
complex
V-manifolds}, Osaka J. Math. 16 (1979), 151-159.
\item{[Kaw 2]} {\bsc T. Kawasaki}, {\ninesl The signature theorem for
V-manifolds}, Topology 17 (1978), 75-83.
\item{[Kaw 3]} {\bsc T. Kawasaki}, {\ninesl The index of elliptic operators
over V-manifolds}, Nagoya Math. J. 84 (1981), 135-157.
\item{[K-O]} {\bsc S. Kobayashi and T. Ochiai}, {\ninesl Characterizations of
complex projective spaces and hyperquadrics}, J. Math.  Kyoto. Univ. 13 (1973),
31-47.
\item{[Kob]} {\bsc S.  Kobayashi}, {\ninesl On compact K\"ahler manifolds with
positive Ricci tensor}, Ann. Math. 74 (1961), 381-385.
\item{[Kon]} {\bsc  M. Konishi}, {\ninesl  On manifolds with
Sasakian 3-structure over quaternion K\"ahlerian manifolds}, Kodai
Math. Sem. Reps. 26 (1975), 194-200.
\item{[Kra]} {\bsc V. Kraines}, {\ninesl 
 Topology of quaternionic manifolds}, Trans. Amer. Math.
Soc. 122 (1966), 357-367.
\item{[Kru 1]} {\bsc B. Kruggel}, {\ninesl A homotopy classification of certain
7-manifolds}, Trans. Amer. Math.
Soc. 349 (1997), 2827-2843.
\item{[Kru 2]} {\bsc B. Kruggel}, {\ninesl Kreck-Stolz invariants, normal
invariants and the homotopy classification of generalized Wallach spaces},
preprint.
\item{[Kru 3]} {\bsc B. Kruggel}, {\ninesl Diffeomorphism and
homeomorphism classification of Eschenburg spaces}, Quat. J. Math. 
Oxford Ser. (2), to appear (1998).
\item{[K-S 1]} {\bsc M. Kreck and S. Stolz}, {\ninesl A diffeomorphism
classification of $7$-dimensional homogeneous Einstein manifolds with
$SU(3)\times SU(2)\times U(1)$ symmetry}, Ann. Math. 127 (1988), 373-388.
\item{[K-S 2]} {\bsc M. Kreck and S. Stolz}, {\ninesl Some nondiffeomorphic
homeomorphic homogeneous $7$-manifolds with positive sectional curvature},
J. Differential Geom. 33 (1991), 465-486.
\item{[Kuo]} {\bsc Y.-Y. Kuo}, {\ninesl  On almost contact
3-structure}, T\^ohoku Math. J. 22 (1970), 325-332.
\item{[Kuo-Tach]} {\bsc Y.-Y. Kuo and S. Tachibana}, {\ninesl  On the 
distribution appeared in contact 3-structure}, 
Taita J. Math. 2 (1970), 17-24.
\item{[K-W]} {\bsc I. R. Klebanov and E. Witten}, {\ninesl
Superconformal Field Theory on Threebranes at a Calabi-Yau Singularity},
preprint, July 1998; hep-th/9807080.
\item{[Le 1]} {\bsc C. LeBrun}, {\ninesl
A finiteness theorem for quaternionic-K\"ahler
manifolds with positive scalar curvature}, Contemp. Math.
154 (1994), 89-101.
\item{[Le 2]} {\bsc C. LeBrun}, {\ninesl
A rigidity theorem for quaternionic-K\"ahler
manifolds,} Proc. Amer. Math. Soc. 103 (1988), 1205-1208.
\item{[Le 3]} {\bsc C. LeBrun}, {\ninesl
Fano manifolds, contact structures, and quaternionic geometry},
Int. J. Math. 6 (1995), 115-127.
\item{[Le-Sal]} {\bsc C. LeBrun and S. M. Salamon}, {\ninesl Strong rigidity
of positive quaternion-K\"ahler manifolds}, Invent. Math. 118 (1994), 109-132.
\item{[L-R]} {\bsc U. Lindstr\"om and M. Ro\v cek}, {\ninesl
Scalar tensor duality and $N=1,2$ non-linear $\sigma$-models},
Nucl. Phys. B222 (1983), 285-308.
\item{[Mal]} {\bsc J. Maldacena}, {\ninesl The large $N$ limit of
superconformal field theories and supergravity}, preprint, 1997;
hep-th/9711200.
\item{[Ma-Ro]} {\bsc S. Marchiafava and G. Romani}, {\ninesl Sui fibrati con
struttura quaternioniale generalizzata}, Ann. Mat. Pura Appl. 107 (1976),
131-157.
\item{[Mil]} {\bsc R. J. Milgram}, {\ninesl The classification of Aloff-Wallach
manifolds and their generalizations}, to appear in the volume in honor of the
60th birthday of C. T. C. Wall.
\item{[Mi-Mo]} {\bsc Y. Miyaoka and S. Mori}, {\ninesl A numerical criterion
for uniruledness}, Ann.  Math. 124 (1986), 65-69.  
\item{[Mol]} {\bsc P. Molino}, {\ninesl Riemannian foliations}, Progress in
Mathematics 73, Birkh\"auser, Boston, 1988.  
\item{[Mo-Mu]} {\bsc S. Mori and S. Mukai}, {\ninesl Classification of Fano
3-folds with $\scriptstyle{B_2\geq 2.}$}, Manuscripta Math. 36 (1981), 147-162.
\item{[Moo-Sch]} {\bsc C.C. Moore and C. Schochet}, {\ninesl Global Analysis on
Foliated Spaces}, Springer-Verlag, New York, 1988.
\item{[Mor]} {\bsc A. Moroianu},
{\ninesl Parallel and Killing spinors on ${\rm Spin}^c$-manifolds}, Comm. Math.
Phys. 187 (1997), 417-427.  
\item{[M-P]} {\bsc D. R. Morrison and M. R. Plesser},
{\ninesl Non-Spherical Horizons, I}, preprint, October 1998;
hep-th/9810201.
\item{[N]} {\bsc Y. Nagatomo}, {\ninesl 
Rigidity of $c_1$-self-dual connections on
quaternionic K\"ahler manifolds}, J. Math. Phys. 33 (1992), 4020-4025.
\item{[O'N]} {\bsc B. O'Neill}, {\ninesl Semi-Riemannian Geometry},
Pure and Applied Math. 103, Academic Press, New York !983.
\item{[Or-Pi]} {\bsc L. Ornea and P. Piccinni}, 
{\ninesl Locally conformally K\"ahler structures in
quaternionic geometry}, Trans. Am. Math. Soc.
349 (1997), 641-655.
\item{[O-T]} {\bsc K. Oh and R. Tatar}, {\ninesl Three Dimensional
SCFT from M2 Branes at Conical Singularities}, preprint, October 1998;
hep-th/9810244.
\item{[Pe-Po 1]} {\bsc H. Pedersen and Y. S. Poon}, {\ninesl
Deformations of hypercomplex
Structures}, J. reine 
angew. Math., 499 (1998), 81-99.
\item{[Pe-Po 2]} {\bsc H. Pedersen and Y. S. Poon}, {\ninesl
A note on rigidity of
3-Sasakian manifolds}, Proc. Am. Math. Soc., to appear (1998).
\item{[Pi]} {\bsc P. Piccinni}, {\ninesl 
The geometry of positive locally quaternion K\"ahler
manifolds}, Ann. Global Anal. Geom., 16 (1998), 255-272.
\item{[Po-Sal]} {\bsc  Y. S. Poon and S. Salamon}, {\ninesl 
Eight-dimensional quaternionic K\"ahler manifolds with positive
scalar curvature}, J. Differential Geom. 33 (1990), 363-378. 
\item{[Sal 1]} {\bsc  S. Salamon}, {\ninesl  Quaternionic K\"ahler
manifolds}, Invent. Math. 67 (1982), 143-171.
\item{[Sal 2]} {\bsc  S. Salamon}, {\ninesl  Differential geometry
of quaternionic manifolds}, Ann. Sci. Ec. Norm. Sup. Paris 19
(1986), 31-55.
\item{[Sal 3]} {\bsc  S. Salamon}, {\ninesl The Dirac operator and
quaternionic manifolds,} Proc. International Conference on
Differential Geometry and its Applications, Opava (1992).
\item{[Sal 4]} {\bsc  S. Salamon}, {\ninesl Riemannian geometry and holonomy
groups}, Pitman Research Notes in Mathematics Series, Longman Scientific \&
Technical, Essex UK, 1989.
\item{[Sw]} {\bsc A. F. Swann}, {\ninesl Hyperk\"ahler and
quaternionic K\"ahler geometry}, Math. Ann. 289 (1991), 421-450.
\item{[Rei]} {\bsc B. L. Reinhart},  {\ninesl Foliated manifolds with
bundle-like metrics}, Ann. Math. 69(2) (1959), 119-132.
\item{[Sas 1]} {\bsc  S. Sasaki}, {\ninesl On differentiable manifolds with
certain structures which are closely related to almost contact structure},
T\^ohoku Math. J. 2 (1960), 459-476.  
\item{[Sas 2]} {\bsc  S. Sasaki}, {\ninesl Spherical space forms
with normal contact metric 3-structure}, J. Differential Geom. 6 (1972),
307-315.
\item{[Sat 1]} {\bsc I. Satake}, {\ninesl On a generalization of the notion of
manifold}, Proc. Nat. Acad. Sci. U.S.A 42 (1956), 359-363.
\item{[Sat 2]} {\bsc I. Satake}, {\ninesl The Gauss-Bonnet theorem for 
$V$-manifolds}, J. Math. Soc. Japan V.9 No 4. (1957), 464-476.
\item{[Sm]} {\bsc S. Smale}, {\ninesl On the structure of 5-manifolds},
Ann. Math. 75 (1962), 38-46.
\item{[Tach-Yu]} {\bsc S. Tachibana and W. N. Yu}, {\ninesl On a
Riemannian space admitting more than one Sasakian structure},
T\^ohoku Math. J. 22 (1970), 536-540.
\item{[Tach]} {\bsc S. Tachibana}, {\ninesl On harmonic tensors in
compact Sasakian spaces}, T\^ohoku Math. J. 17 (1965), 271-284.
\item{[Tan 1]} {\bsc S. Tanno}, {\ninesl On the isometry of Sasakian
manifolds}, J. Math. Soc. Japan 22 (1970), 579-590.
\item{[Tan 2]} {\bsc S. Tanno}, {\ninesl Killing vectors on contact
Riemannian manifolds and fiberings related to the Hopf fibrations},
T\^ohoku Math. J. 23 (1971), 313-333.
\item{[Tan 3]} {\bsc S. Tanno}, {\ninesl Sasakian manifolds
with constant $\Phi$-holomorphic sectional curvature},
T\^ohoku Math. J. 23 (1969), 21-38.
\item{[Tan 4]} {\bsc S. Tanno}, {\ninesl Geodesic flows
on $C_L$-manifolds and Einstein metrics on $S^3\times S^2$}, in
{\it Minimal submanifolds and geodesics (Proc.
Japan-United States Sem., Tokyo, 1977)}, pp. 283-292, North Holland,
Amsterdam-New York, 1979.  
\item{[Thu]} {\bsc W.
Thurston}, {\ninesl The Geometry and Topology of 3-Manifolds}, Mimeographed
Notes, Princeton Univ. Chapt. 13 (1979).
\item{[Tho]} {\bsc Thomas}, {\ninesl Almost regular contact manifolds},
J. Differential Geom. 11 (1976), 521-533.
\item{[T-Y]} {\bsc G. Tian and S.-T. Yau}, {\ninesl K\"ahler-Einstein
metrics on complex surfaces with $c_1>0$}, Comm. Math. Phys. 112 (1987),
175-203.
\item{[Ud]} {\bsc C. Udri\c ste}, {\ninesl Structures presque
coquaternioniennes,} Bull. Math. de la Soc. Sci. Math. de Roumanie 12
(1969), 487-507.
\item{[W]} {\bsc H.C. Wang}, {\ninesl Closed manifolds with homogeneous complex
structures}, Am. J. Math. 76 (1954), 1-32.
\item{[Wan 1]} {\bsc M. Wang}, {\ninesl Some examples of homogeneous
Einstein manifolds in dimension seven}, Duke Math. J. 49 (1982),
23-28.
\item{[Wan 2]} {\bsc M. Wang}, {\ninesl Parallel Spinors and
Parallel Forms}, Ann. Global Anal. Geom. 7 (1989), 59-68.
\item{[Wan 3]} {\bsc M. Wang}, {\ninesl Parallel Spinors and 
Parallel Forms}, Ann. Global Anal. Geom. 13 (1995), 31-42. 
\item{[Wan 4]} {\bsc M. Wang}, {\ninesl 
Einstein metrics and quaternionic K\"ahler manifolds}, Math. Z. 210 (1992),
305-325.
\item{[Wi]} {\bsc E. Witten}, {\ninesl Anti-de Sitter space and holography},
preprint, 1998; hep-th/9802150.
\item{[Wol]} {\bsc J. A. Wolf}, {\ninesl  Complex homogeneous
contact manifolds and quaternionic symmetric spaces}, J. Math.
Mech. 14 (1965), 1033-1047.
\item{[Y-B]} {\bsc K. Yano and S. Bochner}, {\ninesl Curvature and Betti
numbers}, Annals of Math. Studies 32, Princeton University Press (1953). 
\item{[Y-K]} {\bsc K. Yano and M. Kon}, {\ninesl
Structures on manifolds}, Series in Pure Mathematics 3, 
World Scientific Pub. Co., Singapore, 1984.
\medskip
\bigskip \line{ Department of Mathematics and Statistics
\hfil September 1998} \line{ University of New Mexico \hfil} \line{ Albuquerque, NM
87131 \hfil } \line{ email: cboyer@math.unm.edu, galicki@math.unm.edu
\hfil}
\bigskip \line{ Max-Planck-Institut f\"ur Mathematik
\hfil } \line{ Gottfried Claren Strasse 26 \hfil} \line{ 53225 Bonn, Germany
\hfil } \line{ email: galicki@mpim-bonn.mpg.de
\hfil}
\end